\documentclass[%
 reprint,
 amsmath,amssymb,
 aps,
]{revtex4-2}

\usepackage{graphicx}
\usepackage{dcolumn}
\usepackage{bm}
\usepackage{color}
\usepackage{comment}
\usepackage{siunitx}
\usepackage{ulem}

\begin{document}

\preprint{APS/123-QED}

\newcommand{\uc}[1]{\textcolor{red}{#1}}
\newcommand{\ke}[0]{$\kappa_{\text{eff}}$ }
\newcommand{\ka}[0]{$\kappa_{\text{cum}}$ }

\title{Phonon Mean Free Path Spectroscopy By Raman Thermometry}



\author{Katharina Dudde}
\email{duddekat@uni-bremen.de} 
\author{Mahmoud Elhajhasan}
\author{Guillaume Würsch}
\author{Julian Themann}
\author{Jana Lierath}
\author{Gordon Callsen}
\email{gcallsen@uni-bremen.de}

\affiliation{%
	Institut für Festkörperphysik, Universität Bremen, Otto-Hahn-Allee 1, 28359 Bremen, Germany}%


\author{Dwaipayan Paul}
\author{Nakib H. Protik}

\affiliation{%
	Institut für Physik and Center for the Science of Materials Berlin (CSMB), Humboldt-Universität zu Berlin, 12489 Berlin, Germany}%


\author{Giuseppe Romano}
\affiliation{%
	MIT-IBM Watson AI Lab, 214 Main St., Cambridge, Massachusetts 02141, USA}%


\date{\today}


\begin{abstract}
	
	In this work, we exemplify on a bulk silicon sample that Raman thermometry is capable of phonon mean free path (PMFP) spectroscopy. Our experimental approach is similar to the variation of different characteristic length scales $l_{\text{c}}$ during thermal reflectance measurements in the time or frequency domain (TDTR and FDTR) and transient thermal grating (TTG) spectroscopy. In place of $l_{\text{c}}$, we vary the laser focus spot size ($w_{\text{e}}$) and the light penetration depth ($h_{\alpha}$) during one-laser Raman thermometry (1LRT) measurements, enabling control over the size of the temperature probe volume $V$. For our largest $w_{\text{e}}$ values, the derived effective thermal conductivities \ke converge towards the bulk thermal conductivity $\kappa_{\text{bulk}}$ for silicon, which we confirm by two-laser Raman thermometry and \textit{ab\,initio} theory. However, towards smaller $w_{\text{e}}$ values, we observe a pronounced increase for the \ke values, which amounts up to a factor of 5.3 at \SI{293}{\kelvin} and even 8.3 at \SI{200}{\kelvin}. We mainly assign this phenomenon to quasi-ballistic phonon transport and discuss any prominent impact of other factors. As a result, we can compare our measured $\kappa_{\text{eff}}(w_{\text{e}})$ trends with the thermal accumulation function $\kappa_{\text{cum}}$ and its dependence on the phonon mean free path $l_{\text{ph}}$, which we derive from \textit{ab\,initio} solutions of the linearized phonon Boltzmann transport equation (BTE). Since the variation of $w_{\text{e}}$ can be experimentally cumbersome, we also suggest varying $h_{\alpha}(\lambda)$ via the applied Raman laser wavelength $\lambda$ during 1LRT. In this regard, we present proof-of-principle 1LRT measurements, yielding a step-like $\kappa_{\text{eff}}(\lambda)$ trend for four different $\lambda$ values, which we also interpret in terms of quasi-ballistic phonon transport. Interestingly, we find that our $\kappa_{\text{eff}}(w_{\text{e}})$ scaling is opposite to previous TTG results, which can be explained by the actual physical quantity probed. For small $w_{\text{e}}$ or $h_{\alpha}$ values, 1LRT mimics the situation of a local and/or surfacic heat source in a large matrix, which enables probing of real local $\kappa$ values that exceed $\kappa_{\text{bulk}}$. From a theoretical point of view, this was first predicted by Chiloyan \textit{et\,.al.} \cite{Chiloyan2020}, who calculated $\kappa_{\text{eff}}(w_{\text{e}})$ for different initial phonon distributions in good agreement with our data. To show the generality of our findings, we probed $\kappa_{\text{eff}}(w_{\text{e}})$ not only for bulk silicon, but also germanium, which is in-line with our previous findings on GaN \cite{Elhajhasan2023}. Our results shall seed future PMFP spectroscopy based on 1LRT, which can directly be benchmarked against state-of-art theory to probe the effect of, e.g., any nano-structuring by comparison of $\kappa_{\text{cum}}$ trends and not only $\kappa$ values, aiming to test our understanding of the intricate phonon transport physics.

\end{abstract}

\maketitle



\section{\label{sec:intro}Introduction}

Micro- and nano-structured semiconductor devices can benefit strongly from an optimization of their thermal properties, which are key to desired functionalities \cite{Wood1988, CahillReview2014, Funahashi2021}, or supportive regarding device efficiency \cite{HeReview2017, TakiReview2019} and reliability \cite{KuballReview2016, XuBookchapter2023}. However, the requirements for any thermal optimization can be completely opposing in thermoelectric, photonic, and electronic devices, often targeting thermal conductivity $\kappa$ modifications of all materials in use. Naturally, such efforts must consider all limitations given by the particular design of each device which enables its primary functionality as, e.g., a light emitter or transistor. Any reduction in $\kappa$ for thermoelectric devices to increase the figure of merit $ZT$ can be achieved by forming nano-structured solids (e.g., layer stacks \cite{Capinski1999, KohSuperlattices2009, Luckyanova2012, Ravichandran2013}, nano-crystalline material \cite{Ghannam2023}, alloying \cite{Gurunathan2020, Zhang2020, Huang2022, Tran2022}), while increasing $\kappa$ is often a matter of thorough defect management \cite{Capinski1997, Beechem2016, Inyushkin2018, Cai2020, HuRuan2024}. 


Apparently, as long as the phonons represent the primary heat carriers, $\kappa$ is best addressed by tuning the phonon scattering mechanisms that affect their propagation and consequently $\kappa$ \cite{Elhajhasan2023}. However, the propagation of phonons represents one of the most challenging transport phenomena in solid-state physics, which is induced not only by often complicated dispersion relations and numerous scattering processes but also by the fundamentally anharmonic nature of phonons. The precise physical mechanisms that lead to a modification of $\kappa$ based on the aforementioned approaches often remain obscured. Thus, the development of experimental techniques is required in steady synchronization with advancing theory to lift the veil on phonon transport physics \cite{RegnerReview2015}.

\textit{Ab initio} theory can derive the required phonon (ph) dispersion relations and various scattering rates (e.g., ph-ph, ph-electron, ph-defect) that are needed to eventually derive $\kappa$ and even its ph-mode resolved composition \cite{Tian2011, EsfarjaniChenStokes2011, Cheng2021, Elhajhasan2023}. Often, just particle-like phonon transport is considered for this purpose, thus, phonon Boltzmann transport equations (BTE) \cite{ShengBTE2014} are utilized \cite{ShengBTE2014, ProtikElphbolt2022}. However, it is a fundamental challenge to obtain quantitative results, not only due to an inherently large set of ph-scattering mechanisms and thereby the need for, e.g., defect concentrations determined by experiments, but also due to the description of the scattering processes itself \cite{LindsayBroido2012, RavichandranBroido2020}. Hence, it would be beneficial to develop experimental techniques that can bring state-of-the-art theory to a test beyond a simple comparison of $\kappa$. Although it may appear promising when theory and experiment obtain similar absolute values of $\kappa$, it does not mean that the ph-mode-resolved accumulation described by $\kappa_{\text{cum}}$ is correctly described. However, exactly such decoding of the balance of all contributions to $\kappa$ would correspond to a true understanding of phonon transport physics, which can then form the basis for a deliberate tailoring of the thermal material properties. 

When obtained from theory, the thermal accumulation function $\kappa_{\text{cum}}$ is commonly reported over phonon energy ($E_{\text{ph}}$) or phonon mean free path ($l_{\text{ph}}$) \cite{YangDames2013}, while an assignment to a certain set of ph-modes is always still feasible. From an experimental point of view it proves challenging to directly measure $\kappa_{\text{cum}}$ depending on $E_{\text{ph}}$ or $l_{\text{ph}}$ \cite{Regner2013, RegnerReview2015}, while truly ph-mode resolved results at typical device operation temperatures remain a strong motivation for future work. In a recent review article, Regner \textit{et\,al.} \cite{RegnerReview2015} have outlined the particular relevance of $\kappa_{\text{cum}}(l_{\text{ph}})$ as being experimentally accessible, although never directly measurable, which, however, is typical for most thermal characterization techniques. Currently, there are three main benchtop experimental techniques available to probe $\kappa_{\text{cum}}(l_{\text{ph}})$, which are transient thermal grating (TTG) spectroscopy \cite{JohnsonMinnichChen2013, Johnson2012, Johnson2011} as well as time- \cite{Minnich2011, Cheng2021, Hu2015, Highland2007, DingChen2014, JiangKoh2016} and frequency domain \cite{Regner2013, RegnerMalen2013, Freedman2013, Schmidt2009} thermal reflectance (TDTR and FDTR) measurements. All of these techniques carry their own particular advantages and disadvantages for the desired phonon mean free path (PMFP) spectroscopy, which mainly encompasses the role of a metal transducer, the achievable spatial resolution, and the overall level of experimental complication. Generally, all three experimental techniques determine effective thermal conductivities $\kappa_{\text{eff}}$, e.g., based on the heat diffusion equation (generalization of Fourier's law) depending on characteristic experimental length scales $l_{\text{c}}$ \cite{RegnerReview2015, ZengCollins2015}. Towards the upper limits of $l_{\text{c}}$, the measured $\kappa_{\text{eff}}(l_{c})$ trends can - for the most simplistic case of bulk materials - converge against bulk thermal conductivities $\kappa_{\text{bulk}}$, mimicking a broadened step function, which is also known as the suppression function \cite{Maznev2011, Minnich2012, YangDames2013, Regner2014}. For TTG $l_{\text{c}}$ is given by the thermal grating period $l_{\text{g}}$ (TTG), while for TDTR and FDTR measurements the heat and probe laser spot sizes $w_{\text{e}}$ and/or the thermal penetration depth $l_{\text{p}}$ can be varied \cite{RegnerReview2015}. In a best-case scenario, the measured $\kappa_{\text{eff}}(l_{c})$ trends can eventually be benchmarked against the theoretically derived $\kappa_{\text{cum}}(l_{\text{ph}})$ trends to test our understanding of phonon transport physics.

In this work, we demonstrate that one-laser Raman thermometry (1LRT), a well-known and from an experimental point of view rather simple technique \cite{BeechemInvArticle2007, BeechemReview2015}, is capable of PMFP spectroscopy. To establish our technique, we first analyze highly resistive bulk silicon (111) and silicon membranes before confirming our findings also on bulk germanium. We show that our measured $\kappa_{\text{eff}}$ values strongly depend on the laser spot size $w_{\text{e}}$ (here defined as the radius where the laser intensity has dropped to $1/e$-level) and the used Raman wavelength $\lambda$. Varying $\lambda$ provides precise and convenient control over the light penetration depth $h_{\alpha}$, which in turn affects the depth of the heated (and probed) volume similar to $l_{\text{p}}$ in TDTR \cite{KohCahill2007} and FDTR \cite{Freedman2013} measurements. When performing Raman thermometry with a continuous wave (cw) laser, we measure local temperature rises $T_{\text{rise}}$ via the shift $\nu$ of the optical phonon mode of silicon, which are smaller than the $T_{\text{rise}}$ values one would expect according to Fourier's law. When we apply such a diffusive model to our measurements (full numerical description of the experimental situation similar to Refs. \cite{Elhajhasan2023, Seemann2024}), we extract $\kappa_{\text{eff}}$ values for bulk silicon at ambient temperatures $T_{\text{amb}}$ of \SI{293}{\kelvin} and \SI{200}{\kelvin} that exceed $\kappa_{\text{bulk}}$ by up to a factor of 8.3 (is a factor of 5.3 at \SI{293}{\kelvin}) for our smallest $w_{\text{e}}$ value (i.e., \SI{0.34(1)}{\micro\meter}). 


In a simple picture, we explain our measured $\kappa_{\text{eff}}(w_{\text{e}})$ trends by phonons that escape from our temperature probe volume during our steady-state Raman thermometry, which is a sign of quasi-ballistic phonon transport. Towards smaller $w_{\text{e}}$ values, 1LRT probes increasingly lower $T_{\text{rise}}$ values and consequently $\kappa_{\text{eff}} \, \gtrsim \, \kappa_{\text{bulk}}$ is derived from Fourier's law. Only for our largest $w_{\text{e}}$ values (i.e., \SI{16.7(4)}{\micro\meter}) we start to approach $\kappa_{\text{bulk}}$ as reported in the literature, which we also confirm by two-laser Raman thermometry (2LRT, \cite{Hsu2LRToriginal2011, Reparaz2014, Elhajhasan2023, Seemann2024}) and our \textit{ab initio} solution of the linearized phonon Boltzmann transport equation. Interestingly, for TTG, FDTR, and TDTR measurements an increasingly non-diffusive thermal transport (e.g. ballistic transport) is accompanied by lower measured $\kappa_{\text{eff}}$ values \cite{RegnerReview2015}, which is opposite for 1LRT measurements for the range of $w_{\text{e}}$ values employed in this study. Furthermore, our measured variation of $\kappa_{\text{eff}}$ is large compared to most alternative PMFP spectroscopy \cite{KohCahill2007, JohnsonMinnichChen2013, RegnerMalen2013, DingChen2014, WilsonCahill2014, Vermeersch2015}, making 1LRT measurements a sensitive option for exploring the impact of $l_{\text{ph}}$ on $\kappa$.

In a careful step-by-step approach, we discuss potentially alternative explanations for our measurements related to, e.g., the build-up of heat-induced stress \cite{BeechemInvArticle2007} and the role of non-thermal phonon distributions \cite{Vallabhaneni2016, Sullivan2017, Wang2020}. We wish to point out that our measured $\kappa_{\text{eff}}(w_{\text{e}})$ trends also oppose previously reported trends on two-dimensional materials like graphene \cite{Sullivan2017} and MoS$_2$ \cite{Wang2020}, which points to a reduced impact of any thermal non-equilibrium for our findings. By comparing our experimental results with the theoretical results of Chiloyan \textit{et\, al.} \cite{Chiloyan2020} - who did not only consider non-thermal phonon distributions but also non-diffusive thermal transport - we can show that our measured $\kappa_{\text{eff}}$ values are a signature for a local heating situation provided by a non-thermal heat source and large $l_{\text{ph}}$ values. In addition, we used density functional theory (DFT) to calculate $\kappa_{\text{cum}}(l_{\text{ph}})$ for bulk silicon, aiming at a first comparison to our experimental data. 

Finally, we can conclude that 1LRT provides a promising perspective to access information about the different $l_{\text{ph}}$ ranges of phonons that contribute to thermal transport, making these measurements an interesting alternative to TTG, TDTR, and FDTR. Overall, it is advantageous that 1LRT does not require a metal transducer and is enabled by comparably simple spectroscopic setups, while the spatial resolution is still high compared to TTG even for our largest $w_{\text{e}}$ values. However, harvesting these advantages of 1LRT comes at some experimental costs related to highly spectrally resolved and well-stabilized Raman spectroscopy (spectrally and mechanically), which explains the scarcity of similar reports in the literature and is not straightforwardly achievable with simple out-of-the-box equipment. Furthermore, our findings shed light on the tendency of 1LRT to overestimate $\kappa_{\text{bulk}}$ values, which is especially true for high-quality samples with higher upper boundaries for $l_{\text{ph}}$ of phonons that still carry significant fractions of heat (i.e., so-called, thermal phonons). During 1LRT measurements, one has to devote great care to control the size of the heated volume in the sample, which means that $w_{\text{e}}$ and the Raman wavelength $\lambda$ of choice (connected to $h_{\alpha}$) are no longer always free parameters as shown in this work. However, what appears as a restriction at first sight is actually the promising pathway toward PMFP spectroscopy by Raman thermometry.

The paper is structured as follows. In Sec.\,\ref{sec:exptSetup}, we first provide our sample details (Sec.\,\ref{sec:samples}), before explaining our experimental setup, all methods, and the details of the Raman thermometry (1LRT and 2LRT) that is described in this work. The 1LRT technique and setup, as well as the required temperature calibration procedure are described in Sec.\,\ref{sec:1LRT}, while the details of the 1LRT measurements, in which either the laser focus spot size $w_{\text{e}}$ or the light penetration depth $h_{\alpha}$ were varied, are given in Secs.\,\ref{sec:vary-we} and \ref{sec:vary-ha}, respectively. It follows a short explanation of the 2LRT setup and the related measurement principle (Sec.\,\ref{sec:2LRT}). Sec.\,\ref{sec:datAn} deals with the extraction of the \ke values from 1LRT measurements on bulk silicon (Sec.\,\ref{sec:COMSOL-bulk}) and silicon membrane samples (Sec. \ref{sec:COMSOL-membrane}) based on finite element modeling to solve the steady-state heat equation. In Sec. \ref{sec:results-var-we} we present our $\kappa_{\text{eff}}(w_{\text{e}})$ trends measured for bulk silicon and discuss any effect of heat-induced stress and non-thermal phonon distributions on these results in Sec.\,\ref{sec:Eff-on-nu}, before we motivate that our results can only be explained when quasi-ballistic phonon transport is considered (Sec.\,\ref{sec:ballist-transp}), which forms the basis for the desired PMFP spectroscopy. The next Sec.\,\ref{sec:comp-theory} presents the $\kappa_{\text{cum}}(l_{\text{ph}})$ trends that we calculated for bulk silicon. Subsequently, we discuss the advantages and limitations of comparing the experimental $\kappa_{\text{eff}}(w_{\text{e}})$ and the theoretical $\kappa_{\text{cum}}(l_{\text{ph}})$ trends by applying a straightforward normalization (Sec.\,\ref{sec:norm-comp}) or by calculating the derivative of both trends (Sec.\,\ref{sec:derivative-comp}).
Our experimental $\kappa_{\text{eff}}(h_{\alpha})$ trends for our silicon membrane samples are presented and discussed in Sec.\,\ref{sec:results-var-ha}, which represents a second possibility for PMFP spectroscopy by 1LRT. Finally, we summarize our work in Sec.\,\ref{sec:sumout}. In addition to the present manuscript, more technical details and supporting information are provided in the Supplemental Material.

\section{\label{sec:exptSetup}Experimental details}

\begin{figure*}[]
	
	\includegraphics[width=\linewidth]{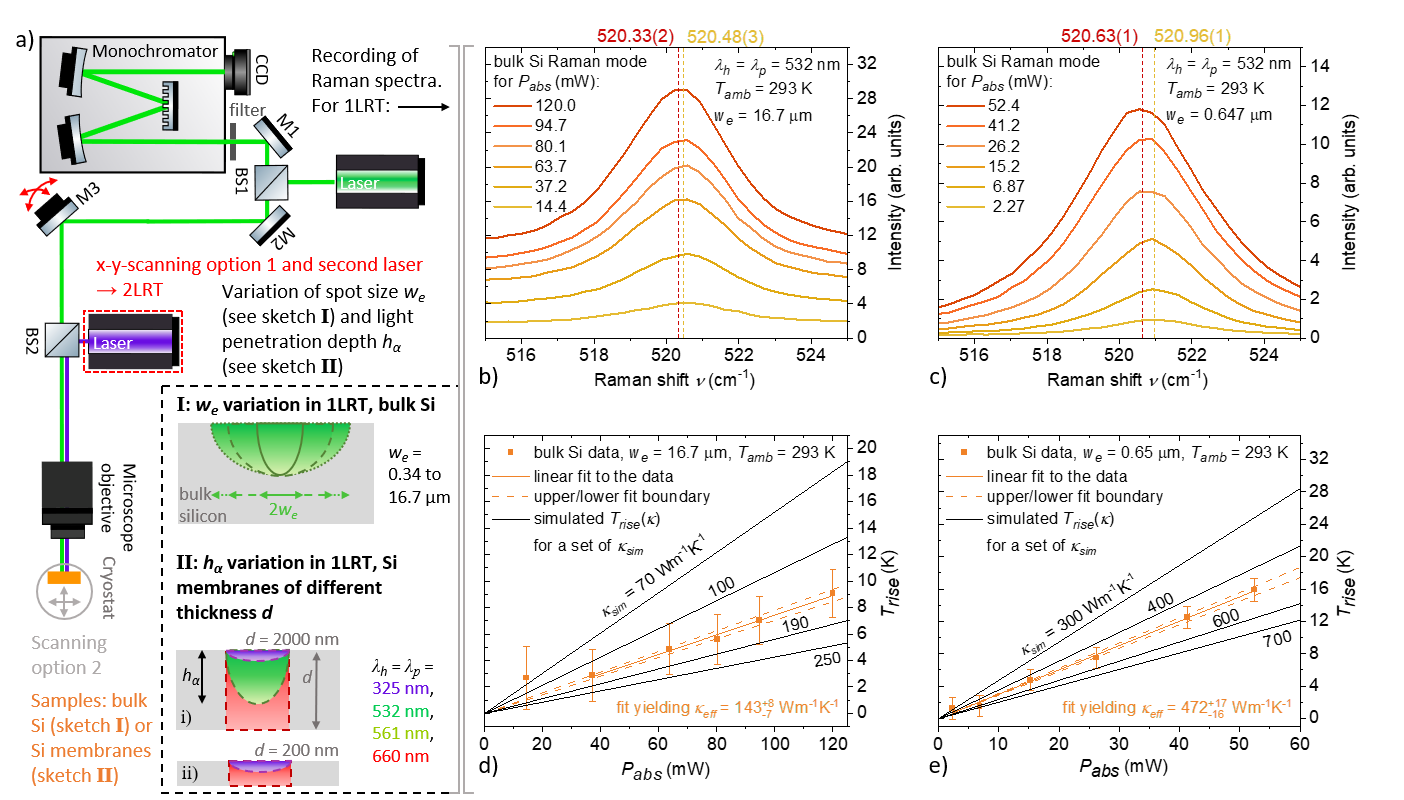}
	
	\caption{a): Scheme of the fully customized Raman setup capable of performing, e.g., 1LRT and 2LRT measurements. For 1LRT measurements the two characteristic length scales $w_{\text{e}}$ and $h_{\alpha}$ can be varied. For our bulk silicon sample $w_{\text{e}}$ is varied between $<\SI{0.5}{\micro\meter}$ and $\approx \SI{17}{\micro\meter}$ (see sketch \MakeUppercase{\romannumeral 1}), while also different ambient temperatures $T_{\text{amb}}$ were targeted (\SI{293}{\kelvin} and \SI{200}{\kelvin}). As denoted, the light penetration depth $h_{\alpha}$ was varied based on four different Raman laser wavelengths $\lambda$, which we applied for our 1LRT on silicon membranes of two different thicknesses $d\,=\,\SI{200}{\nano\meter}$ and $\SI{2000}{\nano\meter}$ at \SI{293}{\kelvin}. A drawing of the corresponding experimental situation is given in sketch \MakeUppercase{\romannumeral 2} of a). For 1LRT the heat ($\lambda_{\text{h}}$) and probe ($\lambda_{\text{p}}$) wavelengths are identical, which changes for 2LRT. In b) and c) we show the most prominent Raman mode of bulk silicon for a 1LRT series of absorbed powers $P_{\text{abs}}$ at \SI{293}{\kelvin} ($\lambda_h = \lambda_p = \SI{532}{\nano\meter}$). Power series b) and c) were recorded for largely deviating $w_{\text{e}}$ values. In d) and e) we show the resulting trends of the rising temperatures $T_{\text{rise}}$ over $P_{\text{abs}}$, which correspond to power series b) and c), respectively. Based on the measured Raman shift $\nu$ and a suitable temperature calibration $T(\nu)$ we can obtain $T_{\text{rise}}$ values that linearly increase with $P_{\text{abs}}$. The experimental $T_{\text{rise}}(P_{\text{abs}})$ slopes (orange solid lines) are compared to our numerical simulations (examples shown as black solid lines) based on a set of thermal conductivities $\kappa_{\text{sim}}$, yielding different experimental thermal conductivities $\kappa_{\text{eff}}$ for deviating $w_{\text{e}}$ values. The orange dashed lines illustrate the confidence interval for our linearly fitted data, which yields asymmetric error bars for $\kappa_{\text{eff}}$ (sub- and superscript of $\kappa_{\text{eff}}$).}
	\label{fig:expt}
	
\end{figure*}


%
%
\subsection{Bulk and membrane samples}\label{sec:samples}
%
%

Our bulk silicon sample is a non-intentionally doped, one-side polished, float-zone silicon wafer (two inch) with a specific resistance of $>\,$\SI{8000}{\ohm\centi\meter} and a (111)-orientation (Crystal GmbH). The two silicon membrane samples are mono-crystalline, (100)-oriented, and exhibit a thickness $d$ of either $200\,\text{nm}$ or $2000\,\text{nm}$ (Silson). The bulk germanium sample mostly presented in the Supplemental Material of this article is (110)-oriented, one-side polished, and n-type doped (dopant: Sb, specific resistivity of $>\,$\SI{30}{\ohm\centi\meter}) float-zone germanium wafer (provider: Crystal GmbH, Germany).



%
%
\subsection{One-laser-Raman thermometry}\label{sec:1LRT}
%
%

Our fully customized Raman setup is shown schematically in Fig. \ref{fig:expt}a. A more detailed drawing can be found in the Supplemental Material of Ref. \cite{Elhajhasan2023}. For 1LRT measurements, we use one laser (green in the scheme) that locally heats the sample and probes the temperature simultaneously. Thus, for 1LRT the same heat and probe laser wavelengths $\lambda_{\text{h}}\,=\,\lambda_{\text{p}}$ are used for Raman spectroscopy, a situation that changes for 2LRT. As shown in Fig. \ref{fig:expt}a the light beam is guided via suited beam splitters (BS) and mirrors (M) towards a microscope objective of choice, which is optimized for the ultraviolet (UV) and/or visible (VIS) spectral range (see Sec.\,\ref{sec:vary-we} for details). This microscope objective is mounted on a three-axes stage (piezo- and motor-driven) to enable precise and reproducible focus control with a step size below the depth of view of our microscope objectives (we apply a step size of down to 100\,nm), which is key to the results of this work. The Raman signal from the sample is collected in back-scattering configuration and is routed towards the single monochromator as depicted in Fig. \ref{fig:expt}a. A suitable Raman filter is installed in front of the monochromator to suppress the Rayleigh-scattered laser light, enabling the recording of Raman spectra (Stokes-side) on a charge-coupled device (CCD) that was nitrogen-cooled (Roper Scientific, SPEC-10, 2048\,$\times$\,512 pixel, UV-enhanced).

In the present work, it is important to maximize the spectral and partially interconnected mechanical stability of the Raman setup in use, which is one of several reasons why we developed our own customized setup. First, the focus position of the microscope objective must be stable over extended time periods because of the long integration times that are required for the series of spectra in need. After extended stability tests, we consider that a focus position stability of $\approx\,\pm\,50\,\text{nm}$ over up to 8\,h is sufficient for the desired spectroscopy. In a best-case scenario, any lateral drift over the sample surface is also suppressed to a similar degree (less relevant for 1LRT due to extended homogeneous sample surfaces, but relevant for 2LRT). For our 1LRT and 2LRT measurements we aim to resolve changes of the Raman shift $\nu$ down to $\approx\,0.02\,\text{cm}^{-1}$, which is not achieved by extremely high-resolution equipment but a rather solid temperature stability of key laboratory components as, e.g. explained in Ref. \cite{Fukura2006}. Our single monochromator (iHR 550 from Horiba, focal length of 55\,cm) was equipped with a 1800\,l/mm grating (holographic, 450\,-\,850\,nm), which enables a spectral resolution of $\approx\,1\,\text{cm}^{-1}$ in the VIS range (first grating order). Because our laboratory's temperature stability over 8\,h only reaches $\approx\,\pm\,0.5\,^{\circ}\text{C}$ in a best-case scenario, we also actively temperature stabilized key components of our setup to better than $\approx\,\pm\,0.1\,^{\circ}\text{C}$ by raising their temperature by $\approx\,2\,^{\circ}\text{C}$ over the laboratory temperature and apply suitable heaters, sensors, and regulation circuitry.

For any 1LRT analysis we measure, so-called power series of Raman spectra. The power of the heat laser is varied in suitable steps, and for each step a Raman spectrum is recorded. During the acquisition of the Raman spectra, the sample is placed in the vacuum chamber (evacuated to $\leq\,$1\,$\times$\,10$^{-6}$\,mBar) of a closed-cycle He-cryostat (Montana Instruments), allowing the stabilization of $T_{\text{amb}}$ at 293\,K or 200\,K. Inside of the He-cryostat a three-axes low-temperature compatible piezo-stage is used for positioning the sample. If the reflectivity and transmissivity of the optical elements and the sample are known, the absorbed power $P_{\text{abs}}$ or even the power that leads to the sample heating $P_{\text{heat}}$ can be calculated from the measured laser power. The equality of $P_{\text{heat}}$ and $P_{\text{abs}}$ holds for the case of silicon (indirect semiconductor) at $T_{\text{amb}} \geq \SI{200}{\kelvin}$, while for strongly luminescent samples with high quantum efficiencies even at $T_{\text{amb}}\,=\,293\,\text{K}$ this equality is not always fulfilled \cite{Seemann2024}. A detailed analysis of the power balance in this case can be found in the Supplemental Material of Ref.\cite{Elhajhasan2023}.

For our silicon samples, we monitor the Raman active optical phonon mode close to the $\Gamma$-point of the Brillouin zone, which is situated around $\nu\,=\,\SI{520}{\per\centi\meter}$ (bulk silicon) in our Raman spectra, cf.  Figs. \ref{fig:expt}b and c. This Raman mode redshifts with increasing temperature on the relative scale of the Raman spectrum (Stokes side). 
For a known temperature calibration $T(\nu)$ the local temperature $T$ of the sample can be extracted from the Raman shift $\nu$ as a function of $P_{\text{abs}}$. We record individual $T(\nu)$ trends in a customized vacuum heat stage (Instec) for most of our samples. For $T_{\text{amb}}\,=\,293\,\text{K}$ we often found that we cannot simply use already published $T(\nu)$ trends \cite{Menendez1984} as frequently applied in literature, meaning that, e.g., for our silicon membranes we recorded high-resolution $T(\nu)$ data with temperature steps down to \SI{2}{\kelvin}. For our bulk silicon sample at $T_{\text{amb}} = \SI{293}{\kelvin}$ we also measured $T(\nu)$ in a heat stage, while for $T_{\text{amb}} = \SI{200}{\kelvin}$ we used literature data \cite{Menendez1984} of sufficient quality. For further details regarding the temperature calibration, see S-Sec. I in the Supplemental Material.

\subsubsection{Variation of the laser focus spot size ($w_{\text{e}}$)}\label{sec:vary-we}

In a first set of 1LRT measurements on bulk silicon at $T_{\text{amb}} = \SI{293}{\kelvin}$, $w_{\text{e}}$ (radius where the laser intensity has dropped to $1/e$-level) was varied from $\SI{0.34(1)}{\micro\meter}$ to $\SI{16.7(4)}{\micro\meter}$ (note: the number in brackets indicates the error in the last displayed digit). We vary $w_{\text{e}}$ by using microscope objectives with different magnifications (2$\times$\,-\,100$\times$), which means that Raman spectra are always recorded in focus. We refrain from defocusing as an alternative experimental approach to vary $w_{\text{e}}$, because this may change the in-coupling of the Raman-scattered light into the monochromator, and therefore introduces changes to the Raman signal (e.g., variation of Raman mode broadenings) \cite{Xu_OutOfFocus_2013, XuReviewRaman2020}. The corresponding aperture-matched imaging lens in front of our single monochromator is not depicted in the experimental sketch of Fig. \ref{fig:expt}a. Sketch \MakeUppercase{\romannumeral 1} in Fig. \ref{fig:expt}a illustrates how the temperature probe volume depicted in green changes in the sample with $w_{\text{e}}$. Similar 1LRT measurements were repeated at $T_{\text{amb}}\,=\,200\,\text{K}$. The knife-edge method was used to measure $w_{\text{e}}$ as further explained in S-Sec. VIII and in Ref. \cite{Elhajhasan2023}. To resolve sufficiently small, heating-induced changes of $\nu$ we aimed for excellent signal-to-noise ratios as shown in Fig. \ref{fig:expt}b and c, which is - even for silicon known for its strong Raman signal - not straightforward to achieve at our lowest $P_{\text{abs}}$ values with, e.g. a 2$\times$ magnification of a Plan Apo microscope objective (Mitutoyo, infinity corrected) exhibiting a numerical aperture (N.A.) of just 0.055. A similar statement is true for our UV-Raman spectroscopy with $\lambda_{\text{h}}\,=\,\lambda_{\text{p}}\,=\,325\,\text{nm}$, which explains the need for a high mechanical and spectral stability of our customized Raman setup, cf. Sec.\,\ref{sec:1LRT}.


%
%
\subsubsection{Variation of the light penetration depth ($h_{\alpha}$)}\label{sec:vary-ha}

In a second set of 1LRT measurements at $T_{\text{amb}} = \SI{293}{\kelvin}$, the material-dependent light penetration depth $h_{\alpha}$ (defined as the depth where the laser light intensity has dropped to $1/e$-level) was tuned via the wavelength ($\lambda$) of the heat ($\lambda_{\text{h}}$) and the temperature probe laser ($\lambda_{\text{p}}$), where for 1LRT measurements $\lambda_{\text{h}} = \lambda_{\text{p}}$. We performed our $\lambda$-dependent 1LRT analyses on silicon membranes based on four different Raman lasers yielding a variation of $h_{\alpha}$ from $<\SI{10}{\nano\meter}$ to $>\SI{4}{\micro\meter}$ in silicon. Here, the selection of $h_{\alpha}$ values was limited only by the available Raman wavelengths in our setup. A picture of the resulting experimental situation is illustrated in \MakeUppercase{\romannumeral 2} of Fig. \ref{fig:expt}a. Silicon membranes of two different thicknesses $d = \SI{2000}{\nano\meter}$ and $\SI{200}{\nano\meter}$ were used to show that the extraction of \ke is affected by the ratio between $h_{\alpha}$ and $d$.

%
%
\subsection{Two-laser-Raman thermometry}\label{sec:2LRT}
%
%

Our setup is also capable of 2LRT measurements, which will prove useful, e.g., for determining quantitative $\kappa_{\text{bulk}}$ values. Therefore, a second laser (purple in the scheme shown in Fig. \ref{fig:expt}a) is added to heat the sample.
This second laser can be independently positioned on the sample surface with respect to the temperature probe laser (green). Frequently, we perform our 2LRT measurements with $\lambda_h\,=\,325\,\text{nm}$ and $\lambda_p\,=\,532\,\text{nm}$. Mirror M3 in Fig. \ref{fig:expt}a can be rotated around two axes, which in addition to a 4f-lens configuration (not shown in Fig. \ref{fig:expt}a) allows us to scan the probe laser over the sample surface. Positioning the heat laser on the sample surface can be performed via the piezo stage in our cryostat or the three-axes stage that we use for mounting our microscope objectives. As a result, we can record spatially resolved temperature maps $T(x,y)$ while $P_{\text{abs}}$ is kept constant. Consequently, such 2LRT measurements enable a larger temperature probe volume compared to 1LRT measurements, as further explained in Ref.\,\cite{Elhajhasan2023} treating 1LRT and 2LRT on GaN membranes.


\section{Determination of the effective thermal conductivities}\label{sec:datAn} 

Figs. \ref{fig:expt}b and c show 1LRT power series for bulk silicon at $T_{\text{amb}} = \SI{293}{\kelvin}$ recorded with $\lambda_h = \lambda_p = \SI{532}{\nano\meter}$ for two strongly deviating $w_{\text{e}}$ values (i.e. \SI{16.7}{\micro\meter} and \SI{0.647}{\micro\meter}). As a result, Figs. \ref{fig:expt}d and e show the matching trends of $T_{\text{rise}}$ over $P_{\text{abs}}$. To obtain these plots, the Raman mode is fitted with a Voigt function (plus background) considering the instrument broadening of the spectrometer, which we measure for all slit settings in use. From our temperature calibration $T(\nu)$ we know that $T_{\text{rise}}$ depends linearly on $\nu$ for sufficiently low temperature rises, in line with the literature \cite{Reparaz2014, Stoib2014}. We remain within such limits for all our 1LRT measurements. Based on $T(\nu)$ we obtain $T_{\text{rise}}$ values that increase linearly with $P_{\text{abs}}$, cf. Figs. \ref{fig:expt}d and e. From our finite element modeling using the COMSOL Multiphysics\textsuperscript{\circledR} software (solution of the heat transport equation for the given experimental situation, see Secs.\,\ref{sec:COMSOL-bulk} and \ref{sec:COMSOL-membrane}), we obtain a simulated linear $T_{\text{rise}}(P_{\text{abs}})$ trend for each thermal conductivity $\kappa_{\text{sim}}$ that serves as an input value for the simulations. Thus, despite our efforts to obtain a realistic description of the experimental situation, we are still bound to a purely diffusive description of thermal transport if it comes to determining experimental, effective thermal conductivities $\kappa_{\text{eff}}$.

Finally, by comparing the experimental and the simulated $T_{\text{rise}}(P_{\text{abs}})$ trends we extract the desired $\kappa_{\text{eff}}$ values as illustrated in Figs. \ref{fig:expt}d and e. Please note that Figs. \ref{fig:expt}d and e only show a small subset of our simulated $T_{\text{rise}}(P_{\text{abs}})$ trends. Using sufficiently small steps for $\kappa_{\text{sim}}$ in our simulations, we can obtain a best fit to our experimental data. The errors of the \ke values that are derived by this method are determined by considering the uncertainty of the slope of the $T_{\text{rise}}(P_{\text{abs}})$ trends that arises from a least-square linear fit to the experimental data shown in Figs. \ref{fig:expt}d and e. Interestingly, for a set of $w_{\text{e}}$ values (compare inset of Fig. \ref{fig:expt}d vs. e) we obtain strongly deviating \ke values, even though we aim to obtain a realistic model of our experimental situation within the aforementioned limitation of a purely diffusive description of phonon transport. Please see Secs.\,\ref{sec:COMSOL-bulk} and \ref{sec:COMSOL-membrane} for further details regarding our numerical modeling for bulk and membrane samples, respectively.

%
%
\subsection{COMSOL Multiphysics\textsuperscript{\circledR} simulations for 1LRT measurements on a bulk sample}\label{sec:COMSOL-bulk}
%
%

In the simulation for the bulk samples, the following three-dimensional steady-state heat equation is solved
\begin{equation}\label{eq:heat_eq}
	-\frac{Q(x,y,z)}{\kappa_{\text{sim}}} = (\vec{\nabla})^2\,T(x,y,z),
\end{equation}
which allows computing a three-dimensional temperature profile $T(x,y,z)$ for a modeled geometry and a heat source density $Q(x,y,z)$ that is determined by the spatial laser intensity profile. In our model, we aim to consider all relevant experimental quantities like $w_{\text{e}}$, $h_{\alpha}$, $T_{\text{amb}}$, and the surrounding vacuum. Effects related to thermal radiation are neglected due to our low experimental $T_{\text{amb}}$ and $T_{\text{rise}}$ values. Details regarding the boundary conditions and modules used in our model can be found in the Supplemental material of Ref. \cite{Elhajhasan2023}. The heat source density is defined as
\begin{equation}\label{eq:heat_source_density}
	Q(r, z) = \frac{\displaystyle P_{\text{abs}}}{\displaystyle \pi\,w_{\text{e}}^2 h_{\alpha}}e^{-\frac{\displaystyle (r - r_0)^2}{\displaystyle w_{\text{e}}^2}}\,e^{-\frac{\displaystyle z - z_0}{\displaystyle h_{\alpha}}},
\end{equation}
depending on the radial coordinate $r = \sqrt{x^2 + y^2}$, with the distance $r - r_0$ from the heat spot position $r_0$ and the coordinate $z$ (perpendicular to the sample surface positioned at $z_0$). The light penetration depth in bulk silicon at $\lambda\,=\,$\SI{532}{\nano\meter} is assumed to yield $h_{\alpha} = \SI{1.1}{\micro\meter}$ for $T_{\text{amb}} = \SI{293}{\kelvin}$ and $h_{\alpha} = \SI{1.5}{\micro\meter}$ for $T_{\text{amb}} = \SI{200}{\kelvin}$ according to Ref. \cite{Franta2019}. Details about the impact of the absorption coefficient used to calculate $h_{\alpha} (\lambda)$ on the extracted $\kappa_{\text{eff}}$ values can be found in S-Sec. II.

The modeled temperature $T_m$ that corresponds to the temperature probed by 1LRT is a mean of all local temperatures $T(r,z)$ weighted by the laser intensity profile, while also the declining light intensity along the $z$-axis must be considered according to Beer-Lambert's law. We implement this based on the following formula:
\begin{equation}\label{eq:Tm_bulk}
	T_m = \frac{\displaystyle \int\int\int_{V} T(r,z) e^{-\frac{\displaystyle (r - r_0)^2}{\displaystyle w_{\text{e}}^2}}\,e^{-\frac{\displaystyle 2\,(z - z_0)}{\displaystyle h_{\alpha}}}\,dV}{\displaystyle \int\int\int_{V}e^{-\frac{\displaystyle (r - r_0)^2}{\displaystyle w_{\text{e}}^2}}\,e^{-\frac{\displaystyle 2\,(z - z_0)}{\displaystyle h_{\alpha}}}\,dV}
\end{equation}\label{eq:T_m}
$Q$ and $T_m$ are defined in a different way if it comes to the evaluation of 1LRT measurements on our membrane samples, which is described in the next Sec.\,\ref{sec:COMSOL-membrane}.

%
%
\subsection{COMSOL Multiphysics\textsuperscript{\circledR} simulations for 1LRT measurements on a membrane sample}\label{sec:COMSOL-membrane}
%
%

Compared to the extraction of \ke for bulk material (see Sec.\,\ref{sec:COMSOL-bulk}), our COMSOL Multiphysics\textsuperscript{\circledR} model for 1LRT measurements on a membrane sample differs. The main reason for that is experimentally motivated, because we want to vary $\lambda$ and not $w_{\text{e}}$ during our 1LRT measurements on different silicon membranes. For example, for $\lambda = \SI{325}{\nano\meter}$ (case A) the light penetration depth for bulk silicon at room temperature only yields $h_{\alpha} \approx \SI{10}{\nano\meter}$ \cite{Franta2019}. To account for this, a surfacic (two-dimensional) heat source density is assumed in the COMSOL model, where all the heat is absorbed in an infinitesimal shallow volume at the surface. Implementing a three-dimensional heat source for such a surfacic heating situation would be possible at the cost of an extremely fine meshing in our continuum model, which is not practical. For a subset of Raman wavelengths in use ($\lambda =\,$ 532, 561, and \SI{660}{\nano\meter}), the three-dimensional heat source density is implemented in a heated volume with a depth $d$ (case B), which corresponds to the membrane thickness. In both cases A and B, no dependence of $h_{\alpha}$ is implemented, because due to multiple internal reflections the exact $z$-dependence of $Q$ remains challenging to model as the interface roughness must be considered. This approach is in-line with the typical procedure found in literature \cite{ChavezAngel2014}. To determine $P_{\text{abs}}$ for the membranes at each applied value of $\lambda$, the transmitted and reflected laser powers were individually measured for all membrane samples. Thus for the particular case of membranes, also $T_m$ from Eq.\,\ref{eq:T_m} can be defined without any dependence on $h_{\alpha}$. The modeled temperature probe volume $V$ is defined with a depth of $\approx \SI{10}{\nano\meter}$ for $\lambda = \SI{325}{\nano\meter}$ to mimic the surfacic heating and temperature probing for case A, while for all other $\lambda$ values the depth of $V$ equals the membrane thickness $d$ (case B). The explicit formulas for $Q$ and $T_\text{m}$ can be found in S-Sec. III.


\section{\label{sec:results-var-we}Impact of the laser focus spot size on the effective thermal conductivities}

\begin{figure}[]
	
	\includegraphics[width=\linewidth]{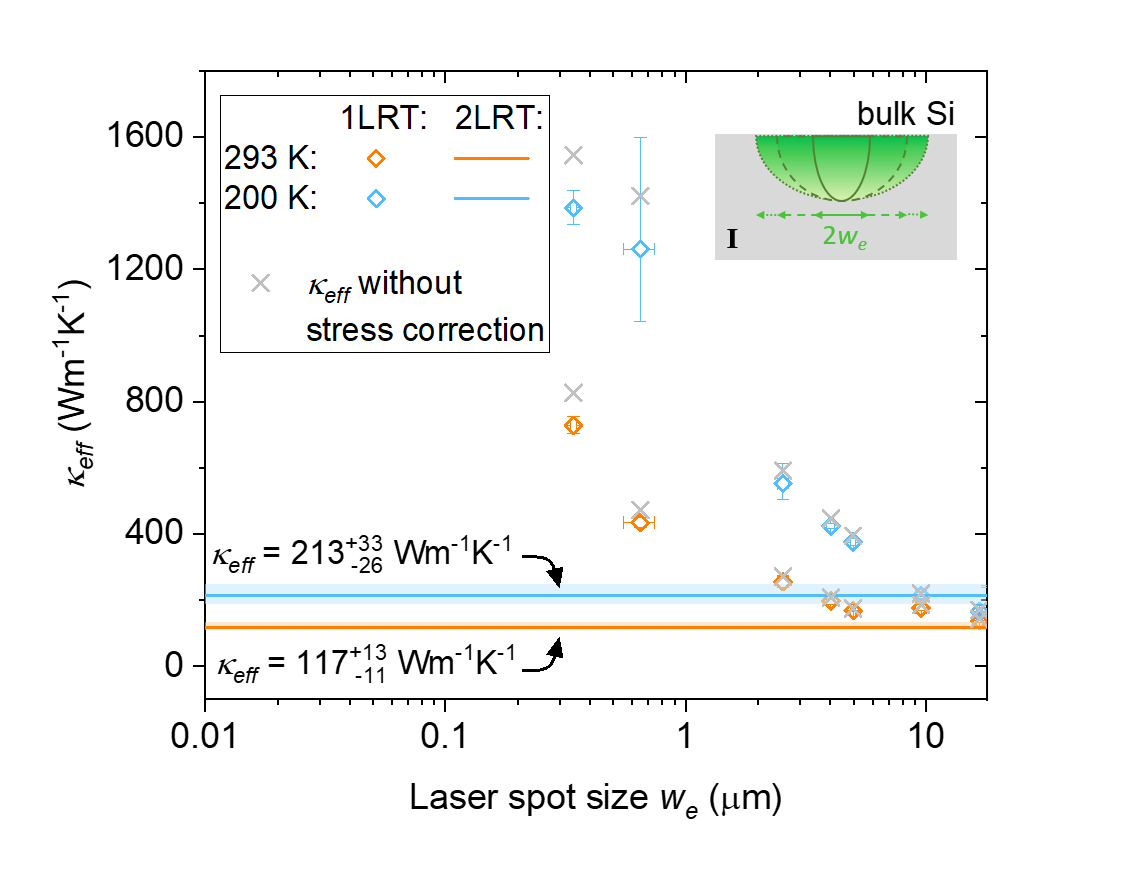}
	
	\caption{Plot of the effective thermal conductivities \ke as a function of $w_{\text{e}}$ extracted from 1LRT measurements on our bulk silicon sample at $T_{\text{amb}} = \SI{293}{\kelvin}$ (orange symbols) or $\SI{200}{\kelvin}$ (blue symbols). Here, we already consider the impact of heat-induced stress as described in Sec.\,\ref{sec:Eff-on-nu}. The gray crosses show \ke values that do not account for the built-up of heat-induced stress during 1LRT. The \ke values resulting from 2LRT are depicted as horizontal lines in matching colors. Please see S-Sec. IV for detailed information about the extraction of \ke from 2LRT. $\kappa_{\text{eff}}(w_{\text{e}})$ clearly declines with rising $w_{\text{e}}$ values, due to increasing relevance of quasi-ballistic phonon transport with decreasing laser spot size $w_{\text{e}}$, which corresponds to the size of the heater in our experiments. The overall $\kappa_{\text{eff}}(w_{\text{e}})$ trend becomes even more pronounced at $T_{\text{amb}} = \SI{200}{\kelvin}$, where the relevance of thermal phonons with larger $l_{\text{ph}}$ values is increased.}
	\label{fig:expt-result}
	
\end{figure}

Fig. \ref{fig:expt-result} shows a summary of our experimental results for \ke in dependence of $w_{\text{e}}$. We extracted these \ke values from 1LRT measurements on our bulk silicon sample at ambient temperatures of $T_{\text{amb}} = \SI{293}{\kelvin}$ and \SI{200}{\kelvin}, respectively. Based on Fig. \ref{fig:expt-result}, we can observe clear trends for $\kappa_{\text{eff}}(w_{\text{e}})$. For higher values of $w_{\text{e}}$, the measured \ke values converge toward the bulk thermal conductivity $\kappa_{\text{bulk}}$ of silicon at a given $T_{\text{amb}}$, which we can determine by our 2LRT measurements. Most interestingly, for our smallest $w_{\text{e}}$ value of $\SI{0.34(1)}{\micro\meter}$ the 1LRT analyses yield \ke values that exceed $\kappa_{\text{bulk}}$ by approximately a factor of 5.3 at $T_{\text{amb}} = \SI{293}{\kelvin}$ and even up to 8.3 at \SI{200}{\kelvin}, despite our best efforts to model the experimental situation, cf. Sec.\,\ref{sec:datAn}.

The 2LRT measurements should be independent of the $w_{\text{e}}$ value of the heating laser, and yield $\kappa_{\text{eff}}(\SI{293}{\kelvin}) = 117^{+13}_{-11}$\,\SI{}{\watt\per\meter\per\kelvin} and $\kappa_{\text{eff}}(\SI{200}{\kelvin}) = 213^{+26}_{-33}$\,\SI{}{\watt\per\meter\per\kelvin} (regarding the asymmetry of the error bars, see Ref. \cite{Elhajhasan2023}). These are reasonable $\kappa_{\text{bulk}}$ values compared, e.g., to Ref. \cite{Inyushkin2018} that reports $\kappa_{\text{bulk}}$ values for bulk silicon with a natural isotopic composition, yielding \SI{143}{\watt\per\meter\per\kelvin} and \SI{252}{\watt\per\meter\per\kelvin} at $T_{\text{amb}} = \SI{293}{\kelvin}$ and \SI{200}{\kelvin}, respectively. Generally, Raman thermometry is not the most precise method of choice if it comes to determining the values of $\kappa_{\text{bulk}}$ as also expressed by the given error bars. However, the fact that \ke exceeds $\kappa_{\text{bulk}}$ by a factor of 5.3 ($\SI{293}{\kelvin}$) to 8.3 ($\SI{200}{\kelvin}$) for 1LRT measurements with small laser spots ($w_{\text{e}}$) remains an interesting finding that needs to be clarified. In the end, it is exactly this finding that will form the basis for the desired phonon mean free path spectroscopy based on Raman thermometry.

The main limiting factor for our 1LRT analysis is given by the purely diffusive heat transport model used to analyze the measured $T_{\text{rise}}(P_{\text{abs}})$ trends that yield \ke as described in Sec. \ref{sec:exptSetup}. To understand the detailed physics of the extracted $\kappa_{\text{eff}}(w_{\text{e}})$ trends shown in Fig. \ref{fig:expt-result}, we must consider all relevant contributions that affect our 1LRT measurements by affecting the determination of $T_{\text{rise}}$ through the measured Raman shifts $\nu$, which is further detailed in Secs.\,\ref{sec:Eff-on-nu} and \ref{sec:ballist-transp}.

%
%
\subsection{Effects on the Raman shift $\nu$ in an optical heating situation}\label{sec:Eff-on-nu}
%
%

There are three important effects that have an impact on $\nu$ upon laser-induced heating, which are the heat-induced stress that can build up in, e.g., a bulk material (I.), the thermal expansion itself (II.), and the anharmonic effects related to phonon-phonon interactions (III.). In the following, we will carefully evaluate the role of these effects in the light of our experimentally found $\kappa_{\text{eff}}(w_{\text{e}})$ trends from Fig.\,\ref{fig:expt-result}.

I. Heating a bulk material with a laser leads to a build-up of thermally induced stress \cite{BeechemInvArticle2007,BeechemReview2015} resulting in an often complicated and nonsymmetric, so-called triaxial stress situation. The locally heated volume cannot freely expand in all directions when being surrounded by colder material, which impedes thermal expansion and consequently causes the build-up of compressive stress. For bulk silicon (and germanium), it is known that compressive stress leads to an increase in the energy of the optical phonon (the so-called blueshift, see Ref. \cite{Mernagh1991}), which counteracts its decreasing energy (the so-called redshift) with increasing temperature caused by thermal expansion and anharmonic interactions. As a result, such heat-induced stress can lead to an overestimation of \ke due to an underestimation of $T_{\text{rise}}$. Since heat-induced stress can be expected to increase with decreasing $w_{\text{e}}$, this phenomenon could in principle explain our measured $\kappa_{\text{eff}}(w_{\text{e}})$ trends from Fig.\,\ref{fig:expt-result}. However, in the experimental trends shown in Fig. \ref{fig:expt-result} we already corrected the extracted \ke values for any effect of heat-induced stress. This correction leads to a maximum reduction of 15\% for the \ke value measured with the smallest $w_{\text{e}}$ values (see gray crosses in Fig. \ref{fig:expt-result}). By our numerical stress-correction procedure that we implemented in COMSOL Multiphysics\textsuperscript{\circledR} (see S-Sec. V for a detailed description), we obtain an upper boundary approximation for heat-induced stress relying on phonon pressure coefficients from the literature \cite{Mernagh1991}. In turn, the \ke values in Fig. \ref{fig:expt-result} that are corrected for heat-induced stress (orange and blue symbols) correspond to a lower boundary approximation. Thus, we can conclude that our measured particular dependencies $\kappa_{\text{eff}}(w_{\text{e}})$ are not predominantly caused by thermally induced stress.

II. Thermal expansion in bulk silicon leads to a redshift of the optical Raman mode over temperature \cite{Menendez1984}. At room temperature, this effect is not negligible, but about a factor 8 weaker than the redshift caused by anharmonic effects \cite{Menendez1984}. Thus, our directly measured $\nu$ changes, which we translate into $T_{\text{rise}}$ values and subsequently, based on their $P_{\text{abs}}$ dependence, into $\kappa_{\text{eff}}$ values, are predominantly caused by anharmonic effects, which are discussed in the following.

III. The aforementioned anharmonic effects comprise the 3-ph and 4-ph scattering processes that, e.g., describe how an optical phonon decays into two or three acoustic phonons. Based on 1LRT measurements on 2D materials such as graphene, MoS$_2$, and MoSe$_{2}$ it was shown that the populations of the different phonon branches are in non-equilibrium inside of the laser spot \cite{Sullivan2017, Wang2020}. A thermal non-equilibrium generally means that the phonon occupation numbers of different phonon modes do not correspond to one and the same equilibrium temperature. Thus, different phonon modes can be assigned to different temperatures probed by experimental approaches that access the phonon occupation numbers and/or the phonon scattering rates among the different phonon modes \cite{Sullivan2017}: The thermal non-equilibrium of the phonon modes is especially relevant for temperature measurements based on the intensity ratio of the Stokes and Anti-Stokes peaks, because this ratio is directly related to the phonon occupation numbers. But the thermal non-equilibrium of the phonon modes can also affect temperature measurements based on the Raman shift $\nu$ \cite{Wang2020}, because the anharmonic coupling that impacts $\nu$ is not only determined by scattering rates, but also linked to the populations of the interacting phonon modes. Thus, for our 1LRT analysis based on $\nu$, the critical point for our deduction of temperatures is the application of a global temperature calibration $T(\nu)$ to a local heating situation.

Generally, any thermal non-equilibrium between phonon modes is - in agreement with a multi-temperature model elaborated by Vallabhaneni \textit{et al.} in Ref. \cite{Vallabhaneni2016} - related to an underestimation of the thermal conductivity for small laser spot sizes $w_{\text{e}}$ during 1LRT measurements on the aforementioned 2D materials if diffusive phonon transport can be assumed. This underestimation arises from a temperature that is probed via the optical phonons, which is higher than the temperature of the acoustic phonons and the (equilibrium) lattice temperature. When comparing bulk silicon to 2D materials like graphene it shall be noted that for graphene the maximum difference in the phonon specific temperatures appears between the acoustic transverse and longitudinal (TA and LA) phonons and the flexural acoustic (ZA) phonons, which is explained by the restricted selection rules for the scattering of ZA phonons with all other charge and heat carriers \cite{Sullivan2017}. Self-evidently, in silicon there are no flexural modes, so we can already expect the effect of a local non-equilibrium temperature to be less pronounced than in graphene. 

However, even in bulk silicon, a non-equilibrium between the $\Gamma$-point optical phonons and the LA and TA phonons can be expected to exist for the present situation of optical heating due to deviating electron-ph and ph-ph scattering rates. The photo-excited electrons predominantly scatter with optical phonons, which can then decay into two or more acoustic phonons. The time constant of electrons that decay via optical phonon scattering is of about \SI{0.1}{\pico\second}, while the time constant for an optical phonon to decay into acoustic phonons is of about \SI{10}{\pico\second} in silicon \cite{Pop2001}. So indeed, one would expect to create a non-thermal phonon distribution when heating with a laser of $\lambda_h = \SI{532}{\nano\meter}$ in bulk silicon. Compared to previous findings on 2D materials \cite{Vallabhaneni2016,Sullivan2017,Wang2020}, our measured trend for bulk silicon is inverted, as our \ke is higher for small $w_{\text{e}}$ as shown in Fig. \ref{fig:expt-result}. Furthermore, we observe absolute changes of \ke on the order of a factor of 5.3\,-\,8.3 (for 293\,K and 200\,K), which surpasses the magnitude of the (inverted) changes of \ke reported for MoS$_{2}$ \cite{Wang2020}. Since we already consider the effect of heat-induced stress (I.) in Fig.\,\ref{fig:expt-result} and neither thermal expansion (II.) nor anharmonic effects (III.) in combination with non-equilibrium phonon distributions are sufficient to explain our measurements, we must consider another effect in Sec.\,\ref{sec:ballist-transp} to explain our experimental findings.

%
%
\subsection{The effect of quasi-ballistic phonon propagation in a 1LRT experiment}\label{sec:ballist-transp}
%
%

To explain our experimental findings from Fig.\,\ref{fig:expt-result}, we suggest quasi-ballistic phonon propagation as the main physical origin for our measured $\kappa_{\text{eff}}(w_{\text{e}})$ trends. This explanation may appear as counterintuitive at first sight, since, e.g., based on TTG spectroscopy lower \ke values were measured for smaller grating periods $l_{\text{g}}$, which was in turn related to quasi-ballistic phonon transport and a corresponding breakdown of Fourier's law \cite{JohnsonMinnichChen2013}. However, we measure significantly larger \ke values for smaller laser spot sizes $w_{\text{e}}$, which does not contradict previous experimental findings by TTG, as a different quantity is measured. A similar observation is true for FDTR and TDTR measurements depending on $w_{\text{e}}$. However, in the following we will focus our discussion on a comparison of TTG and our 1LRT measurements, because both techniques do not necessarily rely on metal transducers and hereby induced further experimental complications \cite{RegnerReview2015}.

In a TTG experiment \cite{JohnsonMinnichChen2013}, \ke is calculated from the decay rate $\gamma$ of a thermal grating. This rate $\gamma$ describes how fast heat is transported between the maxima and the minima of this periodic grating. Diffusive phonon transport always contributes to heat transport from one of these temperature maxima (heat source) to an adjacent temperature minimum (heat sink), while this is not necessarily the case for quasi-ballistic phonon transport. For the latter case the phonons can be transported to different heat maxima and minima, which slows the decay of the thermal grating. Thus, pronounced non-diffusive heat transport leads to lower $\gamma$ values for small grating periods compared to a diffusive model. Since $\kappa_{\text{eff}} \propto \gamma$ holds, one obtains lower \ke values than predicted by the diffusive model. In other words, for TTG measurements the measured quantity is the heat transport, which can be less than expected for a purely diffusive phonon propagation, leading to the determination of lower \ke values for smaller $l_{\text{g}}$ values. 

In contrary, the measured quantity during an 1LRT experiment is a local temperature rise $T_{\text{rise}}$.
All these local temperatures are real temperatures from the perspective of an optical $\Gamma$-point phonon of silicon and provide a suitable approximation for the local lattice temperature as soon as a thermal equilibrium between all phonons is established (see, Sec.\,\ref{sec:Eff-on-nu}). The measured $T_{\text{rise}}$ value for a given $P_{\text{abs}}$ is lower the more heat is transported away from the heat source. Thus, we measure lower local temperatures than predicted by the diffusive model when the amount of non-diffusively transported phonons increases. Since $\kappa_{\text{eff}} \propto \left(\partial T_{\text{rise}}/\partial P_{\text{abs}}\right)^{-1}$ holds (cf., e.g., Ref. \cite{Lax1977}), we obtain higher \ke values for small spot sizes $w_{\text{e}}$. As an additional consequence, our $\kappa_{\text{eff}}(w_{\text{e}})$ trend is even more pronounced at $T_{\text{amb}} = \SI{200}{\kelvin}$, because at lower $T_{\text{amb}}$ values the $l_{\text{ph}}$ values of the phonons are increased.

The theoretical work of Chiloyan \textit{et al.} \cite{Chiloyan2020} also reports effective thermal conductivity values for silicon that can exceed $\kappa_{\text{bulk}}$ for a local heat source as soon as its size approaches $l_{\text{ph}}$ values that are relevant for thermal transport. The authors of Ref. \cite{Chiloyan2020} do not only account for a local non-equilibrium of the phonons (see, Sec.\,\ref{sec:Eff-on-nu}), but also for non-diffusive heat transport. Chiloyan \textit{et al.} solve the steady-state phonon BTE for different heat source geometries and different phonon populations generated by these heat sources. Most interesting for our work is their three-dimensional Gaussian heat source with various heat spot radii $w_{\text{e}}$ for single-crystalline silicon at room temperature. This scenario models our 1LRT experiment in a first approximation, because in two dimensions the heat laser spot intensity has a radial Gaussian intensity distribution, while the intensity decays with Beer-Lambert's law inside of the sample. In Ref. \cite{Chiloyan2020} it was found, that when the generated phonon distribution is non-thermal and also non-diffusive transport effects are taken into consideration, $T_{\text{rise}}$ will be lower than predicted with Fourier's law, yielding \ke values that exceed the bulk thermal conductivity.

\begin{figure*}[]
	
	\includegraphics[width=\linewidth]{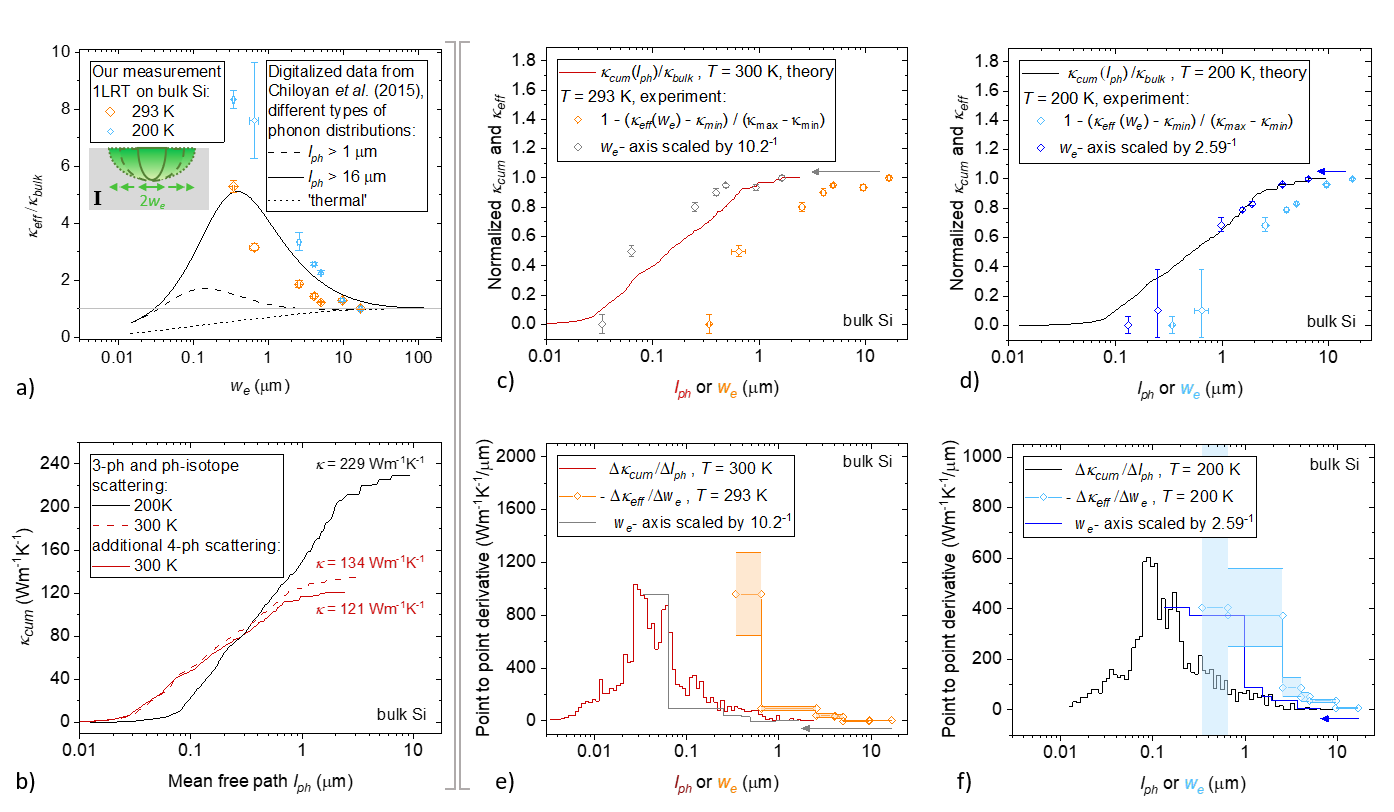}
	
	\caption{a) \ke normalized to the bulk value calculated for single-crystalline bulk silicon at room temperature based on the steady-state phonon BTE for a 3D Gaussian heat source for various radii $w_{\text{e}}$ and for three different generated phonon distributions. This theoretical results originate from Ref.\,\cite{Chiloyan2020} and treat a thermal phonon distribution (short-dashed line) and two phonon distributions with either $l_{\text{ph}} > \SI{1}{\micro\meter}$ (long-dashed lines) or $l_{\text{ph}} > \SI{16}{\micro\meter}$ (solid line). In comparison we plot our 1LRT data normalized to $\kappa_{\text{eff}}(w_{\text{e}}^{\text{max}})$ for bulk silicon at $T = \SI{293}{\kelvin}$ (orange symbols) and $T = \SI{200}{\kelvin}$ (blue symbols). b) Accumulated thermal conductivity \ka from our ab-initio DFT calculations for bulk silicon at \SI{300}{\kelvin} (red) and \SI{200}{\kelvin} (black). These calculations account for ph-isotope scattering, 3-ph, and 4-ph scattering (the latter only needed for $T_{\text{amb}} = \SI{300}{\kelvin}$). The $\kappa_{\text{cum}}(l_{\text{ph}})$ data shown by solid lines is used for the following comparison between theory and experiment. c) Comparison of the normalized theoretical accumulated $\kappa_{\text{cum}}(l_{\text{ph}})$ (solid red line) with the normalized experimental $\kappa_{\text{eff}}(w_{\text{e}})$ trend (orange symbols). The theory values are normalized by the corresponding calculated bulk value indicated in b). The normalization of the experimental data is described in the main text. For better comparison, the abscissa of the experimental data can also be scaled by factor $10.2^{-1}$ (gray symbols). d) The same comparison as described for c), but for $T_{\text{amb}} = \SI{200}{\kelvin}$. Again, a scaling of the experimental data can be applied for a better comparison. e) Comparison of the point-to-point derivative for the theoretical (solid red line) and the experimental data set (connected orange symbols) recorded at $T_{\text{amb}} = \SI{293}{\kelvin}$. Again, the same scaling as described for c) can be applied (solid gray line). f) Similar comparison as shown in e), but for $T_{\text{amb}} = \SI{200}{\kelvin}$.}
	
	\label{fig:theo-expt-comp}
	
\end{figure*}

In Fig. \ref{fig:theo-expt-comp}a we show a comparison between our experimental findings from Fig. \ref{fig:expt-result} and the theoretical results from Ref. \cite{Chiloyan2020}. Here, the theoretical \ke values are normalized to the corresponding $\kappa_{\text{bulk}}$ value of silicon and are plotted over the heat spot size $w_{\text{e}}$ for three different phonon distributions: A thermal phonon distribution (short-dashed line), where a local thermal equilibrium is reached, and a non-thermal distribution, where all $l_{\text{ph}}$ values exceed a threshold value of either \SI{1}{\micro\meter} (long-dashed line) or \SI{16}{\micro\meter} (solid line). The experimental values are normalized to the \ke value measured for the largest $w_{\text{e}}$ value, which corresponds to our best experimental approximation of $\kappa_{\text{bulk}}$, cf. Sec.\,\ref{sec:results-var-we}. Interestingly, our measured $\kappa_{\text{eff}}(w_{\text{e}})$ trend for $T = \SI{293}{\kelvin}$ is situated between the two theoretical trends for the phonon distributions with $l_{\text{ph}} > \SI{1}{\micro\meter}$ and with $l_{\text{ph}} > \SI{16}{\micro\meter}$, while our experimental trend for $T = \SI{200}{\kelvin}$ even exceeds the theoretical predictions towards smaller $w_{\text{e}}$ values as shown in Fig. \ref{fig:theo-expt-comp}a. Thus, our comparison from Fig. \ref{fig:theo-expt-comp}a strengthens the argument that our $\kappa_{\text{eff}}(w_{\text{e}})$ trend can be explained by considering non-diffusive heat transport in bulk silicon. 

We also repeated the 1LRT series with varying $w_{\text{e}}$ values at $T_{\text{amb}} = \SI{293}{\kelvin}$ for a bulk germanium sample to show the general relevance of our findings. Here, we also measured increasing \ke values for smaller $w_{\text{e}}$ values, while the whole measured $\kappa_{\text{eff}}(w_{\text{e}})$ trend significantly deviates from our findings for silicon. In addition, in two previous studies on GaN membranes \cite{Elhajhasan2023,Seemann2024} we already found that \ke depends on $V$, which was varied by performing 1LRT and 2LRT measurements. Thus, our main experimental finding that \ke depends on $V$ (tunable through $w_{\text{e}}$ and also $h_{\alpha}$, see Sec.\,\ref{sec:results-var-ha}) has now been confirmed in three different semiconductor materials, namely silicon, germanium, and GaN. Our $w_{\text{e}}$-dependent 1LRT results for germanium can be found in S-Sec. VI of the Supplemental Material. Furthermore, our findings are of practical relevance for Raman thermometry of high-quality samples even at room temperature and above. Quantitative 1LRT measurements with small laser spot sizes, which are often needed to achieve a sufficiently high spatial resolution, can be challenging as soon as phonons with large $l_{\text{ph}}$ values contribute significantly to thermal transport. Thus, an experimental compromise must be found between the spatial resolution during Raman thermometry and the need to encompass all thermal phonons with an upper boundary of $l_{\text{ph}}$. Thus, experimental findings of high \ke values by Raman thermometry do not necessarily point towards, e.g., a high sample quality, but can also indicate the negligence of an important fraction of thermal phonons with high $l_{\text{ph}}$ values during the Raman-based temperature probing.




\section{Comparison to theory via the effective and the accumulated thermal conductivity}\label{sec:comp-theory}

We now compare the experimental $\kappa_{\text{eff}}(w_{\text{e}})$ trends from Fig.\,\ref{fig:expt-result} for bulk silicon to our \textit{ab initio} theory. We calculate the accumulated thermal conductivity $\kappa_{\text{cum}}$ with respect to the phonon mean free path $l_{\text{ph}}$. The latter is computed by solving the steady-state, momentum resolved, linearized phonon Boltzmann transport equation (BTE) as implemented in the \texttt{elphbolt} package \cite{ProtikElphbolt2022}. We considered 3- and 4-phonon, phonon-isotope, and phonon-thin-film scatterings. The detailed workflow related to the inclusion of these scattering mechanisms has recently been detailed in Ref.\,\cite{Elhajhasan2023} for wurtzite GaN. For the calculation here, we used a norm-conserving pseudopotential with the Perdew-Zunger (local density approximation) exchange-correlation \cite{perdew1981self}. The 2nd order force constants were generated with density functional perturbation theory using \texttt{Quantum Espresso} \cite{giannozzi2009quantum, giannozzi2017advanced} using a $6 \times 6 \times 6$ $\mathbf{q}$-mesh. The 3rd order force constants were generated using \texttt{Quantum Espresso} and \texttt{thirdorder.py} \cite{li2014shengbte}. For this, we used displaced supercells with $250$ atoms with a $\Gamma$-point sampling and a $6$-nearest neighbor cutoff. To assess the importance of the 4-phonon scattering on transport, we computed its effect on the thermal conductivity at $300$ K. For this, we first computed the 4th order force constants combining \texttt{Quantum Espresso} and \texttt{fourthorder.py} \cite{han2022fourphonon}. Here, too, we used displaced supercells with $250$ atoms with a $\Gamma$-point sampling. For this case, we used a 2-nearest neighbor cutoff. The \texttt{FourPhonon} \cite{han2022fourphonon} code was then used to compute the 4-phonon scattering rates on a $12\times12\times12$ $\mathbf{q}$-mesh, which were then passed into \texttt{elphbolt}, where they were trilinearly interpolated on to the $36 \times 36 \times 36$ $\mathbf{q}$ transport mesh. The phonon-isotope scattering was computed within the Tamura model \cite{tamura_isotope_1983}, which is based on the 1st-Born approximation on top of a virtual crystal approximation with respect to the natural isotopic mix of silicon. While it is possible to go beyond this level of the phonon-isotope scattering theory, it has been argued that for silicon, the Tamura model should suffice for phonon thermal conductivity calculations \cite{Protik2024}. The phonon-thin-film scattering was included using the large Knudsen number (ballistic) limit of the Fuchs-Sondheimer theory~\cite{Ziman2001} and assuming zero specularity of the scattering interface.

We first compare our experimental data to such theoretical $\kappa_{\text{cum}}(l_{\text{ph}})$ trends as they are frequently reported in literature \cite{EsfarjaniChenStokes2011, GargBook2014, WangHunag2014, ZengCollins2015, JiangKoh2016, Xu_2019_kappa-cum, Cheng2021}. 
In Fig.\,\ref{fig:theo-expt-comp}b the theoretical $\kappa_{\text{cum}}(l_{\text{ph}})$ trends are shown for bulk silicon at $T_{\text{amb}}\,=\,\SI{300}{\kelvin}$ and \SI{200}{\kelvin}, respectively. At \SI{200}{\kelvin}, the 3-ph scattering processes as well as ph-isotope scattering are considered (solid black line). For $T_{\text{amb}} = \SI{300}{\kelvin}$ we show $\kappa_{\text{cum}}(l_{\text{ph}})$ with (solid red line) and without (dashed red line) the additional consideration of 4-ph processes. Already at $T_{\text{amb}}\,=\,\SI{300}{\kelvin}$ the consideration of these additionally resistive 4-ph scattering processes only leads to a minor reduction of the converged $\kappa$ value (values are denoted in Fig.\,\ref{fig:theo-expt-comp}b) by $\approx 10\%$. Thus, at \SI{200}{\kelvin} the effect of such 4-ph scattering can be expected to be even weaker due to the lower phonon population, which renders the consideration of 3-ph scattering sufficient for the present study. The convergence of the \ka values yields our theoretical $\kappa_{\text{bulk}}$ values at \SI{300}{\kelvin} (\SI{121}{\watt\per\meter\per\kelvin}) and \SI{200}{\kelvin} (\SI{229}{\watt\per\meter\per\kelvin}), which are also in good agreement with our 2LRT results shown in Fig. \ref{fig:expt-result}. Furthermore, a good agreement is also reached for our \ke results based on 1LRT for the largest available $w_{\text{e}}$ values, which already agreed well with our 2LRT results as shown in Fig.\,\ref{fig:expt-result}.

In a next step, the theoretical $\kappa_{\text{cum}}(l_{\text{ph}})$ trends will be compared to the $\kappa_{\text{eff}}(w_{\text{e}})$ trends from our 1LRT experiments based on two different methods shown in Figs. \ref{fig:theo-expt-comp}c and d (comparison by normalization, detailed in Sec.\,\ref{sec:norm-comp}) and Figs. \ref{fig:theo-expt-comp}e and f (comparison by derivative, detailed in Sec.\,\ref{sec:derivative-comp}), respectively.

%
\subsection{Comparison by normalization}\label{sec:norm-comp}
%

For Figs.\,\ref{fig:theo-expt-comp}c and d, the theoretical trends were normalized by the calculated convergence value of $\kappa_{\text{cum}}$. To normalize our experimental $\kappa_{\text{eff}}(w_{\text{e}})$ trends, we apply the following formula:
\begin{equation}
	\kappa_{\text{eff}}^{\text{norm}}(w_{\text{e}}) = 1 - \frac{\displaystyle \kappa_{\text{eff}}(w_{\text{e}}) - \kappa_{\text{min} }}{\displaystyle \kappa_{\text{max}} - \kappa_{\text{min}}}
	\label{eq:k-eff-norm}
\end{equation}
With $\kappa_{\text{max}}$ being the maximum measured \ke and $\kappa_{\text{min}}$ being the minimum \ke measured for the highest $w_{\text{e}}$ value. Eq.\,\ref{eq:k-eff-norm} can easily be motivated from an experimental point of view since the included fraction describes the common subtraction of a baseline and normalization to a maximal value of a dataset. Thus, Eq.\,\ref{eq:k-eff-norm} enables a first straightforward comparison of $\kappa_{\text{eff}}(w_{\text{e}})$ and $\kappa_{\text{cum}}(l_{\text{ph}})$ for bulk silicon as shown in Fig.\,\ref{fig:theo-expt-comp}c (room temperature) and Fig.\,\ref{fig:theo-expt-comp}d  (\SI{200}{\kelvin}). Even though this simple comparison provides a first reasonable agreement between experiment and theory, it still comes with the following drawbacks:

i) This normalization yields $\kappa_{\text{eff}}^{\text{norm}}(w_{\text{e}})$ values between 0 and 1, which assumes that the $\kappa_{\text{bulk}}$ value can be probed by 1LRT with the largest $w_{\text{e}}$ value. Furthermore, for the smallest $w_{\text{e}}$ value this normalization comprises the assumption that the phonon transport is purely non-diffusive.
The first assumption is reasonable, as we can see from Fig. \ref{fig:expt-result} that the \ke values for the largest $w_{\text{e}}$ values agree well with our 2LRT and theoretical results. However, it is clear that with our smallest experimentally realizable value of $w_{\text{e}} = \SI{0.34(1)}{\micro\meter}$, a finite amount of thermal transport is still diffusive. If only non-diffusive transport would be present (as for an infinitely small heat source), no $T_{\text{rise}}$ could be measured by 1LRT, which would translate to an infinite \ke value. Nevertheless, this remains a rather artificial limitation, since smaller heat spots are optically only achievable by tip-enhanced techniques \cite{Reparaz2013_TipEnhRaman}, while our measured $T_{\text{rise}}$ values already reach the temperature resolution limit of Raman thermometry.

ii) A main deficiency of our comparison are the different abscissas when plotting $\kappa_{\text{eff}}(w_{\text{e}})$ and $\kappa_{\text{cum}}(l_{\text{ph}})$ in one and the same plot as shown in Fig.\,\ref{fig:theo-expt-comp}c (room temperature) and d (\SI{200}{\kelvin}). To translate between $w_{\text{e}}$ and $l_{\text{ph}}$, we would already need full information about the $l_{\text{ph}}$ values of all phonons and their impact on thermal transport. A future direction of this work is to employ a real-space BTE solver, such as \texttt{OpenBTE}~\cite{Romano2021}, which can unravel the required translation between $w_{\text{e}}$ and $l_{\text{ph}}$. In such simulations, various non-thermal heat source models~\cite{Chiloyan2020} can be seeded to reproduce the relationship $\kappa_{\text{eff}}(w_{\text{e}})$ observed in this work. Furthermore, differentiable real-space phonon simulations, such as those used for inverse-design in nanostructures~\cite{romano2022inverse}, may be employed for automatic \ke extraction. As a result, our experimental data $\kappa_{\text{eff}}^{\text{norm}}(w_{\text{e}})$ will form the basis for extracting fully quantitative information about optically induced phonon populations or the effect of any, e.g. nanopatterning. Interestingly, despite the simple normalization from Eq.\,\ref{eq:k-eff-norm}, an even better agreement between experiment and theory can already be achieved by scaling the $w_{\text{e}}$-axis with a simple factor of $10.2^{-1}$ for \SI{293}{\kelvin} and $2.59^{-1}$ for \SI{200}{\kelvin}. As a result, the decline of the \ka and \ke values start at the same abscissa value as shown in Figs.\,\ref{fig:theo-expt-comp}c (gray symbols) and d (blue symbols) and a good match between experiment and theory can be found in light of the simplicity of the present approach. 

%
\subsection{Comparison by derivative}\label{sec:derivative-comp}
%

Another way to compare \ke and \ka without the need to normalize the ordinate is based on comparing the derivatives of $\kappa_{\text{eff}}(w_{\text{e}})$ and $\kappa_{\text{cum}}(l_{\text{ph}})$. Since we have discrete data points, we define the derivatives $\Delta \kappa_{\text{cum}}/\Delta l_{\text{ph}}$ and $\Delta \kappa_{\text{eff}}/\Delta w_{\text{e}}$ between two points as the slope of a straight connecting line. For the limited number of experimental \ke values only adjacent data points are used, while for the \ka values a set of three data points could be used, which yields a moderate smoothing. Following this procedure for our bulk silicon data, we obtain a compilation of step functions in Figs. \ref{fig:theo-expt-comp}e (room temperature) and f (\SI{200}{\kelvin}) for the comparison between $\Delta \kappa_{\text{cum}}/\Delta l_{\text{ph}}$ and $\textbf{-}\Delta \kappa_{\text{eff}}/\Delta w_{\text{e}}$. The shaded regions in Figs. \ref{fig:theo-expt-comp}e and f indicate the experimental error bars obtained from error propagation calculations. 

The difficulty in translating between $w_{\text{e}}$ and $l_{\text{ph}}$ on the abscissa remains, and to compare the curves the same scaling factors as in Figs. \ref{fig:theo-expt-comp}c and d are used for the experimental data. Here, the best agreement between experiment and theory is reached for the room temperature data shown in Fig.\,\ref{fig:theo-expt-comp}e. For $T_{\text{amb}} = \SI{200}{\kelvin}$ one can see in Fig.\,\ref{fig:theo-expt-comp}f that the measured slope is too steep compared to the theoretical expectation expressed by the plotted derivative. It should be noted that for $T_{\text{amb}} = \SI{200}{\kelvin}$ the normalization approach from Sec.\,\ref{sec:norm-comp} becomes more realistic because an increasing fraction of the thermal transport becomes quasi-ballistic. Thus, for even lower values of $T_{\text{amb}}$, we expect an increasingly better agreement between our experimental and theoretical data. We wish to note that, independent of the challenges that arise from the comparison between the experimental and theoretical data, our experimental results from Figs.\,\ref{fig:expt-result} and \ref{fig:theo-expt-comp} show that 1LRT is capable of the desired PMFP spectroscopy by varying $w_{\text{e}}$ as the critical parameter $l_{\text{c}}$.



It would be interesting to further reduce the minimal experimental $w_{\text{e}}$ value (i.e., \SI{0.34(1)}{\micro\meter}) that we reached in Fig.\,\ref{fig:theo-expt-comp}. Here, tip-enhanced Raman spectroscopy could be utilized \cite{Reparaz2013_TipEnhRaman}, however, this inflicts severe experimental challenges in determining $P_{\text{abs}}$. Furthermore, it would be beneficial to continuously tune $w_{\text{e}}$ to achieve a better comparison to theory and to resolve changes in the experimental $\kappa_{\text{eff}}(w_{\text{e}})$ trends that are, e.g., caused by a nanopatterning. Thus, in the following Sec.\,\ref{sec:results-var-ha} we will present an alternative approach to any variation of $w_{\text{e}}$, which shall enable a future continuous tuning of the size of the light penetration depth $h_{\alpha}$ and temperature probe volume via the applied Raman laser wavelength $\lambda$. Furthermore, significantly smaller temperature probe volumes can be reached by this method, which enables a pathway to improving the normalization procedure from Sec.\,\ref{sec:norm-comp}.


\section{\label{sec:results-var-ha}Impact of the laser wavelength used for Raman thermometry}

Instead of varying the laser focus spot size $w_{\text{e}}$ during our 1LRT measurements as introduced in Sec.\,\ref{sec:results-var-we}, we can also vary the wavelength $\lambda$ of the laser that we use for the Raman thermometry. In literature, one often finds no mention of the impact of $\lambda$ on the $\kappa_{\text{eff}}$ values one extracts from Raman thermometry. When varying $\lambda$ the light penetration depth $h_ {\alpha}$ is altered, which changes the size of the temperature heat and probe volume similar to the impact of $w_{\text{e}}$. As a result, 1LRT based PMFP spectroscopy should not only be limited to variations of $w_{\text{e}}$ but also $h_ {\alpha}$ should represent a viable tuning parameter as solidified in the following. 

To present the feasibility of our experimental approach, we utilize four different $\lambda$ values for 1LRT at room temperature on a 2000-nm-thick and a 200-nm-thick silicon membrane, cf. Fig. \ref{fig:expt}a sketch \MakeUppercase{\romannumeral 2}. To deduce $h_{\alpha}(\lambda)$, we rely on literature values for the absorption, which were determined by spectroscopic ellipsometry \cite{Franta2019}.

While varying $\lambda$ and therefore $h_{\alpha}$, also $w_{\text{e}}$ can change moderately depending on the microscope objective in use, which we also account for as described in S-Sec. VIII. All corresponding experimental parameters are listed in Tab.\,\ref{tab:summarize_param} and $h_{\alpha}$ was calculated according to Ref. \cite{Franta2019}. For the transition from $\lambda = \SI{325}{\nano\meter}$ to $\lambda = \SI{532}{\nano\meter}$ we can safely assume that the reduction of $w_{\text{e}}$ by $\approx$\,0.3 is negligible compared to the increase in $h_{\alpha}$ by a factor of 110. When transitioning from $\lambda = \SI{532}{\nano\meter}$ to $\lambda = \SI{561}{\nano\meter}$ only $h_{\alpha}$ changes, while the final $\lambda$-step towards $\SI{660}{\nano\meter}$ causes an almost equal increase in $w_{\text{e}}$ and $h_{\alpha}$.


\begin{table} [h]
	\caption{\label{tab:summarize_param}Summary of the experimental parameters regarding our $\lambda$-dependent 1LRT measurements.}
	\begin{ruledtabular}
		\begin{tabular}{r l l l l}
			
			$\lambda$ [nm] &  325 & 532 &  561 & 660\\
			\hline
			$w_{\text{e}}$ [\SI{}{\micro\meter}] & 0.82 & 0.57 & 0.57 & 1.01 \\
			$h_{\alpha}$ [nm] & 10 & 1100 & 1500 & 4200\\ 
			
		\end{tabular}
	\end{ruledtabular}
\end{table}

\begin{figure}[h]
	
	\includegraphics[width=\linewidth]{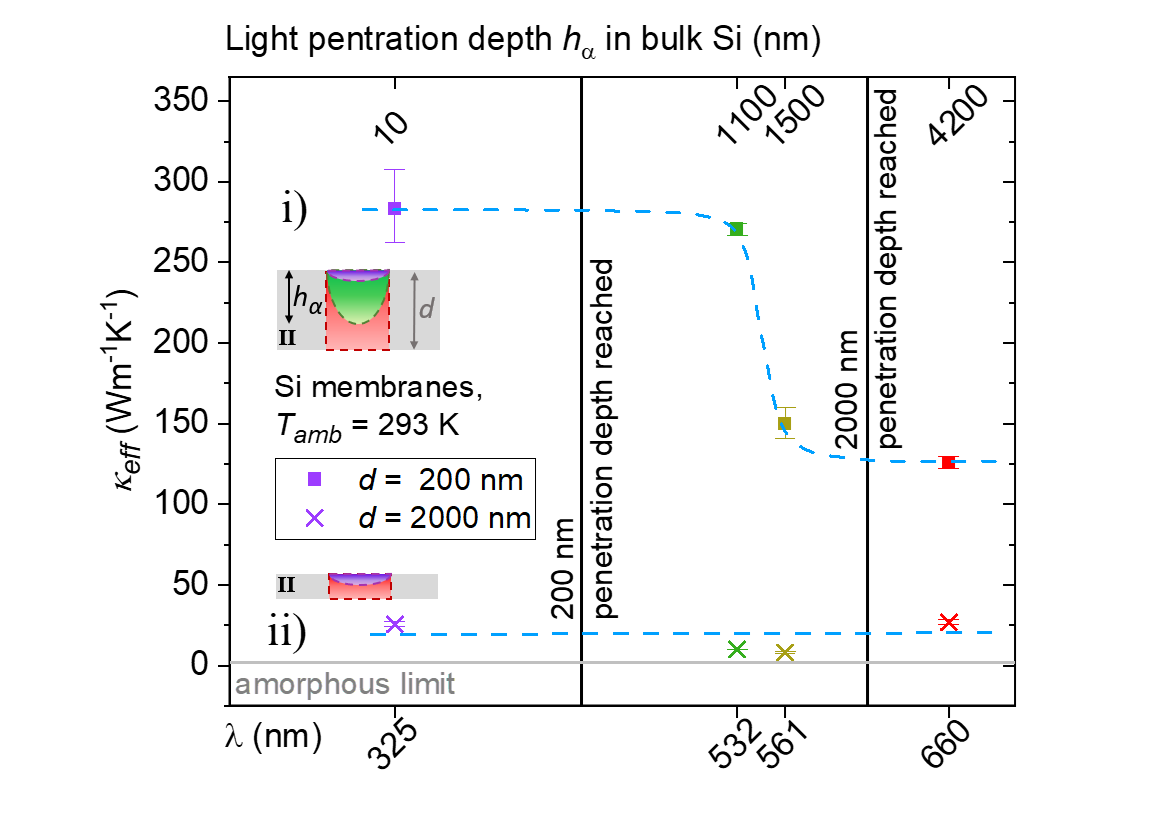}
	
	\caption{Effective thermal conductivity \ke as a function of $h_{\alpha}$ and $\lambda$ extracted from 1LRT measurements at room temperature on silicon membranes with different thickness $d$. Here, crosses indicate our results for a 200-nm-thick silicon membrane, whereas boxes are associated to a 2000-nm-thick silicon membrane. The bottom x-axis shows $\lambda$ on a linear scale and the corresponding $h_{\alpha}$ values are indicated on the top x-axis and calculated based on data taken from Ref. \cite{Franta2019} for bulk silicon at room temperature. The dashed blue lines serve as a guide-to-the-eye to highlight the difference between the results for $d = \SI{200}{\nano\meter}$ and $d = \SI{2000}{\nano\meter}$. The two sketches in the inset illustrate the experimental 1LRT situation for the different $d$ and $\lambda$ values following our sketch from Fig.\,\ref{fig:expt}a. The $w_{\text{e}}$ and $h_{\alpha}$ values are resumed in Tab.\ref{tab:summarize_param}.}
	
	\label{fig:var-ha}
	
\end{figure}

Fig.\,\ref{fig:var-ha} shows the \ke values resulting from the analysis of our 1LRT measurements as described in Sec. \ref{sec:COMSOL-membrane}. Again, it can be seen that the size of the temperature probe volume, which we control via $\lambda$ and therefore $h_ {\alpha}$, has a strong impact on the measured \ke values. For our thick silicon membrane with $d = \SI{2000}{\nano\meter}$, a clear $h_{\alpha}$ dependence is visible. The \ke values seem to mimic a step-function as illustrated by the guide-to-the-eye (dashed blue line). The highest \ke value of $283_{-21}^{+24}$\,\SI{}{\watt\per\meter\per\kelvin} is measured for $\lambda = \SI{325}{\nano\meter}$, which yields $h_{\alpha}\,\approx\,\SI{10}{\nano\meter}$ in silicon. Similar to Fig.\,\ref{fig:expt-result}, we observe \ke values that exceed a realistic $\kappa$ for a thick silicon membrane, which should not surpass their bulk equivalents $\kappa_{\text{bulk}}$. Compared to the membrane thickness $d$, the heating and temperature probing with a UV laser ($\lambda = \SI{325}{\nano\meter}$) occurs close to the sample surface, mimicking a surfacic heating situation. Thus, our experimental observation of \ke$(h_{\alpha}\,\approx\,\SI{10}{\nano\meter})\,>\,\kappa_{\text{bulk}}$ suggests that we again probe quasi-ballistic phonon transport, here, mostly perpendicular to the plane of the membrane. 

When $h_{\alpha}$ is further increased, we observe a reduction in \ke as shown in Fig. \ref{fig:var-ha}, which also includes a sketch of the experimental situation for $d = \SI{2000}{\nano\meter}$ (i). The observed reduction of \ke becomes most significant for $\lambda = \SI{561}{\nano\meter}$ where $h_{\alpha}$ yields $\approx 0.7\,d$. As soon as $\lambda = \SI{660}{\nano\meter}$ is reached, $h_{\alpha}$ exceeds $d$ of the thicker silicon membrane, while also $w_{\text{e}}$ reaches a maximum, cf. Tab.\,\ref{tab:summarize_param}.

With a Raman laser wavelength of \SI{660}{\nano\meter} we measure our smallest value for \ke$\,=\,126_{-4}^{+4}\SI{}{\watt\per\meter\per\kelvin}$ at the bottom of the aforementioned step-function, which is in reasonable agreement with the literature \cite{ChavezAngel2014,Cuffe2015} and our own calculations yielding \SI{91}{\watt\per\meter\per\kelvin} (see S-Sec.\,VII for further details). However, it is interesting that Chávez-Ángel \textit{et\,al.} report $\kappa\,=\,\SI{147}{\watt\per\meter\per\kelvin}$ for a 1.5-\SI{}{\micro\meter}-thick silicon membrane, which is close to $\kappa_{\text{bulk}}$ for silicon. As they performed 1LRT measurements with $\lambda = \SI{514.5}{\nano\meter}$ the resulting $h_{\alpha}$ value is located in the critical region of the step-function illustrated in Fig. \ref{fig:var-ha}. Thus, higher measured $\kappa$ values are not necessarily always a sign of any superior sample quality, but can also already be the signature of quasi-ballistic phonon transport, which must be tested by varying $\lambda$ during 1LRT. As a result, often only effective thermal conductivities \ke are measured by 1LRT as long as the temperature probe volume is not sufficient in size in regard to, e.g., the silicon membrane thickness $d$. We wish to note that the particular ratio between $w_{\text{e}}$ (e.g., controlled by the N.A. of the microscope objective, the beam diameter, the mode profile) and $h_{\alpha}$ will control the position and broadening of the step-function sketched in Fig.\,\ref{fig:var-ha}. As a consequence, from an experimental point of view it is easier to measure thinner membranes for which $h_{\alpha}$ exceeds $d$, whereas measurements on bulk samples and larger micro-structures require an experimental check of $\kappa_{\text{eff}}(\lambda)$.

Thus, no clear trend for \ke can be observed when probing our 200-nm-thick silicon membrane by 1LRT. Our results fluctuate around a mean value of \SI{18}{\watt\per\meter\per\kelvin} which can be explained by the lower $d/h_{\alpha}$ ratio and the fact that the membrane boundaries generally reduce $l_{\text{ph}}$ as also considered in our calculations presented in S-Sec.\,VII. The value \ke = \SI{18}{\watt\per\meter\per\kelvin} is above the amorphous limit of \SI{1.8}{\watt\per\meter\per\kelvin} \cite{ChavezAngel2014}, but is generally lower than one would expect for a 200-nm-thick silicon membrane. In the literature one finds \ke values obtained from 1LRT on silicon membranes with $d\,=\,200\,\text{nm}$ of \SI{59}{\watt\per\meter\per\kelvin} \cite{ChavezAngel2014} and \SI{70}{\watt\per\meter\per\kelvin} \cite{Cuffe2015}, whereas our own calculations yield \SI{56}{\watt\per\meter\per\kelvin}, cf. S-Sec.\,VII. However, based on our rather simplistic implementation of phonon-boundary scattering and the complete negligence of a membrane roughness, we understand our theoretical value of \SI{56}{\watt\per\meter\per\kelvin} as an upper limit for a 200-nm-thick silicon membrane. Thus, it remains a task for future work to re-assess $\kappa_{\text{eff}}(d)$ for a suitable set of $w_{\text{e}}$ and $h_{\alpha}(\lambda)$, aiming to overcome the limits of standard Raman thermometry often applying high N.A. microscope objectives and a limited set of Raman laser wavelengths.

Our variation of $\lambda$ during 1LRT appears promising for probing quasi-ballistic phonon transport similar to TTG spectroscopy. We wish to note that varying the thermal grating constant $\l_{\text{g}}$, which is the main critical length scale $\l_{\text{c}}$ of TTG for PMFP spectroscopy \cite{RegnerReview2015}, is a non-trivial experimental endeavor, which even motivates the use of deep-UV light sources \cite{Marroux2024, NelsonKnobloch2024}. Interestingly, for 1LRT measurements any variation of $h_{\alpha}$ can be achieved with sub-10-nm precision by varying the Raman laser wavelength $\lambda$, which renders 1LRT promising for PMFP spectroscopy in addition to TTG experiments.

%
%

\section{\label{sec:sumout}Summary and Outlook}

%
%

The present work shows that even at $T_{\text{amb}}\,=293\,\text{K}$ one can measure the effect of quasi-ballistic heat transport in bulk silicon by means of 1LRT. Thus, Raman thermometry is capable of PMFP spectroscopy similar to, e.g., TTG, TDTR, and FDTR analyses based on varying the laser focus spot size on the sample $w_{\text{e}}$ and/or the applied Raman laser wavelength $\lambda$. As a result, $w_{\text{e}}$ and $\lambda$ that dictates the material-dependent light penetration depth $h_{\alpha}$ are no longer always free parameters during Raman thermometry, if one aims to obtain, e.g., quantitative values for $\kappa_{\text{bulk}}$. In other words, the size of the volume in which the heating and the temperature probing occurs must potentially be carefully controlled during 1LRT, which strongly depends on the sample to be investigated. In a first approximation, one can differentiate the following two cases:

Case I: As long as $w_{\text{e}}$ and $h_{\alpha}$ are large compared to the mean free path $l_{\text{ph}}$ of thermal phonons (i.e., phonons that strongly contribute to $\kappa$), 1LRT remains as straightforward as frequently reported in literature and often a theoretical description by Fourier's law can hold. Nevertheless, sufficient care must be devoted to a proper theoretical description of the experimental situation to extract realistic $\kappa$ values, which excludes analytical solutions for most cases and requires numerical solutions of the heat diffusion equation, comprising an adequate modeling of the heat source and temperature probing as shown in this work. One can assume that case I holds for all samples that exhibit comparably low $l_{\text{ph}}$ values for thermal phonons, which can be a matter of, e.g., overall sample quality, a generally high anharmonicity of the material, any nano- and micro-structuring, or alloying - just to name a few possibilities.

Case II: In samples of sufficient quality $w_{\text{e}}$ and $h_{\alpha}$ must be tuned during 1LRT measurements to assure that not just an effective thermal conductivity $\kappa_{\text{eff}}$ is measured, but that the convergence towards, e.g., $\kappa_{\text{bulk}}$ has been reached. However, this is not always just required for bulk materials, but can already be of relevance for micro-structures like, e.g., a thicker silicon membrane as shown in this work. In addition, depending on the material-dependent evolution of $h_{\alpha}(\lambda)$, one can reach characteristic length scales $l_{\text{c}}$ during 1LRT-based PMFP spectroscopy in the sub-1-\SI{}{\micro\meter} regime. Thus, even for nano-structures it can be important to control $\lambda$, which allows the transition from surfacic to volumetric heating during 1LRT measurements.

For quantitative 1LRT results, one always has to ensure first, if case I or II holds for the sample to be examined. If case II holds then PMFP spectroscopy can be addressed, which is from a physical point of view more promising than the derivation of, e.g., $\kappa_{\text{bulk}}$, which can often also be achieved by numerous alternative and better suited experimental techniques. For our high-quality bulk silicon samples, case II holds, while for our membrane samples, we only observe a $\lambda$-dependence for our thicker silicon membrane sample ($d\,=\,2000\,\text{nm}$). Thus, when analyzing our thinner silicon membrane sample ($d\,=\,200\,\text{nm}$) we transition towards case I for the $\lambda$-interval addressed in this study. Thus, any overestimation of $\kappa$ values derived from 1LRT can be linked to an erroneous interpretation of the experimental situation described by case I or II.

The experimental trends for our measured $\kappa_{\text{eff}}$ values (at $T_{\text{amb}}\,=293\,\text{K}$ and $200\,\text{K}$) are in qualitative agreement with theoretical results from literature. Furthermore, we observe a good agreement with our own theoretical derivation of the accumulated thermal conductivity $\kappa_{\text{cum}}(l_{\text{ph}})$ as a function of the phonon mean free path $l_{\text{ph}}$ by means of an \textit{ab initio} solution of the linearized phonon Boltzmann transport equation. Even though it remains challenging to extract quantitative information about $l_{\text{ph}}$ from our 1LRT measurements, it becomes clear that the local temperature rise $T_{\text{rise}}$ is over-predicted by Fourier's law, since both quasi-ballistic transport and non-equilibrium phonon distributions are neglected.

Our 1LRT measurements enable the experimental perspective of a local small heat source, which approaches the situation of a point heat source towards decreasing $w_{\text{e}}$ values. We wish to emphasize that our measured $\kappa_{\text{eff}}$ values that exceed $\kappa_{\text{bulk}}$ of silicon up to a factor of 8.3 at $T_{\text{amb}}\,=200\,\text{K}$ (is a factor of 5.3 at $T_{\text{amb}}\,=293\,\text{K}$) are real $\kappa$ values and not just an experimental artifact for the experimental situation of a local heat source embedded in a large matrix. Interestingly, this experimental situation mimics the situation of many heat sources that one can find in photonic and electronic applications (e.g., the center of a laser cavity, contact region of a transistor). The overall variation of our measured $\kappa_{\text{eff}}(w_{\text{e}}, h_{\alpha})$ values is agreeable large, which renders further PMFP spectroscopy by 1LRT a promising future aspect. Finally, our findings point out that quasi-ballistic phonon transport can challenge 1LRT measurements, while also enabling a strong perspective for PMFP spectroscopy.

In a future scenario, the entire spatial distribution of $T_{\text{rise}}(x,y,z)$ would be modeled for 1LRT and 2LRT measurements based on a differential real-space BTE solver with \textit{ab initio} input from our BTE calculations. As a result, one could directly extract, e.g., $\kappa_{\text{cum}}(l_{\text{ph}})$ from our PMFP spectroscopy based on 1LRT. Furthermore, from an experimental point of view, it would be promising to tune the heating-induced initial phonon distribution via the excitation laser energy to investigate how the initially generated phonon distribution impacts $\kappa_{\text{eff}}(w_{\text{e}})$ trends (see also Fig. \ref{fig:theo-expt-comp}a). As a result, one could clarify the role of potentially non-thermal electron and phonon distributions and extract more information about the particular phonon and herewith $l_{\text{ph}}$ distribution that is generated during the optical excitation of the silicon sample.


\section{\label{sec:ack}Acknowledgments}

N.H.P. acknowledges funding from the ``Deutsche Forschungsgemeinschaft" (DFG, German Research Foundation) for an ``Emmy Noether" research grant (Grant No. 534386252). G.R. acknowledges funding from the MIT-IBM Watson AI Laboratory (Challenge No. 2415). G.R. and G.C. acknowledge funding from the MIT Global Seed Funds. M.E. and G.C. acknowledge funding from the Central Research Development Fund (CRDF) of the University of Bremen for the project ``Joint optical and thermal designs for next generation nanophotonics". The research of K.D., M.E., G.W., J.T., J.L. and G.C. was further funded by the major research instrumentation program of the DFG (Grant No. 511416444). G.C. also acknowledges the MAPEX-CF Grant for Correlated Workflows (Grant No. 40401080) and funding from the MAPEX ``Minor Instrumentation Grant" associated to the APF program ``Materials on Demand" (MI06/25 and MI07/25). The authors acknowledge the Gauss Centre for Supercomputing e.V. (\url{www.gauss-centre.eu}) for funding this project by providing computing time on the GCS Supercomputer JUWELS \cite{JUWELS} at Jülich Supercomputing Centre (JSC).\\\\

\section*{Data Availability Statement}

The data that support the findings of this study are available from the corresponding author upon reasonable request.

\bibliography{MFP_references}

\begin{thebibliography}{87}%
\makeatletter
\providecommand \@ifxundefined [1]{%
 \@ifx{#1\undefined}
}%
\providecommand \@ifnum [1]{%
 \ifnum #1\expandafter \@firstoftwo
 \else \expandafter \@secondoftwo
 \fi
}%
\providecommand \@ifx [1]{%
 \ifx #1\expandafter \@firstoftwo
 \else \expandafter \@secondoftwo
 \fi
}%
\providecommand \natexlab [1]{#1}%
\providecommand \enquote  [1]{``#1''}%
\providecommand \bibnamefont  [1]{#1}%
\providecommand \bibfnamefont [1]{#1}%
\providecommand \citenamefont [1]{#1}%
\providecommand \href@noop [0]{\@secondoftwo}%
\providecommand \href [0]{\begingroup \@sanitize@url \@href}%
\providecommand \@href[1]{\@@startlink{#1}\@@href}%
\providecommand \@@href[1]{\endgroup#1\@@endlink}%
\providecommand \@sanitize@url [0]{\catcode `\\12\catcode `\$12\catcode
  `\&12\catcode `\#12\catcode `\^12\catcode `\_12\catcode `\%12\relax}%
\providecommand \@@startlink[1]{}%
\providecommand \@@endlink[0]{}%
\providecommand \url  [0]{\begingroup\@sanitize@url \@url }%
\providecommand \@url [1]{\endgroup\@href {#1}{\urlprefix }}%
\providecommand \urlprefix  [0]{URL }%
\providecommand \Eprint [0]{\href }%
\providecommand \doibase [0]{https://doi.org/}%
\providecommand \selectlanguage [0]{\@gobble}%
\providecommand \bibinfo  [0]{\@secondoftwo}%
\providecommand \bibfield  [0]{\@secondoftwo}%
\providecommand \translation [1]{[#1]}%
\providecommand \BibitemOpen [0]{}%
\providecommand \bibitemStop [0]{}%
\providecommand \bibitemNoStop [0]{.\EOS\space}%
\providecommand \EOS [0]{\spacefactor3000\relax}%
\providecommand \BibitemShut  [1]{\csname bibitem#1\endcsname}%
\let\auto@bib@innerbib\@empty
\bibitem [{\citenamefont {Chiloyan}\ \emph {et~al.}(2020)\citenamefont
  {Chiloyan}, \citenamefont {Huberman}, \citenamefont {Maznev}, \citenamefont
  {Nelson},\ and\ \citenamefont {Chen}}]{Chiloyan2020}%
  \BibitemOpen
  \bibfield  {author} {\bibinfo {author} {\bibfnamefont {V.}~\bibnamefont
  {Chiloyan}}, \bibinfo {author} {\bibfnamefont {S.}~\bibnamefont {Huberman}},
  \bibinfo {author} {\bibfnamefont {A.~A.}\ \bibnamefont {Maznev}}, \bibinfo
  {author} {\bibfnamefont {K.~A.}\ \bibnamefont {Nelson}},\ and\ \bibinfo
  {author} {\bibfnamefont {G.}~\bibnamefont {Chen}},\ }\bibfield  {title}
  {\bibinfo {title} {Thermal transport exceeding bulk heat conduction due to
  nonthermal micro/nanoscale phonon populations},\ }\href@noop {} {\bibfield
  {journal} {\bibinfo  {journal} {Applied Physics Letters}\ }\textbf {\bibinfo
  {volume} {116}} (\bibinfo {year} {2020})}\BibitemShut {NoStop}%
\bibitem [{\citenamefont {Elhajhasan}\ \emph {et~al.}(2023)\citenamefont
  {Elhajhasan}, \citenamefont {Seemann}, \citenamefont {Dudde}, \citenamefont
  {Vaske}, \citenamefont {Callsen}, \citenamefont {Rousseau}, \citenamefont
  {Weatherley}, \citenamefont {Carlin}, \citenamefont {Butté}, \citenamefont
  {Grandjean}, \citenamefont {Protik},\ and\ \citenamefont
  {Romano}}]{Elhajhasan2023}%
  \BibitemOpen
  \bibfield  {author} {\bibinfo {author} {\bibfnamefont {M.}~\bibnamefont
  {Elhajhasan}}, \bibinfo {author} {\bibfnamefont {W.}~\bibnamefont {Seemann}},
  \bibinfo {author} {\bibfnamefont {K.}~\bibnamefont {Dudde}}, \bibinfo
  {author} {\bibfnamefont {D.}~\bibnamefont {Vaske}}, \bibinfo {author}
  {\bibfnamefont {G.}~\bibnamefont {Callsen}}, \bibinfo {author} {\bibfnamefont
  {I.}~\bibnamefont {Rousseau}}, \bibinfo {author} {\bibfnamefont {T.~F.~K.}\
  \bibnamefont {Weatherley}}, \bibinfo {author} {\bibfnamefont {J.-F.}\
  \bibnamefont {Carlin}}, \bibinfo {author} {\bibfnamefont {R.}~\bibnamefont
  {Butté}}, \bibinfo {author} {\bibfnamefont {N.}~\bibnamefont {Grandjean}},
  \bibinfo {author} {\bibfnamefont {N.~H.}\ \bibnamefont {Protik}},\ and\
  \bibinfo {author} {\bibfnamefont {G.}~\bibnamefont {Romano}},\ }\bibfield
  {title} {\bibinfo {title} {Optical and thermal characterization of a
  group-{III} nitride semiconductor membrane by microphotoluminescence
  spectroscopy and raman thermometry},\ }\href
  {https://doi.org/10.1103/physrevb.108.235313} {\bibfield  {journal} {\bibinfo
   {journal} {Physical Review B}\ }\textbf {\bibinfo {volume} {108}},\ \bibinfo
  {pages} {235313} (\bibinfo {year} {2023})}\BibitemShut {NoStop}%
\bibitem [{\citenamefont {Wood}(1988)}]{Wood1988}%
  \BibitemOpen
  \bibfield  {author} {\bibinfo {author} {\bibfnamefont {C.}~\bibnamefont
  {Wood}},\ }\bibfield  {title} {\bibinfo {title} {Materials for thermoelectric
  energy conversion},\ }\href {https://doi.org/10.1088/0034-4885/51/4/001}
  {\bibfield  {journal} {\bibinfo  {journal} {Reports on Progress in Physics}\
  }\textbf {\bibinfo {volume} {51}},\ \bibinfo {pages} {459} (\bibinfo {year}
  {1988})}\BibitemShut {NoStop}%
\bibitem [{\citenamefont {Cahill}\ \emph {et~al.}(2014)\citenamefont {Cahill},
  \citenamefont {Braun}, \citenamefont {Chen}, \citenamefont {Clarke},
  \citenamefont {Fan}, \citenamefont {Goodson}, \citenamefont {Keblinski},
  \citenamefont {King}, \citenamefont {Mahan}, \citenamefont {Majumdar},
  \citenamefont {Maris}, \citenamefont {Phillpot}, \citenamefont {Pop},\ and\
  \citenamefont {Shi}}]{CahillReview2014}%
  \BibitemOpen
  \bibfield  {author} {\bibinfo {author} {\bibfnamefont {D.~G.}\ \bibnamefont
  {Cahill}}, \bibinfo {author} {\bibfnamefont {P.~V.}\ \bibnamefont {Braun}},
  \bibinfo {author} {\bibfnamefont {G.}~\bibnamefont {Chen}}, \bibinfo {author}
  {\bibfnamefont {D.~R.}\ \bibnamefont {Clarke}}, \bibinfo {author}
  {\bibfnamefont {S.}~\bibnamefont {Fan}}, \bibinfo {author} {\bibfnamefont
  {K.~E.}\ \bibnamefont {Goodson}}, \bibinfo {author} {\bibfnamefont
  {P.}~\bibnamefont {Keblinski}}, \bibinfo {author} {\bibfnamefont {W.~P.}\
  \bibnamefont {King}}, \bibinfo {author} {\bibfnamefont {G.~D.}\ \bibnamefont
  {Mahan}}, \bibinfo {author} {\bibfnamefont {A.}~\bibnamefont {Majumdar}},
  \bibinfo {author} {\bibfnamefont {H.~J.}\ \bibnamefont {Maris}}, \bibinfo
  {author} {\bibfnamefont {S.~R.}\ \bibnamefont {Phillpot}}, \bibinfo {author}
  {\bibfnamefont {E.}~\bibnamefont {Pop}},\ and\ \bibinfo {author}
  {\bibfnamefont {L.}~\bibnamefont {Shi}},\ }\bibfield  {title} {\bibinfo
  {title} {Nanoscale thermal transport. {II}. 2003–2012},\ }\href
  {https://doi.org/10.1063/1.4832615} {\bibfield  {journal} {\bibinfo
  {journal} {Applied Physics Reviews}\ }\textbf {\bibinfo {volume} {1}},\
  \bibinfo {pages} {011305} (\bibinfo {year} {2014})}\BibitemShut {NoStop}%
\bibitem [{\citenamefont {Funahashi~(ed.)}(2021)}]{Funahashi2021}%
  \BibitemOpen
  \bibfield  {author} {\bibinfo {author} {\bibfnamefont {R.}~\bibnamefont
  {Funahashi~(ed.)}},\ }\href@noop {} {\bibinfo {title} {Thermoelectric
  {E}nergy {C}onversion - {T}heories and {M}echanisms, {M}aterials, {D}evices,
  and {A}pplications}} (\bibinfo {year} {2021})\BibitemShut {NoStop}%
\bibitem [{\citenamefont {He}\ \emph {et~al.}(2017)\citenamefont {He},
  \citenamefont {Schierning},\ and\ \citenamefont {Nielsch}}]{HeReview2017}%
  \BibitemOpen
  \bibfield  {author} {\bibinfo {author} {\bibfnamefont {R.}~\bibnamefont
  {He}}, \bibinfo {author} {\bibfnamefont {G.}~\bibnamefont {Schierning}},\
  and\ \bibinfo {author} {\bibfnamefont {K.}~\bibnamefont {Nielsch}},\
  }\bibfield  {title} {\bibinfo {title} {Thermoelectric {D}evices: {A} {R}eview
  of {D}evices, {A}rchitectures, and {C}ontact {O}ptimization},\ }\href@noop {}
  {\bibfield  {journal} {\bibinfo  {journal} {Advanced Materials Technologies}\
  }\textbf {\bibinfo {volume} {3}} (\bibinfo {year} {2017})}\BibitemShut
  {NoStop}%
\bibitem [{\citenamefont {Taki}\ and\ \citenamefont
  {Strassburg}(2019)}]{TakiReview2019}%
  \BibitemOpen
  \bibfield  {author} {\bibinfo {author} {\bibfnamefont {T.}~\bibnamefont
  {Taki}}\ and\ \bibinfo {author} {\bibfnamefont {M.}~\bibnamefont
  {Strassburg}},\ }\bibfield  {title} {\bibinfo {title} {Review — {V}isible
  {LED}s: {M}ore than {E}fficient {L}ight},\ }\href
  {https://doi.org/10.1149/2.0402001jss} {\bibfield  {journal} {\bibinfo
  {journal} {ECS Journal of Solid State Science and Technology}\ }\textbf
  {\bibinfo {volume} {9}},\ \bibinfo {pages} {015017} (\bibinfo {year}
  {2019})}\BibitemShut {NoStop}%
\bibitem [{\citenamefont {Kuball}\ and\ \citenamefont
  {Pomeroy}(2016)}]{KuballReview2016}%
  \BibitemOpen
  \bibfield  {author} {\bibinfo {author} {\bibfnamefont {M.}~\bibnamefont
  {Kuball}}\ and\ \bibinfo {author} {\bibfnamefont {J.~W.}\ \bibnamefont
  {Pomeroy}},\ }\bibfield  {title} {\bibinfo {title} {A {R}eview of {R}aman
  {T}hermography for {E}lectronic and {O}pto-{E}lectronic {D}evice
  {M}easurement {W}ith {S}ubmicron {S}patial and {N}anosecond {T}emporal
  {R}esolution},\ }\href {https://doi.org/10.1109/tdmr.2016.2617458} {\bibfield
   {journal} {\bibinfo  {journal} {IEEE Transactions on Device and Materials
  Reliability}\ }\textbf {\bibinfo {volume} {16}},\ \bibinfo {pages} {667}
  (\bibinfo {year} {2016})}\BibitemShut {NoStop}%
\bibitem [{\citenamefont {Xu}\ \emph {et~al.}(2023)\citenamefont {Xu},
  \citenamefont {Shang}, \citenamefont {Liang}, \citenamefont {Ding},\ and\
  \citenamefont {Ren}}]{XuBookchapter2023}%
  \BibitemOpen
  \bibfield  {author} {\bibinfo {author} {\bibfnamefont {C.}~\bibnamefont
  {Xu}}, \bibinfo {author} {\bibfnamefont {H.}~\bibnamefont {Shang}}, \bibinfo
  {author} {\bibfnamefont {Z.}~\bibnamefont {Liang}}, \bibinfo {author}
  {\bibfnamefont {F.}~\bibnamefont {Ding}},\ and\ \bibinfo {author}
  {\bibfnamefont {Z.}~\bibnamefont {Ren}},\ }\href
  {https://doi.org/10.1002/9781394256389.ch1} {\bibinfo {title} {Reliability
  and {D}urability of {T}hermoelectric {M}aterials and {D}evices: {P}resent
  {S}tatus and {S}trategies for {I}mprovement}} (\bibinfo {year}
  {2023})\BibitemShut {NoStop}%
\bibitem [{\citenamefont {Capinski}\ \emph {et~al.}(1999)\citenamefont
  {Capinski}, \citenamefont {Maris}, \citenamefont {Ruf}, \citenamefont
  {Cardona}, \citenamefont {Ploog},\ and\ \citenamefont
  {Katzer}}]{Capinski1999}%
  \BibitemOpen
  \bibfield  {author} {\bibinfo {author} {\bibfnamefont {W.~S.}\ \bibnamefont
  {Capinski}}, \bibinfo {author} {\bibfnamefont {H.~J.}\ \bibnamefont {Maris}},
  \bibinfo {author} {\bibfnamefont {T.}~\bibnamefont {Ruf}}, \bibinfo {author}
  {\bibfnamefont {M.}~\bibnamefont {Cardona}}, \bibinfo {author} {\bibfnamefont
  {K.}~\bibnamefont {Ploog}},\ and\ \bibinfo {author} {\bibfnamefont {D.~S.}\
  \bibnamefont {Katzer}},\ }\bibfield  {title} {\bibinfo {title}
  {Thermal-conductivity measurements of {G}a{A}s/{A}l{A}s superlattices using a
  picosecond optical pump-and-probe technique},\ }\href
  {https://doi.org/10.1103/physrevb.59.8105} {\bibfield  {journal} {\bibinfo
  {journal} {Physical Review B}\ }\textbf {\bibinfo {volume} {59}},\ \bibinfo
  {pages} {8105} (\bibinfo {year} {1999})}\BibitemShut {NoStop}%
\bibitem [{\citenamefont {Koh}\ \emph {et~al.}(2009)\citenamefont {Koh},
  \citenamefont {Cao}, \citenamefont {Cahill},\ and\ \citenamefont
  {Jena}}]{KohSuperlattices2009}%
  \BibitemOpen
  \bibfield  {author} {\bibinfo {author} {\bibfnamefont {Y.~K.}\ \bibnamefont
  {Koh}}, \bibinfo {author} {\bibfnamefont {Y.}~\bibnamefont {Cao}}, \bibinfo
  {author} {\bibfnamefont {D.~G.}\ \bibnamefont {Cahill}},\ and\ \bibinfo
  {author} {\bibfnamefont {D.}~\bibnamefont {Jena}},\ }\bibfield  {title}
  {\bibinfo {title} {Heat‐{T}ransport {M}echanisms in {S}uperlattices},\
  }\href {https://doi.org/10.1002/adfm.200800984} {\bibfield  {journal}
  {\bibinfo  {journal} {Advanced Functional Materials}\ }\textbf {\bibinfo
  {volume} {19}},\ \bibinfo {pages} {610} (\bibinfo {year} {2009})}\BibitemShut
  {NoStop}%
\bibitem [{\citenamefont {Luckyanova}\ \emph {et~al.}(2012)\citenamefont
  {Luckyanova}, \citenamefont {Garg}, \citenamefont {Esfarjani}, \citenamefont
  {Jandl}, \citenamefont {Bulsara}, \citenamefont {Schmidt}, \citenamefont
  {Minnich}, \citenamefont {Chen}, \citenamefont {Dresselhaus}, \citenamefont
  {Ren}, \citenamefont {Fitzgerald},\ and\ \citenamefont
  {Chen}}]{Luckyanova2012}%
  \BibitemOpen
  \bibfield  {author} {\bibinfo {author} {\bibfnamefont {M.~N.}\ \bibnamefont
  {Luckyanova}}, \bibinfo {author} {\bibfnamefont {J.}~\bibnamefont {Garg}},
  \bibinfo {author} {\bibfnamefont {K.}~\bibnamefont {Esfarjani}}, \bibinfo
  {author} {\bibfnamefont {A.}~\bibnamefont {Jandl}}, \bibinfo {author}
  {\bibfnamefont {M.~T.}\ \bibnamefont {Bulsara}}, \bibinfo {author}
  {\bibfnamefont {A.~J.}\ \bibnamefont {Schmidt}}, \bibinfo {author}
  {\bibfnamefont {A.~J.}\ \bibnamefont {Minnich}}, \bibinfo {author}
  {\bibfnamefont {S.}~\bibnamefont {Chen}}, \bibinfo {author} {\bibfnamefont
  {M.~S.}\ \bibnamefont {Dresselhaus}}, \bibinfo {author} {\bibfnamefont
  {Z.}~\bibnamefont {Ren}}, \bibinfo {author} {\bibfnamefont {E.~A.}\
  \bibnamefont {Fitzgerald}},\ and\ \bibinfo {author} {\bibfnamefont
  {G.}~\bibnamefont {Chen}},\ }\bibfield  {title} {\bibinfo {title} {Coherent
  {P}honon {H}eat {C}onduction in {S}uperlattices},\ }\href
  {https://doi.org/10.1126/science.1225549} {\bibfield  {journal} {\bibinfo
  {journal} {Science}\ }\textbf {\bibinfo {volume} {338}},\ \bibinfo {pages}
  {936} (\bibinfo {year} {2012})}\BibitemShut {NoStop}%
\bibitem [{\citenamefont {Ravichandran}\ \emph {et~al.}(2013)\citenamefont
  {Ravichandran}, \citenamefont {Yadav}, \citenamefont {Cheaito}, \citenamefont
  {Rossen}, \citenamefont {Soukiassian}, \citenamefont {Suresha}, \citenamefont
  {Duda}, \citenamefont {Foley}, \citenamefont {Lee}, \citenamefont {Zhu},
  \citenamefont {Lichtenberger}, \citenamefont {Moore}, \citenamefont {Muller},
  \citenamefont {Schlom}, \citenamefont {Hopkins}, \citenamefont {Majumdar},
  \citenamefont {Ramesh},\ and\ \citenamefont {Zurbuchen}}]{Ravichandran2013}%
  \BibitemOpen
  \bibfield  {author} {\bibinfo {author} {\bibfnamefont {J.}~\bibnamefont
  {Ravichandran}}, \bibinfo {author} {\bibfnamefont {A.~K.}\ \bibnamefont
  {Yadav}}, \bibinfo {author} {\bibfnamefont {R.}~\bibnamefont {Cheaito}},
  \bibinfo {author} {\bibfnamefont {P.~B.}\ \bibnamefont {Rossen}}, \bibinfo
  {author} {\bibfnamefont {A.}~\bibnamefont {Soukiassian}}, \bibinfo {author}
  {\bibfnamefont {S.~J.}\ \bibnamefont {Suresha}}, \bibinfo {author}
  {\bibfnamefont {J.~C.}\ \bibnamefont {Duda}}, \bibinfo {author}
  {\bibfnamefont {B.~M.}\ \bibnamefont {Foley}}, \bibinfo {author}
  {\bibfnamefont {C.-H.}\ \bibnamefont {Lee}}, \bibinfo {author} {\bibfnamefont
  {Y.}~\bibnamefont {Zhu}}, \bibinfo {author} {\bibfnamefont {A.~W.}\
  \bibnamefont {Lichtenberger}}, \bibinfo {author} {\bibfnamefont {J.~E.}\
  \bibnamefont {Moore}}, \bibinfo {author} {\bibfnamefont {D.~A.}\ \bibnamefont
  {Muller}}, \bibinfo {author} {\bibfnamefont {D.~G.}\ \bibnamefont {Schlom}},
  \bibinfo {author} {\bibfnamefont {P.~E.}\ \bibnamefont {Hopkins}}, \bibinfo
  {author} {\bibfnamefont {A.}~\bibnamefont {Majumdar}}, \bibinfo {author}
  {\bibfnamefont {R.}~\bibnamefont {Ramesh}},\ and\ \bibinfo {author}
  {\bibfnamefont {M.~A.}\ \bibnamefont {Zurbuchen}},\ }\bibfield  {title}
  {\bibinfo {title} {Crossover from incoherent to coherent phonon scattering in
  epitaxial oxide superlattices},\ }\href@noop {} {\bibfield  {journal}
  {\bibinfo  {journal} {Nature Materials}\ }\textbf {\bibinfo {volume} {13}},\
  \bibinfo {pages} {168} (\bibinfo {year} {2013})}\BibitemShut {NoStop}%
\bibitem [{\citenamefont {Ghannam}\ \emph {et~al.}(2023)\citenamefont
  {Ghannam}, \citenamefont {Moll}, \citenamefont {Bérardan}, \citenamefont
  {Coulomb}, \citenamefont {Vieira-E-Silva}, \citenamefont {Villeroy},
  \citenamefont {Viennois},\ and\ \citenamefont {Beaudhuin}}]{Ghannam2023}%
  \BibitemOpen
  \bibfield  {author} {\bibinfo {author} {\bibfnamefont {R.}~\bibnamefont
  {Ghannam}}, \bibinfo {author} {\bibfnamefont {A.}~\bibnamefont {Moll}},
  \bibinfo {author} {\bibfnamefont {D.}~\bibnamefont {Bérardan}}, \bibinfo
  {author} {\bibfnamefont {L.}~\bibnamefont {Coulomb}}, \bibinfo {author}
  {\bibfnamefont {A.}~\bibnamefont {Vieira-E-Silva}}, \bibinfo {author}
  {\bibfnamefont {B.}~\bibnamefont {Villeroy}}, \bibinfo {author}
  {\bibfnamefont {R.}~\bibnamefont {Viennois}},\ and\ \bibinfo {author}
  {\bibfnamefont {M.}~\bibnamefont {Beaudhuin}},\ }\bibfield  {title} {\bibinfo
  {title} {Impact of the nanostructuring on the thermal and thermoelectric
  properties of $\alpha$-{S}r{S}i$_2$},\ }\href
  {https://doi.org/10.1016/j.jallcom.2023.171876} {\bibfield  {journal}
  {\bibinfo  {journal} {Journal of Alloys and Compounds}\ }\textbf {\bibinfo
  {volume} {968}},\ \bibinfo {pages} {171876} (\bibinfo {year}
  {2023})}\BibitemShut {NoStop}%
\bibitem [{\citenamefont {Gurunathan}\ \emph {et~al.}(2020)\citenamefont
  {Gurunathan}, \citenamefont {Hanus},\ and\ \citenamefont
  {Snyder}}]{Gurunathan2020}%
  \BibitemOpen
  \bibfield  {author} {\bibinfo {author} {\bibfnamefont {R.}~\bibnamefont
  {Gurunathan}}, \bibinfo {author} {\bibfnamefont {R.}~\bibnamefont {Hanus}},\
  and\ \bibinfo {author} {\bibfnamefont {G.~J.}\ \bibnamefont {Snyder}},\
  }\bibfield  {title} {\bibinfo {title} {Alloy scattering of phonons},\ }\href
  {https://doi.org/10.1039/c9mh01990a} {\bibfield  {journal} {\bibinfo
  {journal} {Materials Horizons}\ }\textbf {\bibinfo {volume} {7}},\ \bibinfo
  {pages} {1452} (\bibinfo {year} {2020})}\BibitemShut {NoStop}%
\bibitem [{\citenamefont {Zhang}\ \emph {et~al.}(2020)\citenamefont {Zhang},
  \citenamefont {Ouyang}, \citenamefont {Cheng}, \citenamefont {Chen},
  \citenamefont {Li},\ and\ \citenamefont {Zhang}}]{Zhang2020}%
  \BibitemOpen
  \bibfield  {author} {\bibinfo {author} {\bibfnamefont {Z.}~\bibnamefont
  {Zhang}}, \bibinfo {author} {\bibfnamefont {Y.}~\bibnamefont {Ouyang}},
  \bibinfo {author} {\bibfnamefont {Y.}~\bibnamefont {Cheng}}, \bibinfo
  {author} {\bibfnamefont {J.}~\bibnamefont {Chen}}, \bibinfo {author}
  {\bibfnamefont {N.}~\bibnamefont {Li}},\ and\ \bibinfo {author}
  {\bibfnamefont {G.}~\bibnamefont {Zhang}},\ }\bibfield  {title} {\bibinfo
  {title} {Size-dependent phononic thermal transport in low-dimensional
  nanomaterials},\ }\href {https://doi.org/10.1016/j.physrep.2020.03.001}
  {\bibfield  {journal} {\bibinfo  {journal} {Physics Reports}\ }\textbf
  {\bibinfo {volume} {860}},\ \bibinfo {pages} {1} (\bibinfo {year}
  {2020})}\BibitemShut {NoStop}%
\bibitem [{\citenamefont {Huang}\ \emph {et~al.}(2022)\citenamefont {Huang},
  \citenamefont {Fan}, \citenamefont {Sang}, \citenamefont {Mei}, \citenamefont
  {Ying}, \citenamefont {Zhang},\ and\ \citenamefont {Long}}]{Huang2022}%
  \BibitemOpen
  \bibfield  {author} {\bibinfo {author} {\bibfnamefont {L.}~\bibnamefont
  {Huang}}, \bibinfo {author} {\bibfnamefont {S.}~\bibnamefont {Fan}}, \bibinfo
  {author} {\bibfnamefont {L.}~\bibnamefont {Sang}}, \bibinfo {author}
  {\bibfnamefont {Y.}~\bibnamefont {Mei}}, \bibinfo {author} {\bibfnamefont
  {L.}~\bibnamefont {Ying}}, \bibinfo {author} {\bibfnamefont {B.}~\bibnamefont
  {Zhang}},\ and\ \bibinfo {author} {\bibfnamefont {H.}~\bibnamefont {Long}},\
  }\bibfield  {title} {\bibinfo {title} {Thermal conductivity and phonon
  scattering of {A}l{G}a{N} nanofilms by elastic theory and {B}oltzmann
  transport equation},\ }\href {https://doi.org/10.1088/1361-6641/ac5293}
  {\bibfield  {journal} {\bibinfo  {journal} {Semiconductor Science and
  Technology}\ }\textbf {\bibinfo {volume} {37}},\ \bibinfo {pages} {055003}
  (\bibinfo {year} {2022})}\BibitemShut {NoStop}%
\bibitem [{\citenamefont {Tran}\ \emph {et~al.}(2022)\citenamefont {Tran},
  \citenamefont {Carrascon}, \citenamefont {Iwaya}, \citenamefont {Monemar},
  \citenamefont {Darakchieva},\ and\ \citenamefont {Paskov}}]{Tran2022}%
  \BibitemOpen
  \bibfield  {author} {\bibinfo {author} {\bibfnamefont {D.~Q.}\ \bibnamefont
  {Tran}}, \bibinfo {author} {\bibfnamefont {R.~D.}\ \bibnamefont {Carrascon}},
  \bibinfo {author} {\bibfnamefont {M.}~\bibnamefont {Iwaya}}, \bibinfo
  {author} {\bibfnamefont {B.}~\bibnamefont {Monemar}}, \bibinfo {author}
  {\bibfnamefont {V.}~\bibnamefont {Darakchieva}},\ and\ \bibinfo {author}
  {\bibfnamefont {P.~P.}\ \bibnamefont {Paskov}},\ }\bibfield  {title}
  {\bibinfo {title} {Thermal conductivity of {A}l$_x${G}a$_{1-x}${N} (0 $\leq x
  \leq$ 1) epitaxial layers},\ }\href
  {https://doi.org/10.1103/physrevmaterials.6.104602} {\bibfield  {journal}
  {\bibinfo  {journal} {Physical Review Materials}\ }\textbf {\bibinfo {volume}
  {6}},\ \bibinfo {pages} {104602} (\bibinfo {year} {2022})}\BibitemShut
  {NoStop}%
\bibitem [{\citenamefont {Capinski}\ \emph {et~al.}(1997)\citenamefont
  {Capinski}, \citenamefont {Maris}, \citenamefont {Bauser}, \citenamefont
  {Silier}, \citenamefont {Asen-Palmer}, \citenamefont {Ruf}, \citenamefont
  {Cardona},\ and\ \citenamefont {Gmelin}}]{Capinski1997}%
  \BibitemOpen
  \bibfield  {author} {\bibinfo {author} {\bibfnamefont {W.~S.}\ \bibnamefont
  {Capinski}}, \bibinfo {author} {\bibfnamefont {H.~J.}\ \bibnamefont {Maris}},
  \bibinfo {author} {\bibfnamefont {E.}~\bibnamefont {Bauser}}, \bibinfo
  {author} {\bibfnamefont {I.}~\bibnamefont {Silier}}, \bibinfo {author}
  {\bibfnamefont {M.}~\bibnamefont {Asen-Palmer}}, \bibinfo {author}
  {\bibfnamefont {T.}~\bibnamefont {Ruf}}, \bibinfo {author} {\bibfnamefont
  {M.}~\bibnamefont {Cardona}},\ and\ \bibinfo {author} {\bibfnamefont
  {E.}~\bibnamefont {Gmelin}},\ }\bibfield  {title} {\bibinfo {title} {Thermal
  conductivity of isotopically enriched {S}i},\ }\href
  {https://doi.org/10.1063/1.119384} {\bibfield  {journal} {\bibinfo  {journal}
  {Applied Physics Letters}\ }\textbf {\bibinfo {volume} {71}},\ \bibinfo
  {pages} {2109} (\bibinfo {year} {1997})}\BibitemShut {NoStop}%
\bibitem [{\citenamefont {Beechem}\ \emph {et~al.}(2016)\citenamefont
  {Beechem}, \citenamefont {McDonald}, \citenamefont {Fuller}, \citenamefont
  {Talin}, \citenamefont {Rost}, \citenamefont {Maria}, \citenamefont
  {Gaskins}, \citenamefont {Hopkins},\ and\ \citenamefont
  {Allerman}}]{Beechem2016}%
  \BibitemOpen
  \bibfield  {author} {\bibinfo {author} {\bibfnamefont {T.~E.}\ \bibnamefont
  {Beechem}}, \bibinfo {author} {\bibfnamefont {A.~E.}\ \bibnamefont
  {McDonald}}, \bibinfo {author} {\bibfnamefont {E.~J.}\ \bibnamefont
  {Fuller}}, \bibinfo {author} {\bibfnamefont {A.~A.}\ \bibnamefont {Talin}},
  \bibinfo {author} {\bibfnamefont {C.~M.}\ \bibnamefont {Rost}}, \bibinfo
  {author} {\bibfnamefont {J.-P.}\ \bibnamefont {Maria}}, \bibinfo {author}
  {\bibfnamefont {J.~T.}\ \bibnamefont {Gaskins}}, \bibinfo {author}
  {\bibfnamefont {P.~E.}\ \bibnamefont {Hopkins}},\ and\ \bibinfo {author}
  {\bibfnamefont {A.~A.}\ \bibnamefont {Allerman}},\ }\bibfield  {title}
  {\bibinfo {title} {Size dictated thermal conductivity of {G}a{N}},\
  }\href@noop {} {\bibfield  {journal} {\bibinfo  {journal} {Journal of Applied
  Physics}\ }\textbf {\bibinfo {volume} {120}} (\bibinfo {year}
  {2016})}\BibitemShut {NoStop}%
\bibitem [{\citenamefont {Inyushkin}\ \emph {et~al.}(2018)\citenamefont
  {Inyushkin}, \citenamefont {Taldenkov}, \citenamefont {Ager}, \citenamefont
  {Haller}, \citenamefont {Riemann}, \citenamefont {Abrosimov}, \citenamefont
  {Pohl},\ and\ \citenamefont {Becker}}]{Inyushkin2018}%
  \BibitemOpen
  \bibfield  {author} {\bibinfo {author} {\bibfnamefont {A.~V.}\ \bibnamefont
  {Inyushkin}}, \bibinfo {author} {\bibfnamefont {A.~N.}\ \bibnamefont
  {Taldenkov}}, \bibinfo {author} {\bibfnamefont {J.~W.}\ \bibnamefont {Ager}},
  \bibinfo {author} {\bibfnamefont {E.~E.}\ \bibnamefont {Haller}}, \bibinfo
  {author} {\bibfnamefont {H.}~\bibnamefont {Riemann}}, \bibinfo {author}
  {\bibfnamefont {N.~V.}\ \bibnamefont {Abrosimov}}, \bibinfo {author}
  {\bibfnamefont {H.-J.}\ \bibnamefont {Pohl}},\ and\ \bibinfo {author}
  {\bibfnamefont {P.}~\bibnamefont {Becker}},\ }\bibfield  {title} {\bibinfo
  {title} {Ultrahigh thermal conductivity of isotopically enriched silicon},\
  }\bibfield  {journal} {\bibinfo  {journal} {Journal of Applied Physics}\
  }\textbf {\bibinfo {volume} {123}},\ \href
  {https://doi.org/10.1063/1.5017778} {10.1063/1.5017778} (\bibinfo {year}
  {2018})\BibitemShut {NoStop}%
\bibitem [{\citenamefont {Cai}\ \emph {et~al.}(2020)\citenamefont {Cai},
  \citenamefont {Scullion}, \citenamefont {Gan}, \citenamefont {Falin},
  \citenamefont {Cizek}, \citenamefont {Liu}, \citenamefont {Edgar},
  \citenamefont {Liu}, \citenamefont {Cowie}, \citenamefont {Santos},\ and\
  \citenamefont {Li}}]{Cai2020}%
  \BibitemOpen
  \bibfield  {author} {\bibinfo {author} {\bibfnamefont {Q.}~\bibnamefont
  {Cai}}, \bibinfo {author} {\bibfnamefont {D.}~\bibnamefont {Scullion}},
  \bibinfo {author} {\bibfnamefont {W.}~\bibnamefont {Gan}}, \bibinfo {author}
  {\bibfnamefont {A.}~\bibnamefont {Falin}}, \bibinfo {author} {\bibfnamefont
  {P.}~\bibnamefont {Cizek}}, \bibinfo {author} {\bibfnamefont
  {S.}~\bibnamefont {Liu}}, \bibinfo {author} {\bibfnamefont {J.~H.}\
  \bibnamefont {Edgar}}, \bibinfo {author} {\bibfnamefont {R.}~\bibnamefont
  {Liu}}, \bibinfo {author} {\bibfnamefont {B.~C.~C.}\ \bibnamefont {Cowie}},
  \bibinfo {author} {\bibfnamefont {E.~J.~G.}\ \bibnamefont {Santos}},\ and\
  \bibinfo {author} {\bibfnamefont {L.~H.}\ \bibnamefont {Li}},\ }\bibfield
  {title} {\bibinfo {title} {Outstanding {T}hermal {C}onductivity of {S}ingle
  {A}tomic {L}ayer {I}sotope-{M}odified {B}oron {N}itride},\ }\href
  {https://doi.org/10.1103/physrevlett.125.085902} {\bibfield  {journal}
  {\bibinfo  {journal} {Physical Review Letters}\ }\textbf {\bibinfo {volume}
  {125}},\ \bibinfo {pages} {085902} (\bibinfo {year} {2020})}\BibitemShut
  {NoStop}%
\bibitem [{\citenamefont {Hu}\ \emph {et~al.}(2024)\citenamefont {Hu},
  \citenamefont {Xu}, \citenamefont {Ruan},\ and\ \citenamefont
  {Bao}}]{HuRuan2024}%
  \BibitemOpen
  \bibfield  {author} {\bibinfo {author} {\bibfnamefont {Y.}~\bibnamefont
  {Hu}}, \bibinfo {author} {\bibfnamefont {J.}~\bibnamefont {Xu}}, \bibinfo
  {author} {\bibfnamefont {X.}~\bibnamefont {Ruan}},\ and\ \bibinfo {author}
  {\bibfnamefont {H.}~\bibnamefont {Bao}},\ }\bibfield  {title} {\bibinfo
  {title} {Defect scattering can lead to enhanced phonon transport at
  nanoscale},\ }\href@noop {} {\bibfield  {journal} {\bibinfo  {journal}
  {Nature Communications}\ }\textbf {\bibinfo {volume} {15}} (\bibinfo {year}
  {2024})}\BibitemShut {NoStop}%
\bibitem [{\citenamefont {Regner}\ \emph {et~al.}(2015)\citenamefont {Regner},
  \citenamefont {Freedman},\ and\ \citenamefont {Malen}}]{RegnerReview2015}%
  \BibitemOpen
  \bibfield  {author} {\bibinfo {author} {\bibfnamefont {K.~T.}\ \bibnamefont
  {Regner}}, \bibinfo {author} {\bibfnamefont {J.~P.}\ \bibnamefont
  {Freedman}},\ and\ \bibinfo {author} {\bibfnamefont {J.~A.}\ \bibnamefont
  {Malen}},\ }\bibfield  {title} {\bibinfo {title} {{A}dvances in {S}tudying
  {P}honon {M}ean {F}ree {P}ath {D}ependent {C}ontributions to {T}hermal
  {C}onductivity},\ }\href {https://doi.org/10.1080/15567265.2015.1045640}
  {\bibfield  {journal} {\bibinfo  {journal} {Nanoscale and Microscale
  Thermophysical Engineering}\ }\textbf {\bibinfo {volume} {19}},\ \bibinfo
  {pages} {183} (\bibinfo {year} {2015})}\BibitemShut {NoStop}%
\bibitem [{\citenamefont {Tian}\ \emph {et~al.}(2011)\citenamefont {Tian},
  \citenamefont {Esfarjani}, \citenamefont {Shiomi}, \citenamefont {Henry},\
  and\ \citenamefont {Chen}}]{Tian2011}%
  \BibitemOpen
  \bibfield  {author} {\bibinfo {author} {\bibfnamefont {Z.}~\bibnamefont
  {Tian}}, \bibinfo {author} {\bibfnamefont {K.}~\bibnamefont {Esfarjani}},
  \bibinfo {author} {\bibfnamefont {J.}~\bibnamefont {Shiomi}}, \bibinfo
  {author} {\bibfnamefont {A.~S.}\ \bibnamefont {Henry}},\ and\ \bibinfo
  {author} {\bibfnamefont {G.}~\bibnamefont {Chen}},\ }\bibfield  {title}
  {\bibinfo {title} {On the importance of optical phonons to thermal
  conductivity in nanostructures},\ }\href@noop {} {\bibfield  {journal}
  {\bibinfo  {journal} {Applied Physics Letters}\ }\textbf {\bibinfo {volume}
  {99}} (\bibinfo {year} {2011})}\BibitemShut {NoStop}%
\bibitem [{\citenamefont {Esfarjani}\ \emph {et~al.}(2011)\citenamefont
  {Esfarjani}, \citenamefont {Chen},\ and\ \citenamefont
  {Stokes}}]{EsfarjaniChenStokes2011}%
  \BibitemOpen
  \bibfield  {author} {\bibinfo {author} {\bibfnamefont {K.}~\bibnamefont
  {Esfarjani}}, \bibinfo {author} {\bibfnamefont {G.}~\bibnamefont {Chen}},\
  and\ \bibinfo {author} {\bibfnamefont {H.~T.}\ \bibnamefont {Stokes}},\
  }\bibfield  {title} {\bibinfo {title} {Heat transport in silicon from
  first-principles calculations},\ }\href
  {https://doi.org/10.1103/physrevb.84.085204} {\bibfield  {journal} {\bibinfo
  {journal} {Physical Review B}\ }\textbf {\bibinfo {volume} {84}},\ \bibinfo
  {pages} {085204} (\bibinfo {year} {2011})}\BibitemShut {NoStop}%
\bibitem [{\citenamefont {Cheng}\ \emph {et~al.}(2021)\citenamefont {Cheng},
  \citenamefont {Lu}, \citenamefont {Shi}, \citenamefont {Tanaka},
  \citenamefont {Protik}, \citenamefont {Wang}, \citenamefont {Iwaya},
  \citenamefont {Takeuchi}, \citenamefont {Kamiyama}, \citenamefont {Akasaki},
  \citenamefont {Amano},\ and\ \citenamefont {Graham}}]{Cheng2021}%
  \BibitemOpen
  \bibfield  {author} {\bibinfo {author} {\bibfnamefont {Z.}~\bibnamefont
  {Cheng}}, \bibinfo {author} {\bibfnamefont {W.}~\bibnamefont {Lu}}, \bibinfo
  {author} {\bibfnamefont {J.}~\bibnamefont {Shi}}, \bibinfo {author}
  {\bibfnamefont {D.}~\bibnamefont {Tanaka}}, \bibinfo {author} {\bibfnamefont
  {N.}~\bibnamefont {Protik}}, \bibinfo {author} {\bibfnamefont
  {S.}~\bibnamefont {Wang}}, \bibinfo {author} {\bibfnamefont {M.}~\bibnamefont
  {Iwaya}}, \bibinfo {author} {\bibfnamefont {T.}~\bibnamefont {Takeuchi}},
  \bibinfo {author} {\bibfnamefont {S.}~\bibnamefont {Kamiyama}}, \bibinfo
  {author} {\bibfnamefont {I.}~\bibnamefont {Akasaki}}, \bibinfo {author}
  {\bibfnamefont {H.}~\bibnamefont {Amano}},\ and\ \bibinfo {author}
  {\bibfnamefont {S.}~\bibnamefont {Graham}},\ }\bibfield  {title} {\bibinfo
  {title} {Quasi-ballistic thermal conduction in 6{H}–{S}i{C}},\ }\href
  {https://doi.org/10.1016/j.mtphys.2021.100462} {\bibfield  {journal}
  {\bibinfo  {journal} {Materials Today Physics}\ }\textbf {\bibinfo {volume}
  {20}},\ \bibinfo {pages} {100462} (\bibinfo {year} {2021})}\BibitemShut
  {NoStop}%
\bibitem [{\citenamefont {Li}\ \emph {et~al.}(2014{\natexlab{a}})\citenamefont
  {Li}, \citenamefont {Carrete}, \citenamefont {A.~Katcho},\ and\ \citenamefont
  {Mingo}}]{ShengBTE2014}%
  \BibitemOpen
  \bibfield  {author} {\bibinfo {author} {\bibfnamefont {W.}~\bibnamefont
  {Li}}, \bibinfo {author} {\bibfnamefont {J.}~\bibnamefont {Carrete}},
  \bibinfo {author} {\bibfnamefont {N.}~\bibnamefont {A.~Katcho}},\ and\
  \bibinfo {author} {\bibfnamefont {N.}~\bibnamefont {Mingo}},\ }\bibfield
  {title} {\bibinfo {title} {Sheng{BTE}: {A} solver of the {B}oltzmann
  transport equation for phonons},\ }\href
  {https://doi.org/10.1016/j.cpc.2014.02.015} {\bibfield  {journal} {\bibinfo
  {journal} {Computer Physics Communications}\ }\textbf {\bibinfo {volume}
  {185}},\ \bibinfo {pages} {1747} (\bibinfo {year}
  {2014}{\natexlab{a}})}\BibitemShut {NoStop}%
\bibitem [{\citenamefont {Protik}\ \emph {et~al.}(2022)\citenamefont {Protik},
  \citenamefont {Li}, \citenamefont {Pruneda}, \citenamefont {Broido},\ and\
  \citenamefont {Ordejón}}]{ProtikElphbolt2022}%
  \BibitemOpen
  \bibfield  {author} {\bibinfo {author} {\bibfnamefont {N.~H.}\ \bibnamefont
  {Protik}}, \bibinfo {author} {\bibfnamefont {C.}~\bibnamefont {Li}}, \bibinfo
  {author} {\bibfnamefont {M.}~\bibnamefont {Pruneda}}, \bibinfo {author}
  {\bibfnamefont {D.}~\bibnamefont {Broido}},\ and\ \bibinfo {author}
  {\bibfnamefont {P.}~\bibnamefont {Ordejón}},\ }\bibfield  {title} {\bibinfo
  {title} {The elphbolt ab initio solver for the coupled electron-phonon
  {B}oltzmann transport equations},\ }\href@noop {} {\bibfield  {journal}
  {\bibinfo  {journal} {npj Computational Materials}\ }\textbf {\bibinfo
  {volume} {8}} (\bibinfo {year} {2022})}\BibitemShut {NoStop}%
\bibitem [{\citenamefont {Lindsay}\ \emph {et~al.}(2012)\citenamefont
  {Lindsay}, \citenamefont {Broido},\ and\ \citenamefont
  {Reinecke}}]{LindsayBroido2012}%
  \BibitemOpen
  \bibfield  {author} {\bibinfo {author} {\bibfnamefont {L.}~\bibnamefont
  {Lindsay}}, \bibinfo {author} {\bibfnamefont {D.~A.}\ \bibnamefont
  {Broido}},\ and\ \bibinfo {author} {\bibfnamefont {T.~L.}\ \bibnamefont
  {Reinecke}},\ }\bibfield  {title} {\bibinfo {title} {Thermal {C}onductivity
  and {L}arge {I}sotope {E}ffect in {G}a{N} from {F}irst {P}rinciples},\ }\href
  {https://doi.org/10.1103/physrevlett.109.095901} {\bibfield  {journal}
  {\bibinfo  {journal} {Physical Review Letters}\ }\textbf {\bibinfo {volume}
  {109}},\ \bibinfo {pages} {095901} (\bibinfo {year} {2012})}\BibitemShut
  {NoStop}%
\bibitem [{\citenamefont {Ravichandran}\ and\ \citenamefont
  {Broido}(2020)}]{RavichandranBroido2020}%
  \BibitemOpen
  \bibfield  {author} {\bibinfo {author} {\bibfnamefont {N.~K.}\ \bibnamefont
  {Ravichandran}}\ and\ \bibinfo {author} {\bibfnamefont {D.}~\bibnamefont
  {Broido}},\ }\bibfield  {title} {\bibinfo {title} {Phonon-{P}honon
  {I}nteractions in {S}trongly {B}onded {S}olids: {S}election {R}ules and
  {H}igher-{O}rder {P}rocesses},\ }\href
  {https://doi.org/10.1103/physrevx.10.021063} {\bibfield  {journal} {\bibinfo
  {journal} {Physical Review X}\ }\textbf {\bibinfo {volume} {10}},\ \bibinfo
  {pages} {021063} (\bibinfo {year} {2020})}\BibitemShut {NoStop}%
\bibitem [{\citenamefont {Yang}\ and\ \citenamefont
  {Dames}(2013)}]{YangDames2013}%
  \BibitemOpen
  \bibfield  {author} {\bibinfo {author} {\bibfnamefont {F.}~\bibnamefont
  {Yang}}\ and\ \bibinfo {author} {\bibfnamefont {C.}~\bibnamefont {Dames}},\
  }\bibfield  {title} {\bibinfo {title} {Mean free path spectra as a tool to
  understand thermal conductivity in bulk and nanostructures},\ }\href
  {https://doi.org/10.1103/physrevb.87.035437} {\bibfield  {journal} {\bibinfo
  {journal} {Physical Review B}\ }\textbf {\bibinfo {volume} {87}},\ \bibinfo
  {pages} {035437} (\bibinfo {year} {2013})}\BibitemShut {NoStop}%
\bibitem [{\citenamefont {Regner}\ \emph
  {et~al.}(2013{\natexlab{a}})\citenamefont {Regner}, \citenamefont {Sellan},
  \citenamefont {Su}, \citenamefont {Amon}, \citenamefont {McGaughey},\ and\
  \citenamefont {Malen}}]{Regner2013}%
  \BibitemOpen
  \bibfield  {author} {\bibinfo {author} {\bibfnamefont {K.~T.}\ \bibnamefont
  {Regner}}, \bibinfo {author} {\bibfnamefont {D.~P.}\ \bibnamefont {Sellan}},
  \bibinfo {author} {\bibfnamefont {Z.}~\bibnamefont {Su}}, \bibinfo {author}
  {\bibfnamefont {C.~H.}\ \bibnamefont {Amon}}, \bibinfo {author}
  {\bibfnamefont {A.~J.}\ \bibnamefont {McGaughey}},\ and\ \bibinfo {author}
  {\bibfnamefont {J.~A.}\ \bibnamefont {Malen}},\ }\bibfield  {title} {\bibinfo
  {title} {Broadband phonon mean free path contributions to thermal
  conductivity measured using frequency domain thermoreflectance},\ }\href@noop
  {} {\bibfield  {journal} {\bibinfo  {journal} {Nature Communications}\
  }\textbf {\bibinfo {volume} {4}} (\bibinfo {year}
  {2013}{\natexlab{a}})}\BibitemShut {NoStop}%
\bibitem [{\citenamefont {Johnson}\ \emph {et~al.}(2013)\citenamefont
  {Johnson}, \citenamefont {Maznev}, \citenamefont {Cuffe}, \citenamefont
  {Eliason}, \citenamefont {Minnich}, \citenamefont {Kehoe}, \citenamefont
  {Sotomayor~Torres}, \citenamefont {Chen},\ and\ \citenamefont
  {Nelson}}]{JohnsonMinnichChen2013}%
  \BibitemOpen
  \bibfield  {author} {\bibinfo {author} {\bibfnamefont {J.~A.}\ \bibnamefont
  {Johnson}}, \bibinfo {author} {\bibfnamefont {A.~A.}\ \bibnamefont {Maznev}},
  \bibinfo {author} {\bibfnamefont {J.}~\bibnamefont {Cuffe}}, \bibinfo
  {author} {\bibfnamefont {J.~K.}\ \bibnamefont {Eliason}}, \bibinfo {author}
  {\bibfnamefont {A.~J.}\ \bibnamefont {Minnich}}, \bibinfo {author}
  {\bibfnamefont {T.}~\bibnamefont {Kehoe}}, \bibinfo {author} {\bibfnamefont
  {C.~M.}\ \bibnamefont {Sotomayor~Torres}}, \bibinfo {author} {\bibfnamefont
  {G.}~\bibnamefont {Chen}},\ and\ \bibinfo {author} {\bibfnamefont {K.~A.}\
  \bibnamefont {Nelson}},\ }\bibfield  {title} {\bibinfo {title} {Direct
  {M}easurement of {R}oom-{T}emperature {N}ondiffusive {T}hermal {T}ransport
  {O}ver {M}icron {D}istances in a {S}ilicon {M}embrane},\ }\href
  {https://doi.org/10.1103/physrevlett.110.025901} {\bibfield  {journal}
  {\bibinfo  {journal} {Physical Review Letters}\ }\textbf {\bibinfo {volume}
  {110}},\ \bibinfo {pages} {025901} (\bibinfo {year} {2013})}\BibitemShut
  {NoStop}%
\bibitem [{\citenamefont {Johnson}\ \emph {et~al.}(2012)\citenamefont
  {Johnson}, \citenamefont {Maznev}, \citenamefont {Bulsara}, \citenamefont
  {Fitzgerald}, \citenamefont {Harman}, \citenamefont {Calawa}, \citenamefont
  {Vineis}, \citenamefont {Turner},\ and\ \citenamefont
  {Nelson}}]{Johnson2012}%
  \BibitemOpen
  \bibfield  {author} {\bibinfo {author} {\bibfnamefont {J.~A.}\ \bibnamefont
  {Johnson}}, \bibinfo {author} {\bibfnamefont {A.~A.}\ \bibnamefont {Maznev}},
  \bibinfo {author} {\bibfnamefont {M.~T.}\ \bibnamefont {Bulsara}}, \bibinfo
  {author} {\bibfnamefont {E.~A.}\ \bibnamefont {Fitzgerald}}, \bibinfo
  {author} {\bibfnamefont {T.~C.}\ \bibnamefont {Harman}}, \bibinfo {author}
  {\bibfnamefont {S.}~\bibnamefont {Calawa}}, \bibinfo {author} {\bibfnamefont
  {C.~J.}\ \bibnamefont {Vineis}}, \bibinfo {author} {\bibfnamefont
  {G.}~\bibnamefont {Turner}},\ and\ \bibinfo {author} {\bibfnamefont {K.~A.}\
  \bibnamefont {Nelson}},\ }\bibfield  {title} {\bibinfo {title}
  {Phase-controlled, heterodyne laser-induced transient grating measurements of
  thermal transport properties in opaque material},\ }\href@noop {} {\bibfield
  {journal} {\bibinfo  {journal} {Journal of Applied Physics}\ }\textbf
  {\bibinfo {volume} {111}} (\bibinfo {year} {2012})}\BibitemShut {NoStop}%
\bibitem [{\citenamefont {Johnson}\ \emph {et~al.}(2011)\citenamefont
  {Johnson}, \citenamefont {Maznev}, \citenamefont {Eliason}, \citenamefont
  {Minnich}, \citenamefont {Collins}, \citenamefont {Chen}, \citenamefont
  {Cuffe}, \citenamefont {Kehoe}, \citenamefont {Sotomayor~Torres},\ and\
  \citenamefont {Nelson}}]{Johnson2011}%
  \BibitemOpen
  \bibfield  {author} {\bibinfo {author} {\bibfnamefont {J.~A.}\ \bibnamefont
  {Johnson}}, \bibinfo {author} {\bibfnamefont {A.~A.}\ \bibnamefont {Maznev}},
  \bibinfo {author} {\bibfnamefont {J.~K.}\ \bibnamefont {Eliason}}, \bibinfo
  {author} {\bibfnamefont {A.}~\bibnamefont {Minnich}}, \bibinfo {author}
  {\bibfnamefont {K.}~\bibnamefont {Collins}}, \bibinfo {author} {\bibfnamefont
  {G.}~\bibnamefont {Chen}}, \bibinfo {author} {\bibfnamefont {J.}~\bibnamefont
  {Cuffe}}, \bibinfo {author} {\bibfnamefont {T.}~\bibnamefont {Kehoe}},
  \bibinfo {author} {\bibfnamefont {C.~M.}\ \bibnamefont {Sotomayor~Torres}},\
  and\ \bibinfo {author} {\bibfnamefont {K.~A.}\ \bibnamefont {Nelson}},\
  }\bibfield  {title} {\bibinfo {title} {Experimental {E}vidence of
  {N}on-{D}iffusive {T}hermal {T}ransport in {S}i and {G}a{A}s},\ }\href@noop
  {} {\bibfield  {journal} {\bibinfo  {journal} {MRS Proceedings}\ }\textbf
  {\bibinfo {volume} {1347}} (\bibinfo {year} {2011})}\BibitemShut {NoStop}%
\bibitem [{\citenamefont {Minnich}\ \emph {et~al.}(2011)\citenamefont
  {Minnich}, \citenamefont {Johnson}, \citenamefont {Schmidt}, \citenamefont
  {Esfarjani}, \citenamefont {Dresselhaus}, \citenamefont {Nelson},\ and\
  \citenamefont {Chen}}]{Minnich2011}%
  \BibitemOpen
  \bibfield  {author} {\bibinfo {author} {\bibfnamefont {A.~J.}\ \bibnamefont
  {Minnich}}, \bibinfo {author} {\bibfnamefont {J.~A.}\ \bibnamefont
  {Johnson}}, \bibinfo {author} {\bibfnamefont {A.~J.}\ \bibnamefont
  {Schmidt}}, \bibinfo {author} {\bibfnamefont {K.}~\bibnamefont {Esfarjani}},
  \bibinfo {author} {\bibfnamefont {M.~S.}\ \bibnamefont {Dresselhaus}},
  \bibinfo {author} {\bibfnamefont {K.~A.}\ \bibnamefont {Nelson}},\ and\
  \bibinfo {author} {\bibfnamefont {G.}~\bibnamefont {Chen}},\ }\bibfield
  {title} {\bibinfo {title} {{T}hermal {C}onductivity {S}pectroscopy
  {T}echnique to {M}easure {P}honon {M}ean {F}ree {P}aths},\ }\href
  {https://doi.org/10.1103/physrevlett.107.095901} {\bibfield  {journal}
  {\bibinfo  {journal} {Physical Review Letters}\ }\textbf {\bibinfo {volume}
  {107}},\ \bibinfo {pages} {095901} (\bibinfo {year} {2011})}\BibitemShut
  {NoStop}%
\bibitem [{\citenamefont {Hu}\ \emph {et~al.}(2015)\citenamefont {Hu},
  \citenamefont {Zeng}, \citenamefont {Minnich}, \citenamefont {Dresselhaus},\
  and\ \citenamefont {Chen}}]{Hu2015}%
  \BibitemOpen
  \bibfield  {author} {\bibinfo {author} {\bibfnamefont {Y.}~\bibnamefont
  {Hu}}, \bibinfo {author} {\bibfnamefont {L.}~\bibnamefont {Zeng}}, \bibinfo
  {author} {\bibfnamefont {A.~J.}\ \bibnamefont {Minnich}}, \bibinfo {author}
  {\bibfnamefont {M.~S.}\ \bibnamefont {Dresselhaus}},\ and\ \bibinfo {author}
  {\bibfnamefont {G.}~\bibnamefont {Chen}},\ }\bibfield  {title} {\bibinfo
  {title} {Spectral mapping of thermal conductivity through nanoscale ballistic
  transport},\ }\href {https://doi.org/10.1038/nnano.2015.109} {\bibfield
  {journal} {\bibinfo  {journal} {Nature Nanotechnology}\ }\textbf {\bibinfo
  {volume} {10}},\ \bibinfo {pages} {701} (\bibinfo {year} {2015})}\BibitemShut
  {NoStop}%
\bibitem [{\citenamefont {Highland}\ \emph {et~al.}(2007)\citenamefont
  {Highland}, \citenamefont {Gundrum}, \citenamefont {Koh}, \citenamefont
  {Averback}, \citenamefont {Cahill}, \citenamefont {Elarde}, \citenamefont
  {Coleman}, \citenamefont {Walko},\ and\ \citenamefont
  {Landahl}}]{Highland2007}%
  \BibitemOpen
  \bibfield  {author} {\bibinfo {author} {\bibfnamefont {M.}~\bibnamefont
  {Highland}}, \bibinfo {author} {\bibfnamefont {B.~C.}\ \bibnamefont
  {Gundrum}}, \bibinfo {author} {\bibfnamefont {Y.~K.}\ \bibnamefont {Koh}},
  \bibinfo {author} {\bibfnamefont {R.~S.}\ \bibnamefont {Averback}}, \bibinfo
  {author} {\bibfnamefont {D.~G.}\ \bibnamefont {Cahill}}, \bibinfo {author}
  {\bibfnamefont {V.~C.}\ \bibnamefont {Elarde}}, \bibinfo {author}
  {\bibfnamefont {J.~J.}\ \bibnamefont {Coleman}}, \bibinfo {author}
  {\bibfnamefont {D.~A.}\ \bibnamefont {Walko}},\ and\ \bibinfo {author}
  {\bibfnamefont {E.~C.}\ \bibnamefont {Landahl}},\ }\bibfield  {title}
  {\bibinfo {title} {Ballistic-phonon heat conduction at the nanoscale as
  revealed by time-resolved {X}-ray diffraction and time-domain
  thermoreflectance},\ }\href {https://doi.org/10.1103/physrevb.76.075337}
  {\bibfield  {journal} {\bibinfo  {journal} {Physical Review B}\ }\textbf
  {\bibinfo {volume} {76}},\ \bibinfo {pages} {075337} (\bibinfo {year}
  {2007})}\BibitemShut {NoStop}%
\bibitem [{\citenamefont {Ding}\ \emph {et~al.}(2014)\citenamefont {Ding},
  \citenamefont {Chen},\ and\ \citenamefont {Minnich}}]{DingChen2014}%
  \BibitemOpen
  \bibfield  {author} {\bibinfo {author} {\bibfnamefont {D.}~\bibnamefont
  {Ding}}, \bibinfo {author} {\bibfnamefont {X.}~\bibnamefont {Chen}},\ and\
  \bibinfo {author} {\bibfnamefont {A.~J.}\ \bibnamefont {Minnich}},\
  }\bibfield  {title} {\bibinfo {title} {Radial quasiballistic transport in
  time-domain thermoreflectance studied using monte carlo simulations},\
  }\href@noop {} {\bibfield  {journal} {\bibinfo  {journal} {Applied Physics
  Letters}\ }\textbf {\bibinfo {volume} {104}} (\bibinfo {year}
  {2014})}\BibitemShut {NoStop}%
\bibitem [{\citenamefont {Jiang}\ \emph {et~al.}(2016)\citenamefont {Jiang},
  \citenamefont {Lindsay},\ and\ \citenamefont {Koh}}]{JiangKoh2016}%
  \BibitemOpen
  \bibfield  {author} {\bibinfo {author} {\bibfnamefont {P.}~\bibnamefont
  {Jiang}}, \bibinfo {author} {\bibfnamefont {L.}~\bibnamefont {Lindsay}},\
  and\ \bibinfo {author} {\bibfnamefont {Y.~K.}\ \bibnamefont {Koh}},\
  }\bibfield  {title} {\bibinfo {title} {Role of low-energy phonons with
  mean-free-paths $>$\,0.8$\,\mu$m in heat conduction in silicon},\ }\href@noop
  {} {\bibfield  {journal} {\bibinfo  {journal} {Journal of Applied Physics}\
  }\textbf {\bibinfo {volume} {119}} (\bibinfo {year} {2016})}\BibitemShut
  {NoStop}%
\bibitem [{\citenamefont {Regner}\ \emph
  {et~al.}(2013{\natexlab{b}})\citenamefont {Regner}, \citenamefont
  {Majumdar},\ and\ \citenamefont {Malen}}]{RegnerMalen2013}%
  \BibitemOpen
  \bibfield  {author} {\bibinfo {author} {\bibfnamefont {K.~T.}\ \bibnamefont
  {Regner}}, \bibinfo {author} {\bibfnamefont {S.}~\bibnamefont {Majumdar}},\
  and\ \bibinfo {author} {\bibfnamefont {J.~A.}\ \bibnamefont {Malen}},\
  }\bibfield  {title} {\bibinfo {title} {Instrumentation of broadband frequency
  domain thermoreflectance for measuring thermal conductivity accumulation
  functions},\ }\href@noop {} {\bibfield  {journal} {\bibinfo  {journal}
  {Review of Scientific Instruments}\ }\textbf {\bibinfo {volume} {84}}
  (\bibinfo {year} {2013}{\natexlab{b}})}\BibitemShut {NoStop}%
\bibitem [{\citenamefont {Freedman}\ \emph {et~al.}(2013)\citenamefont
  {Freedman}, \citenamefont {Leach}, \citenamefont {Preble}, \citenamefont
  {Sitar}, \citenamefont {Davis},\ and\ \citenamefont {Malen}}]{Freedman2013}%
  \BibitemOpen
  \bibfield  {author} {\bibinfo {author} {\bibfnamefont {J.~P.}\ \bibnamefont
  {Freedman}}, \bibinfo {author} {\bibfnamefont {J.~H.}\ \bibnamefont {Leach}},
  \bibinfo {author} {\bibfnamefont {E.~A.}\ \bibnamefont {Preble}}, \bibinfo
  {author} {\bibfnamefont {Z.}~\bibnamefont {Sitar}}, \bibinfo {author}
  {\bibfnamefont {R.~F.}\ \bibnamefont {Davis}},\ and\ \bibinfo {author}
  {\bibfnamefont {J.~A.}\ \bibnamefont {Malen}},\ }\bibfield  {title} {\bibinfo
  {title} {Universal phonon mean free path spectra in crystalline
  semiconductors at high temperature},\ }\href@noop {} {\bibfield  {journal}
  {\bibinfo  {journal} {Scientific Reports}\ }\textbf {\bibinfo {volume} {3}}
  (\bibinfo {year} {2013})}\BibitemShut {NoStop}%
\bibitem [{\citenamefont {Schmidt}\ \emph {et~al.}(2009)\citenamefont
  {Schmidt}, \citenamefont {Cheaito},\ and\ \citenamefont
  {Chiesa}}]{Schmidt2009}%
  \BibitemOpen
  \bibfield  {author} {\bibinfo {author} {\bibfnamefont {A.~J.}\ \bibnamefont
  {Schmidt}}, \bibinfo {author} {\bibfnamefont {R.}~\bibnamefont {Cheaito}},\
  and\ \bibinfo {author} {\bibfnamefont {M.}~\bibnamefont {Chiesa}},\
  }\bibfield  {title} {\bibinfo {title} {A frequency-domain thermoreflectance
  method for the characterization of thermal properties},\ }\bibfield
  {journal} {\bibinfo  {journal} {Review of Scientific Instruments}\ }\textbf
  {\bibinfo {volume} {80}},\ \href {https://doi.org/10.1063/1.3212673}
  {10.1063/1.3212673} (\bibinfo {year} {2009})\BibitemShut {NoStop}%
\bibitem [{\citenamefont {Zeng}\ \emph {et~al.}(2015)\citenamefont {Zeng},
  \citenamefont {Collins}, \citenamefont {Hu}, \citenamefont {Luckyanova},
  \citenamefont {Maznev}, \citenamefont {Huberman}, \citenamefont {Chiloyan},
  \citenamefont {Zhou}, \citenamefont {Huang}, \citenamefont {Nelson},\ and\
  \citenamefont {Chen}}]{ZengCollins2015}%
  \BibitemOpen
  \bibfield  {author} {\bibinfo {author} {\bibfnamefont {L.}~\bibnamefont
  {Zeng}}, \bibinfo {author} {\bibfnamefont {K.~C.}\ \bibnamefont {Collins}},
  \bibinfo {author} {\bibfnamefont {Y.}~\bibnamefont {Hu}}, \bibinfo {author}
  {\bibfnamefont {M.~N.}\ \bibnamefont {Luckyanova}}, \bibinfo {author}
  {\bibfnamefont {A.~A.}\ \bibnamefont {Maznev}}, \bibinfo {author}
  {\bibfnamefont {S.}~\bibnamefont {Huberman}}, \bibinfo {author}
  {\bibfnamefont {V.}~\bibnamefont {Chiloyan}}, \bibinfo {author}
  {\bibfnamefont {J.}~\bibnamefont {Zhou}}, \bibinfo {author} {\bibfnamefont
  {X.}~\bibnamefont {Huang}}, \bibinfo {author} {\bibfnamefont {K.~A.}\
  \bibnamefont {Nelson}},\ and\ \bibinfo {author} {\bibfnamefont
  {G.}~\bibnamefont {Chen}},\ }\bibfield  {title} {\bibinfo {title} {Measuring
  {P}honon {M}ean {F}ree {P}ath {D}istributions by {P}robing {Q}uasiballistic
  {P}honon {T}ransport in {G}rating {N}anostructures},\ }\href@noop {}
  {\bibfield  {journal} {\bibinfo  {journal} {Scientific Reports}\ }\textbf
  {\bibinfo {volume} {5}} (\bibinfo {year} {2015})}\BibitemShut {NoStop}%
\bibitem [{\citenamefont {Maznev}\ \emph {et~al.}(2011)\citenamefont {Maznev},
  \citenamefont {Johnson},\ and\ \citenamefont {Nelson}}]{Maznev2011}%
  \BibitemOpen
  \bibfield  {author} {\bibinfo {author} {\bibfnamefont {A.~A.}\ \bibnamefont
  {Maznev}}, \bibinfo {author} {\bibfnamefont {J.~A.}\ \bibnamefont
  {Johnson}},\ and\ \bibinfo {author} {\bibfnamefont {K.~A.}\ \bibnamefont
  {Nelson}},\ }\bibfield  {title} {\bibinfo {title} {Onset of nondiffusive
  phonon transport in transient thermal grating decay},\ }\href
  {https://doi.org/10.1103/physrevb.84.195206} {\bibfield  {journal} {\bibinfo
  {journal} {Physical Review B}\ }\textbf {\bibinfo {volume} {84}},\ \bibinfo
  {pages} {195206} (\bibinfo {year} {2011})}\BibitemShut {NoStop}%
\bibitem [{\citenamefont {Minnich}(2012)}]{Minnich2012}%
  \BibitemOpen
  \bibfield  {author} {\bibinfo {author} {\bibfnamefont {A.~J.}\ \bibnamefont
  {Minnich}},\ }\bibfield  {title} {\bibinfo {title} {Determining {P}honon
  {M}ean {F}ree {P}aths from {O}bservations of {Q}uasiballistic {T}hermal
  {T}ransport},\ }\href {https://doi.org/10.1103/physrevlett.109.205901}
  {\bibfield  {journal} {\bibinfo  {journal} {Physical Review Letters}\
  }\textbf {\bibinfo {volume} {109}},\ \bibinfo {pages} {205901} (\bibinfo
  {year} {2012})}\BibitemShut {NoStop}%
\bibitem [{\citenamefont {Regner}\ \emph {et~al.}(2014)\citenamefont {Regner},
  \citenamefont {McGaughey},\ and\ \citenamefont {Malen}}]{Regner2014}%
  \BibitemOpen
  \bibfield  {author} {\bibinfo {author} {\bibfnamefont {K.~T.}\ \bibnamefont
  {Regner}}, \bibinfo {author} {\bibfnamefont {A.~J.~H.}\ \bibnamefont
  {McGaughey}},\ and\ \bibinfo {author} {\bibfnamefont {J.~A.}\ \bibnamefont
  {Malen}},\ }\bibfield  {title} {\bibinfo {title} {Analytical interpretation
  of nondiffusive phonon transport in thermoreflectance thermal conductivity
  measurements},\ }\href {https://doi.org/10.1103/physrevb.90.064302}
  {\bibfield  {journal} {\bibinfo  {journal} {Physical Review B}\ }\textbf
  {\bibinfo {volume} {90}},\ \bibinfo {pages} {064302} (\bibinfo {year}
  {2014})}\BibitemShut {NoStop}%
\bibitem [{\citenamefont {Beechem}\ \emph {et~al.}(2007)\citenamefont
  {Beechem}, \citenamefont {Graham}, \citenamefont {Kearney}, \citenamefont
  {Phinney},\ and\ \citenamefont {Serrano}}]{BeechemInvArticle2007}%
  \BibitemOpen
  \bibfield  {author} {\bibinfo {author} {\bibfnamefont {T.}~\bibnamefont
  {Beechem}}, \bibinfo {author} {\bibfnamefont {S.}~\bibnamefont {Graham}},
  \bibinfo {author} {\bibfnamefont {S.~P.}\ \bibnamefont {Kearney}}, \bibinfo
  {author} {\bibfnamefont {L.~M.}\ \bibnamefont {Phinney}},\ and\ \bibinfo
  {author} {\bibfnamefont {J.~R.}\ \bibnamefont {Serrano}},\ }\bibfield
  {title} {\bibinfo {title} {Invited {A}rticle: {S}imultaneous mapping of
  temperature and stress in microdevices using micro-{R}aman spectroscopy},\
  }\bibfield  {journal} {\bibinfo  {journal} {Review of Scientific
  Instruments}\ }\textbf {\bibinfo {volume} {78}},\ \href
  {https://doi.org/10.1063/1.2738946} {10.1063/1.2738946} (\bibinfo {year}
  {2007})\BibitemShut {NoStop}%
\bibitem [{\citenamefont {Beechem}\ \emph {et~al.}(2015)\citenamefont
  {Beechem}, \citenamefont {Yates},\ and\ \citenamefont
  {Graham}}]{BeechemReview2015}%
  \BibitemOpen
  \bibfield  {author} {\bibinfo {author} {\bibfnamefont {T.}~\bibnamefont
  {Beechem}}, \bibinfo {author} {\bibfnamefont {L.}~\bibnamefont {Yates}},\
  and\ \bibinfo {author} {\bibfnamefont {S.}~\bibnamefont {Graham}},\
  }\bibfield  {title} {\bibinfo {title} {Invited {R}eview {A}rticle: {E}rror
  and uncertainty in {R}aman thermal conductivity measurements},\ }\href@noop
  {} {\bibfield  {journal} {\bibinfo  {journal} {Review of Scientific
  Instruments}\ }\textbf {\bibinfo {volume} {86}} (\bibinfo {year}
  {2015})}\BibitemShut {NoStop}%
\bibitem [{\citenamefont {Koh}\ and\ \citenamefont
  {Cahill}(2007)}]{KohCahill2007}%
  \BibitemOpen
  \bibfield  {author} {\bibinfo {author} {\bibfnamefont {Y.~K.}\ \bibnamefont
  {Koh}}\ and\ \bibinfo {author} {\bibfnamefont {D.~G.}\ \bibnamefont
  {Cahill}},\ }\bibfield  {title} {\bibinfo {title} {Frequency dependence of
  the thermal conductivity of semiconductor alloys},\ }\href
  {https://doi.org/10.1103/physrevb.76.075207} {\bibfield  {journal} {\bibinfo
  {journal} {Physical Review B}\ }\textbf {\bibinfo {volume} {76}},\ \bibinfo
  {pages} {075207} (\bibinfo {year} {2007})}\BibitemShut {NoStop}%
\bibitem [{\citenamefont {Seemann}\ \emph {et~al.}(2024)\citenamefont
  {Seemann}, \citenamefont {Elhajhasan}, \citenamefont {Themann}, \citenamefont
  {Dudde}, \citenamefont {Würsch}, \citenamefont {Lierath}, \citenamefont
  {Ciers}, \citenamefont {Haglund}, \citenamefont {Protik}, \citenamefont
  {Romano}, \citenamefont {Butté}, \citenamefont {Carlin}, \citenamefont
  {Grandjean},\ and\ \citenamefont {Callsen}}]{Seemann2024}%
  \BibitemOpen
  \bibfield  {author} {\bibinfo {author} {\bibfnamefont {W.}~\bibnamefont
  {Seemann}}, \bibinfo {author} {\bibfnamefont {M.}~\bibnamefont {Elhajhasan}},
  \bibinfo {author} {\bibfnamefont {J.}~\bibnamefont {Themann}}, \bibinfo
  {author} {\bibfnamefont {K.}~\bibnamefont {Dudde}}, \bibinfo {author}
  {\bibfnamefont {G.}~\bibnamefont {Würsch}}, \bibinfo {author} {\bibfnamefont
  {J.}~\bibnamefont {Lierath}}, \bibinfo {author} {\bibfnamefont
  {J.}~\bibnamefont {Ciers}}, \bibinfo {author} {\bibfnamefont {{\r
  A}.}~\bibnamefont {Haglund}}, \bibinfo {author} {\bibfnamefont {N.~H.}\
  \bibnamefont {Protik}}, \bibinfo {author} {\bibfnamefont {G.}~\bibnamefont
  {Romano}}, \bibinfo {author} {\bibfnamefont {R.}~\bibnamefont {Butté}},
  \bibinfo {author} {\bibfnamefont {J.-F.}\ \bibnamefont {Carlin}}, \bibinfo
  {author} {\bibfnamefont {N.}~\bibnamefont {Grandjean}},\ and\ \bibinfo
  {author} {\bibfnamefont {G.}~\bibnamefont {Callsen}},\ }\href
  {https://doi.org/10.48550/ARXIV.2410.12515} {\bibinfo {title} {Thermal
  analysis of {G}a{N}-based photonic membranes for optoelectronics}} (\bibinfo
  {year} {2024})\BibitemShut {NoStop}%
\bibitem [{\citenamefont {Hsu}\ \emph {et~al.}(2011)\citenamefont {Hsu},
  \citenamefont {Pettes}, \citenamefont {Aykol}, \citenamefont {Chang},
  \citenamefont {Hung}, \citenamefont {Theiss}, \citenamefont {Shi},\ and\
  \citenamefont {Cronin}}]{Hsu2LRToriginal2011}%
  \BibitemOpen
  \bibfield  {author} {\bibinfo {author} {\bibfnamefont {I.-K.}\ \bibnamefont
  {Hsu}}, \bibinfo {author} {\bibfnamefont {M.~T.}\ \bibnamefont {Pettes}},
  \bibinfo {author} {\bibfnamefont {M.}~\bibnamefont {Aykol}}, \bibinfo
  {author} {\bibfnamefont {C.-C.}\ \bibnamefont {Chang}}, \bibinfo {author}
  {\bibfnamefont {W.-H.}\ \bibnamefont {Hung}}, \bibinfo {author}
  {\bibfnamefont {J.}~\bibnamefont {Theiss}}, \bibinfo {author} {\bibfnamefont
  {L.}~\bibnamefont {Shi}},\ and\ \bibinfo {author} {\bibfnamefont {S.~B.}\
  \bibnamefont {Cronin}},\ }\bibfield  {title} {\bibinfo {title} {Direct
  observation of heat dissipation in individual suspended carbon nanotubes
  using a two-laser technique},\ }\href@noop {} {\bibfield  {journal} {\bibinfo
   {journal} {Journal of Applied Physics}\ }\textbf {\bibinfo {volume} {110}}
  (\bibinfo {year} {2011})}\BibitemShut {NoStop}%
\bibitem [{\citenamefont {Reparaz}\ \emph {et~al.}(2014)\citenamefont
  {Reparaz}, \citenamefont {Chavez-Angel}, \citenamefont {Wagner},
  \citenamefont {Graczykowski}, \citenamefont {Gomis-Bresco}, \citenamefont
  {Alzina},\ and\ \citenamefont {Torres}}]{Reparaz2014}%
  \BibitemOpen
  \bibfield  {author} {\bibinfo {author} {\bibfnamefont {J.~S.}\ \bibnamefont
  {Reparaz}}, \bibinfo {author} {\bibfnamefont {E.}~\bibnamefont
  {Chavez-Angel}}, \bibinfo {author} {\bibfnamefont {M.~R.}\ \bibnamefont
  {Wagner}}, \bibinfo {author} {\bibfnamefont {B.}~\bibnamefont
  {Graczykowski}}, \bibinfo {author} {\bibfnamefont {J.}~\bibnamefont
  {Gomis-Bresco}}, \bibinfo {author} {\bibfnamefont {F.}~\bibnamefont
  {Alzina}},\ and\ \bibinfo {author} {\bibfnamefont {C.~M.~S.}\ \bibnamefont
  {Torres}},\ }\bibfield  {title} {\bibinfo {title} {A novel contactless
  technique for thermal field mapping and thermal conductivity determination:
  {T}wo-{L}aser {R}aman {T}hermometry},\ }\href
  {https://doi.org/10.1063/1.4867166} {\bibfield  {journal} {\bibinfo
  {journal} {Review of Scientific Instruments}\ }\textbf {\bibinfo {volume}
  {85}},\ \bibinfo {pages} {034901} (\bibinfo {year} {2014})}\BibitemShut
  {NoStop}%
\bibitem [{\citenamefont {Wilson}\ and\ \citenamefont
  {Cahill}(2014)}]{WilsonCahill2014}%
  \BibitemOpen
  \bibfield  {author} {\bibinfo {author} {\bibfnamefont {R.~B.}\ \bibnamefont
  {Wilson}}\ and\ \bibinfo {author} {\bibfnamefont {D.~G.}\ \bibnamefont
  {Cahill}},\ }\bibfield  {title} {\bibinfo {title} {Anisotropic failure of
  fourier theory in time-domain thermoreflectance experiments},\ }\bibfield
  {journal} {\bibinfo  {journal} {Nature Communications}\ }\textbf {\bibinfo
  {volume} {5}},\ \href {https://doi.org/10.1038/ncomms6075}
  {10.1038/ncomms6075} (\bibinfo {year} {2014})\BibitemShut {NoStop}%
\bibitem [{\citenamefont {Vermeersch}\ \emph {et~al.}(2015)\citenamefont
  {Vermeersch}, \citenamefont {Mohammed}, \citenamefont {Pernot}, \citenamefont
  {Koh},\ and\ \citenamefont {Shakouri}}]{Vermeersch2015}%
  \BibitemOpen
  \bibfield  {author} {\bibinfo {author} {\bibfnamefont {B.}~\bibnamefont
  {Vermeersch}}, \bibinfo {author} {\bibfnamefont {A.~M.~S.}\ \bibnamefont
  {Mohammed}}, \bibinfo {author} {\bibfnamefont {G.}~\bibnamefont {Pernot}},
  \bibinfo {author} {\bibfnamefont {Y.~R.}\ \bibnamefont {Koh}},\ and\ \bibinfo
  {author} {\bibfnamefont {A.}~\bibnamefont {Shakouri}},\ }\bibfield  {title}
  {\bibinfo {title} {Superdiffusive heat conduction in semiconductor alloys.
  {II}. {T}runcated {L}évy formalism for experimental analysis},\ }\href
  {https://doi.org/10.1103/physrevb.91.085203} {\bibfield  {journal} {\bibinfo
  {journal} {Physical Review B}\ }\textbf {\bibinfo {volume} {91}},\ \bibinfo
  {pages} {085203} (\bibinfo {year} {2015})}\BibitemShut {NoStop}%
\bibitem [{\citenamefont {Vallabhaneni}\ \emph {et~al.}(2016)\citenamefont
  {Vallabhaneni}, \citenamefont {Singh}, \citenamefont {Bao}, \citenamefont
  {Murthy},\ and\ \citenamefont {Ruan}}]{Vallabhaneni2016}%
  \BibitemOpen
  \bibfield  {author} {\bibinfo {author} {\bibfnamefont {A.~K.}\ \bibnamefont
  {Vallabhaneni}}, \bibinfo {author} {\bibfnamefont {D.}~\bibnamefont {Singh}},
  \bibinfo {author} {\bibfnamefont {H.}~\bibnamefont {Bao}}, \bibinfo {author}
  {\bibfnamefont {J.}~\bibnamefont {Murthy}},\ and\ \bibinfo {author}
  {\bibfnamefont {X.}~\bibnamefont {Ruan}},\ }\bibfield  {title} {\bibinfo
  {title} {Reliability of {R}aman measurements of thermal conductivity of
  single-layer graphene due to selective electron-phonon coupling: {A}
  first-principles study},\ }\href {https://doi.org/10.1103/physrevb.93.125432}
  {\bibfield  {journal} {\bibinfo  {journal} {Physical Review B}\ }\textbf
  {\bibinfo {volume} {93}},\ \bibinfo {pages} {125432} (\bibinfo {year}
  {2016})}\BibitemShut {NoStop}%
\bibitem [{\citenamefont {Sullivan}\ \emph {et~al.}(2017)\citenamefont
  {Sullivan}, \citenamefont {Vallabhaneni}, \citenamefont {Kholmanov},
  \citenamefont {Ruan}, \citenamefont {Murthy},\ and\ \citenamefont
  {Shi}}]{Sullivan2017}%
  \BibitemOpen
  \bibfield  {author} {\bibinfo {author} {\bibfnamefont {S.}~\bibnamefont
  {Sullivan}}, \bibinfo {author} {\bibfnamefont {A.}~\bibnamefont
  {Vallabhaneni}}, \bibinfo {author} {\bibfnamefont {I.}~\bibnamefont
  {Kholmanov}}, \bibinfo {author} {\bibfnamefont {X.}~\bibnamefont {Ruan}},
  \bibinfo {author} {\bibfnamefont {J.}~\bibnamefont {Murthy}},\ and\ \bibinfo
  {author} {\bibfnamefont {L.}~\bibnamefont {Shi}},\ }\bibfield  {title}
  {\bibinfo {title} {Optical {G}eneration and {D}etection of {L}ocal
  {N}onequilibrium {P}honons in {S}uspended {G}raphene},\ }\href
  {https://doi.org/10.1021/acs.nanolett.7b00110} {\bibfield  {journal}
  {\bibinfo  {journal} {Nano Letters}\ }\textbf {\bibinfo {volume} {17}},\
  \bibinfo {pages} {2049} (\bibinfo {year} {2017})}\BibitemShut {NoStop}%
\bibitem [{\citenamefont {Wang}\ \emph {et~al.}(2020)\citenamefont {Wang},
  \citenamefont {Zobeiri}, \citenamefont {Xie}, \citenamefont {Wang},
  \citenamefont {Zhang},\ and\ \citenamefont {Yue}}]{Wang2020}%
  \BibitemOpen
  \bibfield  {author} {\bibinfo {author} {\bibfnamefont {R.}~\bibnamefont
  {Wang}}, \bibinfo {author} {\bibfnamefont {H.}~\bibnamefont {Zobeiri}},
  \bibinfo {author} {\bibfnamefont {Y.}~\bibnamefont {Xie}}, \bibinfo {author}
  {\bibfnamefont {X.}~\bibnamefont {Wang}}, \bibinfo {author} {\bibfnamefont
  {X.}~\bibnamefont {Zhang}},\ and\ \bibinfo {author} {\bibfnamefont
  {Y.}~\bibnamefont {Yue}},\ }\bibfield  {title} {\bibinfo {title}
  {Distinguishing {O}ptical and {A}coustic {P}honon {T}emperatures and {T}heir
  {E}nergy {C}oupling {F}actor under {P}hoton {E}xcitation in nm 2{D}
  {M}aterials},\ }\href@noop {} {\bibfield  {journal} {\bibinfo  {journal}
  {Advanced Science}\ }\textbf {\bibinfo {volume} {7}} (\bibinfo {year}
  {2020})}\BibitemShut {NoStop}%
\bibitem [{\citenamefont {Fukura}\ \emph {et~al.}(2006)\citenamefont {Fukura},
  \citenamefont {Mizukami}, \citenamefont {Odake},\ and\ \citenamefont
  {Kagi}}]{Fukura2006}%
  \BibitemOpen
  \bibfield  {author} {\bibinfo {author} {\bibfnamefont {S.}~\bibnamefont
  {Fukura}}, \bibinfo {author} {\bibfnamefont {T.}~\bibnamefont {Mizukami}},
  \bibinfo {author} {\bibfnamefont {S.}~\bibnamefont {Odake}},\ and\ \bibinfo
  {author} {\bibfnamefont {H.}~\bibnamefont {Kagi}},\ }\bibfield  {title}
  {\bibinfo {title} {Factors {D}etermining the {S}tability, {R}esolution, and
  {P}recision of a {C}onventional {R}aman {S}pectrometer},\ }\href
  {https://doi.org/10.1366/000370206778062165} {\bibfield  {journal} {\bibinfo
  {journal} {Applied Spectroscopy}\ }\textbf {\bibinfo {volume} {60}},\
  \bibinfo {pages} {946} (\bibinfo {year} {2006})}\BibitemShut {NoStop}%
\bibitem [{\citenamefont {Men{\'{e}}ndez}\ and\ \citenamefont
  {Cardona}(1984)}]{Menendez1984}%
  \BibitemOpen
  \bibfield  {author} {\bibinfo {author} {\bibfnamefont {J.}~\bibnamefont
  {Men{\'{e}}ndez}}\ and\ \bibinfo {author} {\bibfnamefont {M.}~\bibnamefont
  {Cardona}},\ }\bibfield  {title} {\bibinfo {title} {Temperature dependence of
  the first-order {R}aman scattering by phonons in {S}i, {G}e, and
  $\alpha$-{S}n: {A}nharmonic effects},\ }\href
  {https://doi.org/10.1103/physrevb.29.2051} {\bibfield  {journal} {\bibinfo
  {journal} {Physical Review B}\ }\textbf {\bibinfo {volume} {29}},\ \bibinfo
  {pages} {2051} (\bibinfo {year} {1984})}\BibitemShut {NoStop}%
\bibitem [{\citenamefont {Xu}\ \emph {et~al.}(2013)\citenamefont {Xu},
  \citenamefont {Tang}, \citenamefont {Yue},\ and\ \citenamefont
  {Wang}}]{Xu_OutOfFocus_2013}%
  \BibitemOpen
  \bibfield  {author} {\bibinfo {author} {\bibfnamefont {S.}~\bibnamefont
  {Xu}}, \bibinfo {author} {\bibfnamefont {X.}~\bibnamefont {Tang}}, \bibinfo
  {author} {\bibfnamefont {Y.}~\bibnamefont {Yue}},\ and\ \bibinfo {author}
  {\bibfnamefont {X.}~\bibnamefont {Wang}},\ }\bibfield  {title} {\bibinfo
  {title} {Sub‐micron imaging of sub‐surface nanocrystalline structure in
  silicon},\ }\href {https://doi.org/10.1002/jrs.4366} {\bibfield  {journal}
  {\bibinfo  {journal} {Journal of Raman Spectroscopy}\ }\textbf {\bibinfo
  {volume} {44}},\ \bibinfo {pages} {1523} (\bibinfo {year}
  {2013})}\BibitemShut {NoStop}%
\bibitem [{\citenamefont {Xu}\ \emph {et~al.}(2020)\citenamefont {Xu},
  \citenamefont {Fan}, \citenamefont {Wang}, \citenamefont {Zhang},\ and\
  \citenamefont {Wang}}]{XuReviewRaman2020}%
  \BibitemOpen
  \bibfield  {author} {\bibinfo {author} {\bibfnamefont {S.}~\bibnamefont
  {Xu}}, \bibinfo {author} {\bibfnamefont {A.}~\bibnamefont {Fan}}, \bibinfo
  {author} {\bibfnamefont {H.}~\bibnamefont {Wang}}, \bibinfo {author}
  {\bibfnamefont {X.}~\bibnamefont {Zhang}},\ and\ \bibinfo {author}
  {\bibfnamefont {X.}~\bibnamefont {Wang}},\ }\bibfield  {title} {\bibinfo
  {title} {Raman-based {N}anoscale {T}hermal {T}ransport {C}haracterization:
  {A} {C}ritical {R}eview},\ }\href
  {https://doi.org/10.1016/j.ijheatmasstransfer.2020.119751} {\bibfield
  {journal} {\bibinfo  {journal} {International Journal of Heat and Mass
  Transfer}\ }\textbf {\bibinfo {volume} {154}},\ \bibinfo {pages} {119751}
  (\bibinfo {year} {2020})}\BibitemShut {NoStop}%
\bibitem [{\citenamefont {Stoib}\ \emph {et~al.}(2014)\citenamefont {Stoib},
  \citenamefont {Filser}, \citenamefont {Stötzel}, \citenamefont {Greppmair},
  \citenamefont {Petermann}, \citenamefont {Wiggers}, \citenamefont
  {Schierning}, \citenamefont {Stutzmann},\ and\ \citenamefont
  {Brandt}}]{Stoib2014}%
  \BibitemOpen
  \bibfield  {author} {\bibinfo {author} {\bibfnamefont {B.}~\bibnamefont
  {Stoib}}, \bibinfo {author} {\bibfnamefont {S.}~\bibnamefont {Filser}},
  \bibinfo {author} {\bibfnamefont {J.}~\bibnamefont {Stötzel}}, \bibinfo
  {author} {\bibfnamefont {A.}~\bibnamefont {Greppmair}}, \bibinfo {author}
  {\bibfnamefont {N.}~\bibnamefont {Petermann}}, \bibinfo {author}
  {\bibfnamefont {H.}~\bibnamefont {Wiggers}}, \bibinfo {author} {\bibfnamefont
  {G.}~\bibnamefont {Schierning}}, \bibinfo {author} {\bibfnamefont
  {M.}~\bibnamefont {Stutzmann}},\ and\ \bibinfo {author} {\bibfnamefont
  {M.~S.}\ \bibnamefont {Brandt}},\ }\bibfield  {title} {\bibinfo {title}
  {Spatially resolved determination of thermal conductivity by {R}aman
  spectroscopy},\ }\href {https://doi.org/10.1088/0268-1242/29/12/124005}
  {\bibfield  {journal} {\bibinfo  {journal} {Semiconductor Science and
  Technology}\ }\textbf {\bibinfo {volume} {29}},\ \bibinfo {pages} {124005}
  (\bibinfo {year} {2014})}\BibitemShut {NoStop}%
\bibitem [{\citenamefont {Franta}\ \emph {et~al.}(2019)\citenamefont {Franta},
  \citenamefont {Vohánka}, \citenamefont {Čermák}, \citenamefont {Franta},\
  and\ \citenamefont {Ohlídal}}]{Franta2019}%
  \BibitemOpen
  \bibfield  {author} {\bibinfo {author} {\bibfnamefont {D.}~\bibnamefont
  {Franta}}, \bibinfo {author} {\bibfnamefont {J.}~\bibnamefont {Vohánka}},
  \bibinfo {author} {\bibfnamefont {M.}~\bibnamefont {Čermák}}, \bibinfo
  {author} {\bibfnamefont {P.}~\bibnamefont {Franta}},\ and\ \bibinfo {author}
  {\bibfnamefont {I.}~\bibnamefont {Ohlídal}},\ }\bibfield  {title} {\bibinfo
  {title} {Temperature dependent dispersion models applicable in solid state
  physics},\ }\href {https://doi.org/10.2478/jee-2019-0036} {\bibfield
  {journal} {\bibinfo  {journal} {Journal of Electrical Engineering}\ }\textbf
  {\bibinfo {volume} {70}},\ \bibinfo {pages} {1} (\bibinfo {year}
  {2019})}\BibitemShut {NoStop}%
\bibitem [{\citenamefont {Chávez-Ángel}\ \emph {et~al.}(2014)\citenamefont
  {Chávez-Ángel}, \citenamefont {Reparaz}, \citenamefont {Gomis-Bresco},
  \citenamefont {Wagner}, \citenamefont {Cuffe}, \citenamefont {Graczykowski},
  \citenamefont {Shchepetov}, \citenamefont {Jiang}, \citenamefont {Prunnila},
  \citenamefont {Ahopelto}, \citenamefont {Alzina},\ and\ \citenamefont
  {Sotomayor~Torres}}]{ChavezAngel2014}%
  \BibitemOpen
  \bibfield  {author} {\bibinfo {author} {\bibfnamefont {E.}~\bibnamefont
  {Chávez-Ángel}}, \bibinfo {author} {\bibfnamefont {J.~S.}\ \bibnamefont
  {Reparaz}}, \bibinfo {author} {\bibfnamefont {J.}~\bibnamefont
  {Gomis-Bresco}}, \bibinfo {author} {\bibfnamefont {M.~R.}\ \bibnamefont
  {Wagner}}, \bibinfo {author} {\bibfnamefont {J.}~\bibnamefont {Cuffe}},
  \bibinfo {author} {\bibfnamefont {B.}~\bibnamefont {Graczykowski}}, \bibinfo
  {author} {\bibfnamefont {A.}~\bibnamefont {Shchepetov}}, \bibinfo {author}
  {\bibfnamefont {H.}~\bibnamefont {Jiang}}, \bibinfo {author} {\bibfnamefont
  {M.}~\bibnamefont {Prunnila}}, \bibinfo {author} {\bibfnamefont
  {J.}~\bibnamefont {Ahopelto}}, \bibinfo {author} {\bibfnamefont
  {F.}~\bibnamefont {Alzina}},\ and\ \bibinfo {author} {\bibfnamefont {C.~M.}\
  \bibnamefont {Sotomayor~Torres}},\ }\bibfield  {title} {\bibinfo {title}
  {Reduction of the thermal conductivity in free-standing silicon
  nano-membranes investigated by non-invasive {R}aman thermometry},\
  }\href@noop {} {\bibfield  {journal} {\bibinfo  {journal} {APL Materials}\
  }\textbf {\bibinfo {volume} {2}} (\bibinfo {year} {2014})}\BibitemShut
  {NoStop}%
\bibitem [{\citenamefont {Mernagh}\ and\ \citenamefont
  {Liu}(1991)}]{Mernagh1991}%
  \BibitemOpen
  \bibfield  {author} {\bibinfo {author} {\bibfnamefont {T.~P.}\ \bibnamefont
  {Mernagh}}\ and\ \bibinfo {author} {\bibfnamefont {L.-G.}\ \bibnamefont
  {Liu}},\ }\bibfield  {title} {\bibinfo {title} {Pressure dependence of
  {R}aman phonons of some group {IVA} ({C}, {S}i, and {G}e) elements},\ }\href
  {https://doi.org/10.1016/0022-3697(91)90183-z} {\bibfield  {journal}
  {\bibinfo  {journal} {Journal of Physics and Chemistry of Solids}\ }\textbf
  {\bibinfo {volume} {52}},\ \bibinfo {pages} {507} (\bibinfo {year}
  {1991})}\BibitemShut {NoStop}%
\bibitem [{\citenamefont {Pop}\ \emph {et~al.}()\citenamefont {Pop},
  \citenamefont {Banerjee}, \citenamefont {Sverdrup}, \citenamefont {Dutton},\
  and\ \citenamefont {Goodson}}]{Pop2001}%
  \BibitemOpen
  \bibfield  {author} {\bibinfo {author} {\bibfnamefont {E.}~\bibnamefont
  {Pop}}, \bibinfo {author} {\bibfnamefont {K.}~\bibnamefont {Banerjee}},
  \bibinfo {author} {\bibfnamefont {P.}~\bibnamefont {Sverdrup}}, \bibinfo
  {author} {\bibfnamefont {R.}~\bibnamefont {Dutton}},\ and\ \bibinfo {author}
  {\bibfnamefont {K.}~\bibnamefont {Goodson}},\ }\bibfield  {title} {\bibinfo
  {title} {Localized heating effects and scaling of sub-0.18 micron {CMOS}
  devices - \textit{{I}n: {I}nternational {E}lectron {D}evices {M}eeting.
  {T}echnical {D}igest ({C}at. {N}o.01{CH}37224)}}\ }(\bibinfo  {publisher}
  {IEEE})\ pp.\ \bibinfo {pages} {31.1.1--31.1.4}\BibitemShut {NoStop}%
\bibitem [{\citenamefont {Lax}(1977)}]{Lax1977}%
  \BibitemOpen
  \bibfield  {author} {\bibinfo {author} {\bibfnamefont {M.}~\bibnamefont
  {Lax}},\ }\bibfield  {title} {\bibinfo {title} {Temperature rise induced by a
  laser beam},\ }\href {https://doi.org/10.1063/1.324265} {\bibfield  {journal}
  {\bibinfo  {journal} {Journal of Applied Physics}\ }\textbf {\bibinfo
  {volume} {48}},\ \bibinfo {pages} {3919} (\bibinfo {year}
  {1977})}\BibitemShut {NoStop}%
\bibitem [{\citenamefont {Perdew}\ and\ \citenamefont
  {Zunger}(1981)}]{perdew1981self}%
  \BibitemOpen
  \bibfield  {author} {\bibinfo {author} {\bibfnamefont {J.~P.}\ \bibnamefont
  {Perdew}}\ and\ \bibinfo {author} {\bibfnamefont {A.}~\bibnamefont
  {Zunger}},\ }\bibfield  {title} {\bibinfo {title} {{Self-interaction
  correction to density-functional approximations for many-electron systems}},\
  }\href@noop {} {\bibfield  {journal} {\bibinfo  {journal} {Physical review
  B}\ }\textbf {\bibinfo {volume} {23}},\ \bibinfo {pages} {5048} (\bibinfo
  {year} {1981})}\BibitemShut {NoStop}%
\bibitem [{\citenamefont {Giannozzi}\ \emph {et~al.}(2009)\citenamefont
  {Giannozzi}, \citenamefont {Baroni}, \citenamefont {Bonini}, \citenamefont
  {Calandra}, \citenamefont {Car}, \citenamefont {Cavazzoni}, \citenamefont
  {Ceresoli}, \citenamefont {Chiarotti}, \citenamefont {Cococcioni},
  \citenamefont {Dabo} \emph {et~al.}}]{giannozzi2009quantum}%
  \BibitemOpen
  \bibfield  {author} {\bibinfo {author} {\bibfnamefont {P.}~\bibnamefont
  {Giannozzi}}, \bibinfo {author} {\bibfnamefont {S.}~\bibnamefont {Baroni}},
  \bibinfo {author} {\bibfnamefont {N.}~\bibnamefont {Bonini}}, \bibinfo
  {author} {\bibfnamefont {M.}~\bibnamefont {Calandra}}, \bibinfo {author}
  {\bibfnamefont {R.}~\bibnamefont {Car}}, \bibinfo {author} {\bibfnamefont
  {C.}~\bibnamefont {Cavazzoni}}, \bibinfo {author} {\bibfnamefont
  {D.}~\bibnamefont {Ceresoli}}, \bibinfo {author} {\bibfnamefont {G.~L.}\
  \bibnamefont {Chiarotti}}, \bibinfo {author} {\bibfnamefont {M.}~\bibnamefont
  {Cococcioni}}, \bibinfo {author} {\bibfnamefont {I.}~\bibnamefont {Dabo}},
  \emph {et~al.},\ }\bibfield  {title} {\bibinfo {title} {{QUANTUM ESPRESSO: a
  modular and open-source software project for quantum simulations of
  materials}},\ }\href@noop {} {\bibfield  {journal} {\bibinfo  {journal}
  {Journal of physics: Condensed matter}\ }\textbf {\bibinfo {volume} {21}},\
  \bibinfo {pages} {395502} (\bibinfo {year} {2009})}\BibitemShut {NoStop}%
\bibitem [{\citenamefont {Giannozzi}\ \emph {et~al.}(2017)\citenamefont
  {Giannozzi}, \citenamefont {Andreussi}, \citenamefont {Brumme}, \citenamefont
  {Bunau}, \citenamefont {Nardelli}, \citenamefont {Calandra}, \citenamefont
  {Car}, \citenamefont {Cavazzoni}, \citenamefont {Ceresoli}, \citenamefont
  {Cococcioni} \emph {et~al.}}]{giannozzi2017advanced}%
  \BibitemOpen
  \bibfield  {author} {\bibinfo {author} {\bibfnamefont {P.}~\bibnamefont
  {Giannozzi}}, \bibinfo {author} {\bibfnamefont {O.}~\bibnamefont
  {Andreussi}}, \bibinfo {author} {\bibfnamefont {T.}~\bibnamefont {Brumme}},
  \bibinfo {author} {\bibfnamefont {O.}~\bibnamefont {Bunau}}, \bibinfo
  {author} {\bibfnamefont {M.~B.}\ \bibnamefont {Nardelli}}, \bibinfo {author}
  {\bibfnamefont {M.}~\bibnamefont {Calandra}}, \bibinfo {author}
  {\bibfnamefont {R.}~\bibnamefont {Car}}, \bibinfo {author} {\bibfnamefont
  {C.}~\bibnamefont {Cavazzoni}}, \bibinfo {author} {\bibfnamefont
  {D.}~\bibnamefont {Ceresoli}}, \bibinfo {author} {\bibfnamefont
  {M.}~\bibnamefont {Cococcioni}}, \emph {et~al.},\ }\bibfield  {title}
  {\bibinfo {title} {{Advanced capabilities for materials modelling with
  Quantum ESPRESSO}},\ }\href@noop {} {\bibfield  {journal} {\bibinfo
  {journal} {Journal of physics: Condensed matter}\ }\textbf {\bibinfo {volume}
  {29}},\ \bibinfo {pages} {465901} (\bibinfo {year} {2017})}\BibitemShut
  {NoStop}%
\bibitem [{\citenamefont {Li}\ \emph {et~al.}(2014{\natexlab{b}})\citenamefont
  {Li}, \citenamefont {Carrete}, \citenamefont {Katcho},\ and\ \citenamefont
  {Mingo}}]{li2014shengbte}%
  \BibitemOpen
  \bibfield  {author} {\bibinfo {author} {\bibfnamefont {W.}~\bibnamefont
  {Li}}, \bibinfo {author} {\bibfnamefont {J.}~\bibnamefont {Carrete}},
  \bibinfo {author} {\bibfnamefont {N.~A.}\ \bibnamefont {Katcho}},\ and\
  \bibinfo {author} {\bibfnamefont {N.}~\bibnamefont {Mingo}},\ }\bibfield
  {title} {\bibinfo {title} {Shengbte: A solver of the boltzmann transport
  equation for phonons},\ }\href@noop {} {\bibfield  {journal} {\bibinfo
  {journal} {Computer Physics Communications}\ }\textbf {\bibinfo {volume}
  {185}},\ \bibinfo {pages} {1747} (\bibinfo {year}
  {2014}{\natexlab{b}})}\BibitemShut {NoStop}%
\bibitem [{\citenamefont {Han}\ \emph {et~al.}(2022)\citenamefont {Han},
  \citenamefont {Yang}, \citenamefont {Li}, \citenamefont {Feng},\ and\
  \citenamefont {Ruan}}]{han2022fourphonon}%
  \BibitemOpen
  \bibfield  {author} {\bibinfo {author} {\bibfnamefont {Z.}~\bibnamefont
  {Han}}, \bibinfo {author} {\bibfnamefont {X.}~\bibnamefont {Yang}}, \bibinfo
  {author} {\bibfnamefont {W.}~\bibnamefont {Li}}, \bibinfo {author}
  {\bibfnamefont {T.}~\bibnamefont {Feng}},\ and\ \bibinfo {author}
  {\bibfnamefont {X.}~\bibnamefont {Ruan}},\ }\bibfield  {title} {\bibinfo
  {title} {Fourphonon: An extension module to {S}heng{BTE} for computing
  four-phonon scattering rates and thermal conductivity},\ }\href@noop {}
  {\bibfield  {journal} {\bibinfo  {journal} {Computer Physics Communications}\
  }\textbf {\bibinfo {volume} {270}},\ \bibinfo {pages} {108179} (\bibinfo
  {year} {2022})}\BibitemShut {NoStop}%
\bibitem [{\citenamefont {Tamura}(1983)}]{tamura_isotope_1983}%
  \BibitemOpen
  \bibfield  {author} {\bibinfo {author} {\bibfnamefont {S.-i.}\ \bibnamefont
  {Tamura}},\ }\bibfield  {title} {\bibinfo {title} {Isotope scattering of
  dispersive phonons in {Ge}},\ }\href
  {https://doi.org/10.1103/PhysRevB.27.858} {\bibfield  {journal} {\bibinfo
  {journal} {Physical Review B}\ }\textbf {\bibinfo {volume} {27}},\ \bibinfo
  {pages} {858} (\bibinfo {year} {1983})}\BibitemShut {NoStop}%
\bibitem [{\citenamefont {Protik}\ and\ \citenamefont
  {Draxl}(2024)}]{Protik2024}%
  \BibitemOpen
  \bibfield  {author} {\bibinfo {author} {\bibfnamefont {N.~H.}\ \bibnamefont
  {Protik}}\ and\ \bibinfo {author} {\bibfnamefont {C.}~\bibnamefont {Draxl}},\
  }\bibfield  {title} {\bibinfo {title} {Beyond the {T}amura model of
  phonon-isotope scattering},\ }\href@noop {} {\bibfield  {journal} {\bibinfo
  {journal} {Physical Review B}\ }\textbf {\bibinfo {volume} {109}},\ \bibinfo
  {pages} {165201} (\bibinfo {year} {2024})}\BibitemShut {NoStop}%
\bibitem [{\citenamefont {Ziman}(2001)}]{Ziman2001}%
  \BibitemOpen
  \bibfield  {author} {\bibinfo {author} {\bibfnamefont {J.~M.}\ \bibnamefont
  {Ziman}},\ }\href
  {http://dx.doi.org/10.1093/acprof:oso/9780198507796.001.0001} {\emph
  {\bibinfo {title} {{Electrons and Phonons}}}}\ (\bibinfo  {publisher} {Oxford
  University Press},\ \bibinfo {year} {2001})\BibitemShut {NoStop}%
\bibitem [{\citenamefont {Garg}\ \emph {et~al.}(2014)\citenamefont {Garg},
  \citenamefont {Bonini}, \citenamefont {Marzari}, \citenamefont
  {Shindé~(ed.)},\ and\ \citenamefont {Srivastava~(ed.)}}]{GargBook2014}%
  \BibitemOpen
  \bibfield  {author} {\bibinfo {author} {\bibfnamefont {J.}~\bibnamefont
  {Garg}}, \bibinfo {author} {\bibfnamefont {N.}~\bibnamefont {Bonini}},
  \bibinfo {author} {\bibfnamefont {N.}~\bibnamefont {Marzari}}, \bibinfo
  {author} {\bibfnamefont {S.~L.}\ \bibnamefont {Shindé~(ed.)}},\ and\
  \bibinfo {author} {\bibfnamefont {G.~P.}\ \bibnamefont {Srivastava~(ed.)}},\
  }\href {https://doi.org/10.1007/978-1-4614-8651-0} {\bibinfo {title}
  {Length-{S}cale {D}ependent {P}honon {I}nteractions, {C}hapter 4.6 {F}ull
  {I}terative {S}olution}} (\bibinfo {year} {2014})\BibitemShut {NoStop}%
\bibitem [{\citenamefont {Wang}\ and\ \citenamefont
  {Huang}(2014)}]{WangHunag2014}%
  \BibitemOpen
  \bibfield  {author} {\bibinfo {author} {\bibfnamefont {X.}~\bibnamefont
  {Wang}}\ and\ \bibinfo {author} {\bibfnamefont {B.}~\bibnamefont {Huang}},\
  }\bibfield  {title} {\bibinfo {title} {{C}omputational {S}tudy of
  {I}n-{P}lane {P}honon {T}ransport in {S}i {T}hin {F}ilms},\ }\href@noop {}
  {\bibfield  {journal} {\bibinfo  {journal} {Scientific Reports}\ }\textbf
  {\bibinfo {volume} {4}} (\bibinfo {year} {2014})}\BibitemShut {NoStop}%
\bibitem [{\citenamefont {Xu}\ \emph {et~al.}(2019)\citenamefont {Xu},
  \citenamefont {Zhou}, \citenamefont {Liu},\ and\ \citenamefont
  {Chen}}]{Xu_2019_kappa-cum}%
  \BibitemOpen
  \bibfield  {author} {\bibinfo {author} {\bibfnamefont {Q.}~\bibnamefont
  {Xu}}, \bibinfo {author} {\bibfnamefont {J.}~\bibnamefont {Zhou}}, \bibinfo
  {author} {\bibfnamefont {T.-H.}\ \bibnamefont {Liu}},\ and\ \bibinfo {author}
  {\bibfnamefont {G.}~\bibnamefont {Chen}},\ }\bibfield  {title} {\bibinfo
  {title} {Effect of electron-phonon interaction on lattice thermal
  conductivity of {S}i{G}e alloys},\ }\href {https://doi.org/10.1063/1.5108836}
  {\bibfield  {journal} {\bibinfo  {journal} {Applied Physics Letters}\
  }\textbf {\bibinfo {volume} {115}},\ \bibinfo {pages} {023903} (\bibinfo
  {year} {2019})}\BibitemShut {NoStop}%
\bibitem [{\citenamefont {Reparaz}\ \emph {et~al.}(2013)\citenamefont
  {Reparaz}, \citenamefont {Peica}, \citenamefont {Kirste}, \citenamefont
  {Goñi}, \citenamefont {Wagner}, \citenamefont {Callsen}, \citenamefont
  {Alonso}, \citenamefont {Garriga}, \citenamefont {Marcus}, \citenamefont
  {Ronda}, \citenamefont {Berbezier}, \citenamefont {Maultzsch}, \citenamefont
  {Thomsen},\ and\ \citenamefont {Hoffmann}}]{Reparaz2013_TipEnhRaman}%
  \BibitemOpen
  \bibfield  {author} {\bibinfo {author} {\bibfnamefont {J.~S.}\ \bibnamefont
  {Reparaz}}, \bibinfo {author} {\bibfnamefont {N.}~\bibnamefont {Peica}},
  \bibinfo {author} {\bibfnamefont {R.}~\bibnamefont {Kirste}}, \bibinfo
  {author} {\bibfnamefont {A.~R.}\ \bibnamefont {Goñi}}, \bibinfo {author}
  {\bibfnamefont {M.~R.}\ \bibnamefont {Wagner}}, \bibinfo {author}
  {\bibfnamefont {G.}~\bibnamefont {Callsen}}, \bibinfo {author} {\bibfnamefont
  {M.~I.}\ \bibnamefont {Alonso}}, \bibinfo {author} {\bibfnamefont
  {M.}~\bibnamefont {Garriga}}, \bibinfo {author} {\bibfnamefont {I.~C.}\
  \bibnamefont {Marcus}}, \bibinfo {author} {\bibfnamefont {A.}~\bibnamefont
  {Ronda}}, \bibinfo {author} {\bibfnamefont {I.}~\bibnamefont {Berbezier}},
  \bibinfo {author} {\bibfnamefont {J.}~\bibnamefont {Maultzsch}}, \bibinfo
  {author} {\bibfnamefont {C.}~\bibnamefont {Thomsen}},\ and\ \bibinfo {author}
  {\bibfnamefont {A.}~\bibnamefont {Hoffmann}},\ }\bibfield  {title} {\bibinfo
  {title} {Probing local strain and composition in {G}e nanowires by means of
  tip-enhanced {R}aman scattering},\ }\href
  {https://doi.org/10.1088/0957-4484/24/18/185704} {\bibfield  {journal}
  {\bibinfo  {journal} {Nanotechnology}\ }\textbf {\bibinfo {volume} {24}},\
  \bibinfo {pages} {185704} (\bibinfo {year} {2013})}\BibitemShut {NoStop}%
\bibitem [{\citenamefont {Romano}(2021)}]{Romano2021}%
  \BibitemOpen
  \bibfield  {author} {\bibinfo {author} {\bibfnamefont {G.}~\bibnamefont
  {Romano}},\ }\href {https://doi.org/10.48550/ARXIV.2106.02764} {\bibinfo
  {title} {{O}pen{BTE}: a {S}olver for ab-initio {P}honon {T}ransport in
  {M}ultidimensional {S}tructures}} (\bibinfo {year} {2021})\BibitemShut
  {NoStop}%
\bibitem [{\citenamefont {Romano}\ and\ \citenamefont
  {Johnson}(2022)}]{romano2022inverse}%
  \BibitemOpen
  \bibfield  {author} {\bibinfo {author} {\bibfnamefont {G.}~\bibnamefont
  {Romano}}\ and\ \bibinfo {author} {\bibfnamefont {S.~G.}\ \bibnamefont
  {Johnson}},\ }\bibfield  {title} {\bibinfo {title} {Inverse design in
  nanoscale heat transport via interpolating interfacial phonon transmission},\
  }\href@noop {} {\bibfield  {journal} {\bibinfo  {journal} {Structural and
  Multidisciplinary Optimization}\ }\textbf {\bibinfo {volume} {65}},\ \bibinfo
  {pages} {297} (\bibinfo {year} {2022})}\BibitemShut {NoStop}%
\bibitem [{\citenamefont {Cuffe}\ \emph {et~al.}(2015)\citenamefont {Cuffe},
  \citenamefont {Eliason}, \citenamefont {Maznev}, \citenamefont {Collins},
  \citenamefont {Johnson}, \citenamefont {Shchepetov}, \citenamefont
  {Prunnila}, \citenamefont {Ahopelto}, \citenamefont {Sotomayor~Torres},
  \citenamefont {Chen},\ and\ \citenamefont {Nelson}}]{Cuffe2015}%
  \BibitemOpen
  \bibfield  {author} {\bibinfo {author} {\bibfnamefont {J.}~\bibnamefont
  {Cuffe}}, \bibinfo {author} {\bibfnamefont {J.~K.}\ \bibnamefont {Eliason}},
  \bibinfo {author} {\bibfnamefont {A.~A.}\ \bibnamefont {Maznev}}, \bibinfo
  {author} {\bibfnamefont {K.~C.}\ \bibnamefont {Collins}}, \bibinfo {author}
  {\bibfnamefont {J.~A.}\ \bibnamefont {Johnson}}, \bibinfo {author}
  {\bibfnamefont {A.}~\bibnamefont {Shchepetov}}, \bibinfo {author}
  {\bibfnamefont {M.}~\bibnamefont {Prunnila}}, \bibinfo {author}
  {\bibfnamefont {J.}~\bibnamefont {Ahopelto}}, \bibinfo {author}
  {\bibfnamefont {C.~M.}\ \bibnamefont {Sotomayor~Torres}}, \bibinfo {author}
  {\bibfnamefont {G.}~\bibnamefont {Chen}},\ and\ \bibinfo {author}
  {\bibfnamefont {K.~A.}\ \bibnamefont {Nelson}},\ }\bibfield  {title}
  {\bibinfo {title} {Reconstructing phonon mean-free-path contributions to
  thermal conductivity using nanoscale membranes},\ }\href@noop {} {\bibfield
  {journal} {\bibinfo  {journal} {Physical Review B}\ }\textbf {\bibinfo
  {volume} {91}},\ \bibinfo {pages} {245423} (\bibinfo {year}
  {2015})}\BibitemShut {NoStop}%
\bibitem [{\citenamefont {Marroux}\ \emph {et~al.}(2024)\citenamefont
  {Marroux}, \citenamefont {Polishchuk}, \citenamefont {Cannelli},
  \citenamefont {Ingle}, \citenamefont {Mancini}, \citenamefont {Bacellar},
  \citenamefont {Puppin}, \citenamefont {Geneaux}, \citenamefont {Knopp},
  \citenamefont {Foglia}, \citenamefont {Pedersoli}, \citenamefont {Capotondi},
  \citenamefont {Nikolov}, \citenamefont {Bencivenga}, \citenamefont
  {Mincigrucci}, \citenamefont {Masciovecchio},\ and\ \citenamefont
  {Chergui}}]{Marroux2024}%
  \BibitemOpen
  \bibfield  {author} {\bibinfo {author} {\bibfnamefont {H.~J.~B.}\
  \bibnamefont {Marroux}}, \bibinfo {author} {\bibfnamefont {S.}~\bibnamefont
  {Polishchuk}}, \bibinfo {author} {\bibfnamefont {O.}~\bibnamefont
  {Cannelli}}, \bibinfo {author} {\bibfnamefont {R.~A.}\ \bibnamefont {Ingle}},
  \bibinfo {author} {\bibfnamefont {G.~F.}\ \bibnamefont {Mancini}}, \bibinfo
  {author} {\bibfnamefont {C.}~\bibnamefont {Bacellar}}, \bibinfo {author}
  {\bibfnamefont {M.}~\bibnamefont {Puppin}}, \bibinfo {author} {\bibfnamefont
  {R.}~\bibnamefont {Geneaux}}, \bibinfo {author} {\bibfnamefont
  {G.}~\bibnamefont {Knopp}}, \bibinfo {author} {\bibfnamefont
  {L.}~\bibnamefont {Foglia}}, \bibinfo {author} {\bibfnamefont
  {E.}~\bibnamefont {Pedersoli}}, \bibinfo {author} {\bibfnamefont
  {F.}~\bibnamefont {Capotondi}}, \bibinfo {author} {\bibfnamefont {I.~P.}\
  \bibnamefont {Nikolov}}, \bibinfo {author} {\bibfnamefont {F.}~\bibnamefont
  {Bencivenga}}, \bibinfo {author} {\bibfnamefont {R.}~\bibnamefont
  {Mincigrucci}}, \bibinfo {author} {\bibfnamefont {C.}~\bibnamefont
  {Masciovecchio}},\ and\ \bibinfo {author} {\bibfnamefont {M.}~\bibnamefont
  {Chergui}},\ }\bibfield  {title} {\bibinfo {title} {Separation of kinetic
  rate orders in extreme ultraviolet transient grating spectroscopy},\ }\href
  {https://doi.org/10.1088/1361-6455/ad421f} {\bibfield  {journal} {\bibinfo
  {journal} {Journal of Physics B: Atomic, Molecular and Optical Physics}\
  }\textbf {\bibinfo {volume} {57}},\ \bibinfo {pages} {115401} (\bibinfo
  {year} {2024})}\BibitemShut {NoStop}%
\bibitem [{\citenamefont {Nelson}\ \emph {et~al.}(2024)\citenamefont {Nelson},
  \citenamefont {McBennett}, \citenamefont {Culman}, \citenamefont {Beardo},
  \citenamefont {Kapteyn}, \citenamefont {Frey}, \citenamefont {Atkinson},
  \citenamefont {Murnane},\ and\ \citenamefont
  {Knobloch}}]{NelsonKnobloch2024}%
  \BibitemOpen
  \bibfield  {author} {\bibinfo {author} {\bibfnamefont {E.~E.}\ \bibnamefont
  {Nelson}}, \bibinfo {author} {\bibfnamefont {B.}~\bibnamefont {McBennett}},
  \bibinfo {author} {\bibfnamefont {T.~H.}\ \bibnamefont {Culman}}, \bibinfo
  {author} {\bibfnamefont {A.}~\bibnamefont {Beardo}}, \bibinfo {author}
  {\bibfnamefont {H.~C.}\ \bibnamefont {Kapteyn}}, \bibinfo {author}
  {\bibfnamefont {M.~H.}\ \bibnamefont {Frey}}, \bibinfo {author}
  {\bibfnamefont {M.~R.}\ \bibnamefont {Atkinson}}, \bibinfo {author}
  {\bibfnamefont {M.~M.}\ \bibnamefont {Murnane}},\ and\ \bibinfo {author}
  {\bibfnamefont {J.~L.}\ \bibnamefont {Knobloch}},\ }\bibfield  {title}
  {\bibinfo {title} {Tabletop deep-ultraviolet transient grating for ultrafast
  nanoscale carrier-transport measurements in ultrawide-band-gap materials},\
  }\href {https://doi.org/10.1103/physrevapplied.22.054007} {\bibfield
  {journal} {\bibinfo  {journal} {Physical Review Applied}\ }\textbf {\bibinfo
  {volume} {22}},\ \bibinfo {pages} {054007} (\bibinfo {year}
  {2024})}\BibitemShut {NoStop}%
\bibitem [{\citenamefont {{J\"{u}lich Supercomputing Centre}}(2021)}]{JUWELS}%
  \BibitemOpen
  \bibfield  {author} {\bibinfo {author} {\bibnamefont {{J\"{u}lich
  Supercomputing Centre}}},\ }\bibfield  {title} {\bibinfo {title} {{JUWELS
  Cluster and Booster: Exascale Pathfinder with Modular Supercomputing
  Architecture at Juelich Supercomputing Centre}},\ }\href
  {http://dx.doi.org/10.17815/jlsrf-7-183} {\bibfield  {journal} {\bibinfo
  {journal} {Journal of large-scale research facilities}\ }\textbf {\bibinfo
  {volume} {7}} (\bibinfo {year} {2021})}\BibitemShut {NoStop}%
\end{thebibliography}%


\begin{thebibliography}{10}%
\makeatletter
\providecommand \@ifxundefined [1]{%
 \@ifx{#1\undefined}
}%
\providecommand \@ifnum [1]{%
 \ifnum #1\expandafter \@firstoftwo
 \else \expandafter \@secondoftwo
 \fi
}%
\providecommand \@ifx [1]{%
 \ifx #1\expandafter \@firstoftwo
 \else \expandafter \@secondoftwo
 \fi
}%
\providecommand \natexlab [1]{#1}%
\providecommand \enquote  [1]{``#1''}%
\providecommand \bibnamefont  [1]{#1}%
\providecommand \bibfnamefont [1]{#1}%
\providecommand \citenamefont [1]{#1}%
\providecommand \href@noop [0]{\@secondoftwo}%
\providecommand \href [0]{\begingroup \@sanitize@url \@href}%
\providecommand \@href[1]{\@@startlink{#1}\@@href}%
\providecommand \@@href[1]{\endgroup#1\@@endlink}%
\providecommand \@sanitize@url [0]{\catcode `\\12\catcode `\$12\catcode
  `\&12\catcode `\#12\catcode `\^12\catcode `\_12\catcode `\%12\relax}%
\providecommand \@@startlink[1]{}%
\providecommand \@@endlink[0]{}%
\providecommand \url  [0]{\begingroup\@sanitize@url \@url }%
\providecommand \@url [1]{\endgroup\@href {#1}{\urlprefix }}%
\providecommand \urlprefix  [0]{URL }%
\providecommand \Eprint [0]{\href }%
\providecommand \doibase [0]{http://dx.doi.org/}%
\providecommand \selectlanguage [0]{\@gobble}%
\providecommand \bibinfo  [0]{\@secondoftwo}%
\providecommand \bibfield  [0]{\@secondoftwo}%
\providecommand \translation [1]{[#1]}%
\providecommand \BibitemOpen [0]{}%
\providecommand \bibitemStop [0]{}%
\providecommand \bibitemNoStop [0]{.\EOS\space}%
\providecommand \EOS [0]{\spacefactor3000\relax}%
\providecommand \BibitemShut  [1]{\csname bibitem#1\endcsname}%
\let\auto@bib@innerbib\@empty
\bibitem [{\citenamefont {Men{\'{e}}ndez}\ and\ \citenamefont
  {Cardona}(1984)}]{Menendez1984}%
  \BibitemOpen
  \bibfield  {author} {\bibinfo {author} {\bibfnamefont {Jos{\'{e}}}\
  \bibnamefont {Men{\'{e}}ndez}}\ and\ \bibinfo {author} {\bibfnamefont
  {Manuel}\ \bibnamefont {Cardona}},\ }\bibfield  {title} {\enquote {\bibinfo
  {title} {Temperature dependence of the first-order {R}aman scattering by
  phonons in {S}i, {G}e, and $\alpha$-{S}n: {A}nharmonic effects},}\ }\href
  {\doibase 10.1103/physrevb.29.2051} {\bibfield  {journal} {\bibinfo
  {journal} {Physical Review B}\ }\textbf {\bibinfo {volume} {29}},\ \bibinfo
  {pages} {2051--2059} (\bibinfo {year} {1984})}\BibitemShut {NoStop}%
\bibitem [{\citenamefont {Franta}\ \emph {et~al.}(2019)\citenamefont {Franta},
  \citenamefont {Vohánka}, \citenamefont {Čermák}, \citenamefont {Franta},\
  and\ \citenamefont {Ohlídal}}]{Franta2019}%
  \BibitemOpen
  \bibfield  {author} {\bibinfo {author} {\bibfnamefont {Daniel}\ \bibnamefont
  {Franta}}, \bibinfo {author} {\bibfnamefont {Jiří}\ \bibnamefont
  {Vohánka}}, \bibinfo {author} {\bibfnamefont {Martin}\ \bibnamefont
  {Čermák}}, \bibinfo {author} {\bibfnamefont {Pavel}\ \bibnamefont
  {Franta}}, \ and\ \bibinfo {author} {\bibfnamefont {Ivan}\ \bibnamefont
  {Ohlídal}},\ }\bibfield  {title} {\enquote {\bibinfo {title} {Temperature
  dependent dispersion models applicable in solid state physics},}\ }\href
  {\doibase 10.2478/jee-2019-0036} {\bibfield  {journal} {\bibinfo  {journal}
  {Journal of Electrical Engineering}\ }\textbf {\bibinfo {volume} {70}},\
  \bibinfo {pages} {1--15} (\bibinfo {year} {2019})}\BibitemShut {NoStop}%
\bibitem [{\citenamefont {Aspnes}\ and\ \citenamefont
  {Studna}(1983)}]{Aspnes1983}%
  \BibitemOpen
  \bibfield  {author} {\bibinfo {author} {\bibfnamefont {D.~E.}\ \bibnamefont
  {Aspnes}}\ and\ \bibinfo {author} {\bibfnamefont {A.~A.}\ \bibnamefont
  {Studna}},\ }\bibfield  {title} {\enquote {\bibinfo {title} {Dielectric
  functions and optical parameters of {S}i, {G}e, {G}a{P}, {G}a{A}s, {G}a{S}b,
  {I}n{P}, {I}n{A}s, and {I}n{S}b from 1.5 to 6.0 e{V}},}\ }\href {\doibase
  10.1103/physrevb.27.985} {\bibfield  {journal} {\bibinfo  {journal} {Physical
  Review B}\ }\textbf {\bibinfo {volume} {27}},\ \bibinfo {pages} {985--1009}
  (\bibinfo {year} {1983})}\BibitemShut {NoStop}%
\bibitem [{\citenamefont {Chávez-Ángel}\ \emph {et~al.}(2014)\citenamefont
  {Chávez-Ángel}, \citenamefont {Reparaz}, \citenamefont {Gomis-Bresco},
  \citenamefont {Wagner}, \citenamefont {Cuffe}, \citenamefont {Graczykowski},
  \citenamefont {Shchepetov}, \citenamefont {Jiang}, \citenamefont {Prunnila},
  \citenamefont {Ahopelto}, \citenamefont {Alzina},\ and\ \citenamefont
  {Sotomayor~Torres}}]{ChavezAngel2014}%
  \BibitemOpen
  \bibfield  {author} {\bibinfo {author} {\bibfnamefont {E.}~\bibnamefont
  {Chávez-Ángel}}, \bibinfo {author} {\bibfnamefont {J.~S.}\ \bibnamefont
  {Reparaz}}, \bibinfo {author} {\bibfnamefont {J.}~\bibnamefont
  {Gomis-Bresco}}, \bibinfo {author} {\bibfnamefont {M.~R.}\ \bibnamefont
  {Wagner}}, \bibinfo {author} {\bibfnamefont {J.}~\bibnamefont {Cuffe}},
  \bibinfo {author} {\bibfnamefont {B.}~\bibnamefont {Graczykowski}}, \bibinfo
  {author} {\bibfnamefont {A.}~\bibnamefont {Shchepetov}}, \bibinfo {author}
  {\bibfnamefont {H.}~\bibnamefont {Jiang}}, \bibinfo {author} {\bibfnamefont
  {M.}~\bibnamefont {Prunnila}}, \bibinfo {author} {\bibfnamefont
  {J.}~\bibnamefont {Ahopelto}}, \bibinfo {author} {\bibfnamefont
  {F.}~\bibnamefont {Alzina}}, \ and\ \bibinfo {author} {\bibfnamefont {C.~M.}\
  \bibnamefont {Sotomayor~Torres}},\ }\bibfield  {title} {\enquote {\bibinfo
  {title} {Reduction of the thermal conductivity in free-standing silicon
  nano-membranes investigated by non-invasive {R}aman thermometry},}\
  }\href@noop {} {\bibfield  {journal} {\bibinfo  {journal} {APL Materials}\
  }\textbf {\bibinfo {volume} {2}} (\bibinfo {year} {2014})}\BibitemShut
  {NoStop}%
\bibitem [{\citenamefont {Glassbrenner}\ and\ \citenamefont
  {Slack}(1964)}]{Glassbrenner1964}%
  \BibitemOpen
  \bibfield  {author} {\bibinfo {author} {\bibfnamefont {C.~J.}\ \bibnamefont
  {Glassbrenner}}\ and\ \bibinfo {author} {\bibfnamefont {Glen~A.}\
  \bibnamefont {Slack}},\ }\bibfield  {title} {\enquote {\bibinfo {title}
  {Thermal {C}onductivity of {S}ilicon and {G}ermanium from 3$^\circ${K} to the
  {M}elting {P}oint},}\ }\href {\doibase 10.1103/physrev.134.a1058} {\bibfield
  {journal} {\bibinfo  {journal} {Physical Review}\ }\textbf {\bibinfo {volume}
  {134}},\ \bibinfo {pages} {A1058--A1069} (\bibinfo {year}
  {1964})}\BibitemShut {NoStop}%
\bibitem [{\citenamefont {Mernagh}\ and\ \citenamefont
  {Liu}(1991)}]{Mernagh1991}%
  \BibitemOpen
  \bibfield  {author} {\bibinfo {author} {\bibfnamefont {Terrence~P.}\
  \bibnamefont {Mernagh}}\ and\ \bibinfo {author} {\bibfnamefont {Lin-Gun}\
  \bibnamefont {Liu}},\ }\bibfield  {title} {\enquote {\bibinfo {title}
  {Pressure dependence of {R}aman phonons of some group {IVA} ({C}, {S}i, and
  {G}e) elements},}\ }\href {\doibase 10.1016/0022-3697(91)90183-z} {\bibfield
  {journal} {\bibinfo  {journal} {Journal of Physics and Chemistry of Solids}\
  }\textbf {\bibinfo {volume} {52}},\ \bibinfo {pages} {507--512} (\bibinfo
  {year} {1991})}\BibitemShut {NoStop}%
\bibitem [{\citenamefont {Lax}(1977)}]{Lax1977}%
  \BibitemOpen
  \bibfield  {author} {\bibinfo {author} {\bibfnamefont {M.}~\bibnamefont
  {Lax}},\ }\bibfield  {title} {\enquote {\bibinfo {title} {Temperature rise
  induced by a laser beam},}\ }\href {\doibase 10.1063/1.324265} {\bibfield
  {journal} {\bibinfo  {journal} {Journal of Applied Physics}\ }\textbf
  {\bibinfo {volume} {48}},\ \bibinfo {pages} {3919--3924} (\bibinfo {year}
  {1977})}\BibitemShut {NoStop}%
\bibitem [{\citenamefont {Jellison}(1992)}]{Jellison1992}%
  \BibitemOpen
  \bibfield  {author} {\bibinfo {author} {\bibfnamefont {G.E.}\ \bibnamefont
  {Jellison}},\ }\bibfield  {title} {\enquote {\bibinfo {title} {Optical
  functions of {G}a{A}s, {G}a{P}, and {G}e determined by two-channel
  polarization modulation ellipsometry},}\ }\href {\doibase
  10.1016/0925-3467(92)90022-f} {\bibfield  {journal} {\bibinfo  {journal}
  {Optical Materials}\ }\textbf {\bibinfo {volume} {1}},\ \bibinfo {pages}
  {151--160} (\bibinfo {year} {1992})}\BibitemShut {NoStop}%
\bibitem [{\citenamefont {Nunley}\ \emph {et~al.}(2016)\citenamefont {Nunley},
  \citenamefont {Fernando}, \citenamefont {Samarasingha}, \citenamefont {Moya},
  \citenamefont {Nelson}, \citenamefont {Medina},\ and\ \citenamefont
  {Zollner}}]{Nunley2016}%
  \BibitemOpen
  \bibfield  {author} {\bibinfo {author} {\bibfnamefont {Timothy~Nathan}\
  \bibnamefont {Nunley}}, \bibinfo {author} {\bibfnamefont {Nalin~S.}\
  \bibnamefont {Fernando}}, \bibinfo {author} {\bibfnamefont {Nuwanjula}\
  \bibnamefont {Samarasingha}}, \bibinfo {author} {\bibfnamefont {Jaime~M.}\
  \bibnamefont {Moya}}, \bibinfo {author} {\bibfnamefont {Cayla~M.}\
  \bibnamefont {Nelson}}, \bibinfo {author} {\bibfnamefont {Amber~A.}\
  \bibnamefont {Medina}}, \ and\ \bibinfo {author} {\bibfnamefont {Stefan}\
  \bibnamefont {Zollner}},\ }\bibfield  {title} {\enquote {\bibinfo {title}
  {Optical constants of germanium and thermally grown germanium dioxide from
  0.5 to 6.6 e{V} via a multisample ellipsometry investigation},}\ }\href@noop
  {} {\bibfield  {journal} {\bibinfo  {journal} {Journal of Vacuum Science \&
  Technology B, Nanotechnology and Microelectronics: Materials, Processing,
  Measurement, and Phenomena}\ }\textbf {\bibinfo {volume} {34}} (\bibinfo
  {year} {2016})}\BibitemShut {NoStop}%
\bibitem [{\citenamefont {Elhajhasan}\ \emph {et~al.}(2023)\citenamefont
  {Elhajhasan}, \citenamefont {Seemann}, \citenamefont {Dudde}, \citenamefont
  {Vaske}, \citenamefont {Callsen}, \citenamefont {Rousseau}, \citenamefont
  {Weatherley}, \citenamefont {Carlin}, \citenamefont {Butté}, \citenamefont
  {Grandjean}, \citenamefont {Protik},\ and\ \citenamefont
  {Romano}}]{Elhajhasan2023}%
  \BibitemOpen
  \bibfield  {author} {\bibinfo {author} {\bibfnamefont {Mahmoud}\ \bibnamefont
  {Elhajhasan}}, \bibinfo {author} {\bibfnamefont {Wilken}\ \bibnamefont
  {Seemann}}, \bibinfo {author} {\bibfnamefont {Katharina}\ \bibnamefont
  {Dudde}}, \bibinfo {author} {\bibfnamefont {Daniel}\ \bibnamefont {Vaske}},
  \bibinfo {author} {\bibfnamefont {Gordon}\ \bibnamefont {Callsen}}, \bibinfo
  {author} {\bibfnamefont {Ian}\ \bibnamefont {Rousseau}}, \bibinfo {author}
  {\bibfnamefont {Thomas F.~K.}\ \bibnamefont {Weatherley}}, \bibinfo {author}
  {\bibfnamefont {Jean-François}\ \bibnamefont {Carlin}}, \bibinfo {author}
  {\bibfnamefont {Raphaël}\ \bibnamefont {Butté}}, \bibinfo {author}
  {\bibfnamefont {Nicolas}\ \bibnamefont {Grandjean}}, \bibinfo {author}
  {\bibfnamefont {Nakib~H.}\ \bibnamefont {Protik}}, \ and\ \bibinfo {author}
  {\bibfnamefont {Giuseppe}\ \bibnamefont {Romano}},\ }\bibfield  {title}
  {\enquote {\bibinfo {title} {Optical and thermal characterization of a
  group-{III} nitride semiconductor membrane by microphotoluminescence
  spectroscopy and {R}aman thermometry},}\ }\href {\doibase
  10.1103/physrevb.108.235313} {\bibfield  {journal} {\bibinfo  {journal}
  {Physical Review B}\ }\textbf {\bibinfo {volume} {108}},\ \bibinfo {pages}
  {235313} (\bibinfo {year} {2023})}\BibitemShut {NoStop}%
\end{thebibliography}%

\end{document}


\title[]{SUPPLEMENTAL MATERIAL: \\Phonon Mean Free Path Spectroscopy By Raman Thermometry
}

\author{Katharina~Dudde*}
\author{Mahmoud~Elhajhasan}
\author{Guillaume~Würsch}
\author{Julian~Themann}
\author{Jana~Lierath}
\author{Gordon~Callsen{\textsuperscript{\textdagger}}}

\affiliation{Institut für Festk\"orperphysik, Universit\"at Bremen, Otto-Hahn-Allee 1, 28359 Bremen, Germany}

\author{Dwaipayan Paul}
\author{Nakib H. Protik}

\affiliation{Institut für Physik and Center for the Science of Materials Berlin (CSMB), Humboldt-Universität zu Berlin, 12489 Berlin, Germany}%

\author{Giuseppe Romano}

\affiliation{MIT-IBM Watson AI Lab, 214 Main St., Cambridge, Massachusetts 02141, USA}

\author{(*duddekat@uni-bremen.de, \textsuperscript{\textdagger}gcallsen@uni-bremen.de)}
\affiliation{}

\date{\today}

\maketitle

%
%
\section{\label{S-sec:Tcal}Temperature calibration of the Raman mode shift}
%
%

The temperature ($T$) calibration factor of $\frac{\displaystyle \Delta T}{\displaystyle \Delta \nu} = \SI{-0.0229(1)}{\per\centi\meter\per\kelvin}$ for the Raman mode (with frequency $\nu$) of the optical phonon at the $\Gamma$-point in bulk silicon was measured in a heat stage in a temperature range of $\SI{293}{\kelvin} \leq T \leq \SI{573}{\kelvin}$.
For the measurements at \SI{200}{\kelvin} we used the data from Ref. \cite{Menendez1984} yielding a coefficient of $\frac{\displaystyle \Delta T}{\displaystyle \Delta \nu} = \SI{-0.0217(5)}{\per\centi\meter\per\kelvin}$.\\
From the temperature calibration measurement for the silicon membranes in the heat stage, we obtain $\frac{\displaystyle \Delta T}{\displaystyle \Delta \nu} = \SI{-0.0432(13)}{\per\centi\meter\per\kelvin}$ for the 200-nm-thick membrane, which is valid for a temperature range of $\SI{293}{\kelvin} \leq T \leq \SI{673}{\kelvin}$.
For the 2000-nm-thick membrane these calibration measurements yield a shift of $\frac{\displaystyle \Delta T}{\displaystyle \Delta \nu} = \SI{-0.0252(4)}{\per\centi\meter\per\kelvin}$ for $\SI{293}{\kelvin} \leq T \leq \SI{323}{\kelvin}$, which is much closer to the bulk silicon value.
The temperature calibration for the 200-nm-thick membrane was recorded in steps of \SI{50}{\kelvin}, while for the 2000-nm-thick membrane temperature steps of \SI{2}{\kelvin}.\\
We wish to note that for all temperature calibration measurements, we also measure the system-specific Gaussian broadening that impacts the Raman spectra.
This effect can be encountered by fitting Voigt functions to the Raman spectra.
This way, the natural Lorentzian mode shape is considered along with its Gaussian broadening induced by our detection system.
For these Voigt fits, the Gaussian width is a fixed parameter, which we determined by measuring the emission lines of a mercury lamp.
Therefore, we can apply the sample-specific temperature calibration to any of our recorded Raman spectra, independent of the measurement apparatus in use. \\
For the bulk germanium $\Gamma$-point Raman mode we measured a calibration factor of $\frac{\displaystyle \Delta T}{\displaystyle \Delta \nu} = \SI{-0.0183(1)}{\per\centi\meter\per\kelvin}$, which is valid in a temperature range of $\SI{293}{\kelvin} \leq T \leq \SI{473}{\kelvin}$.

%
%
\section{\label{S-sec:ha-choice}The impact of the light penetration depth on the extraction of the effective thermal condictivity}
%
%

\begin{figure}
    \includegraphics[width=12cm]{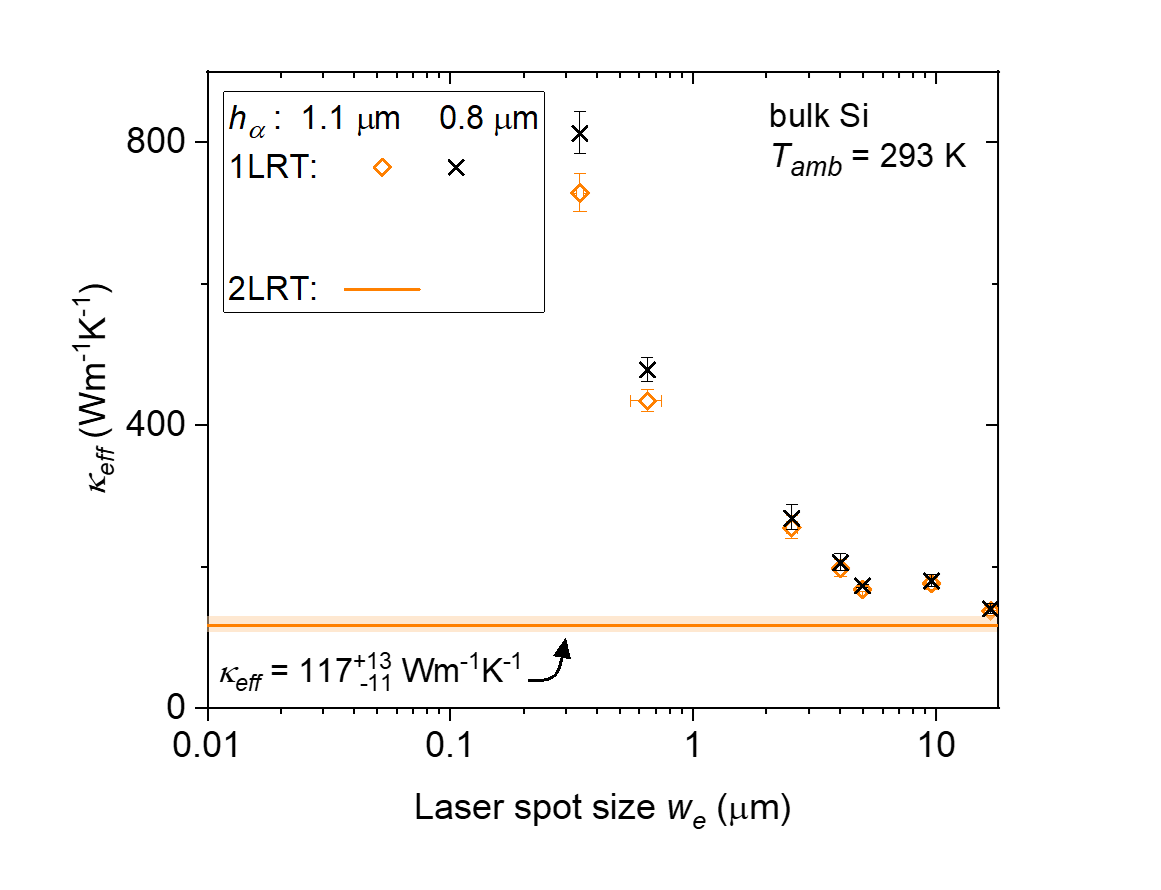}
    
    \caption{\textit{Comparison of the \ke values depending on the laser spot size $w_{\text{e}}$ for bulk silicon at $T_{\text{amb}} = \SI{293}{\kelvin}$ for two different $h_{\alpha}$ values, which are required as input parameters for the COMSOL Multiphysics\textsuperscript{\circledR} model.
    A similar trend is observed for $\kappa_{\text{eff}}(w_{\text{e}})$, no matter if $h_{\alpha} = \SI{1.1}{\micro\meter}$ (orange data points) based on Ref. \cite{Franta2019} or $h_{\alpha} = \SI{0.8}{\micro\meter}$ (black crosses) based on Ref. \cite{Aspnes1983} are utilized.
    The \ke value extracted from the 2LRT measurement (horizontal line) is not affected, because $h_{\alpha}$ plays no role in our extraction of \ke from 2LRT measurements (see S-Sec. \ref{S-sec:2LRT}).}}
    \label{S-fig:ha-choice}
\end{figure}

In our main work we use data from literature to calculate the light penetration depth $h_{\alpha}$ for our silicon samples at various ambient temperatures $T_{\text{amb}}$ for a known Raman laser wavelength $\lambda$.
We use the data published in Ref. \cite{Franta2019} for crystalline float zone silicon, since this data is available not only for $T_{\text{amb}} = \SI{293}{\kelvin}$ but also at $T_{\text{amb}} = \SI{200}{\kelvin}$.
However, other data that can be found for silicon at room temperature in, e.g., Ref. \cite{Aspnes1983} yield deviating $h_{\alpha}$ values for a fixed $\lambda$ value.
For instance, for $T_{\text{amb}} = \SI{293}{\kelvin}$ and $\lambda = \SI{532}{\nano\meter}$ one can obtain $h_{\alpha} \approx \SI{1.1}{\micro\meter}$ \cite{Franta2019} or $h_{\alpha} \approx \SI{0.8}{\micro\meter}$ \cite{Aspnes1983} for crystalline silicon.
Since $h_{\alpha}$ is an important parameter for our data analysis by the COMSOL Multiphysics\textsuperscript{\circledR} software package, we compare the resulting \ke values, which we obtain when either $h_{\alpha} = \SI{1.1}{\micro\meter}$ or $h_{\alpha} = \SI{0.8}{\micro\meter}$ are utilized.
A summary of the resulting \ke values can be found in S-Fig. \ref{S-fig:ha-choice}.\\
The \ke values reduce, if the higher $h_{\alpha}$ value is assumed for the modeling.
With 13\,\% this reduction is maximized for the smallest $w_{\text{e}}$ value, while for $w_{\text{e}} > h_{\alpha}$ this reduction is $<$ 6\,\%.
Certainly, the choice of $h_{\alpha}$ is only of relevance for the determination of \ke as soon as the value of $h_{\alpha}$ approaches the laser spot size $w_{\text{e}}$ (defined as the radius of the focal laser spot, where the intensity has dropped to $1/e$-level).
Thus, the overall trend for $\kappa_{\text{eff}}(w_{\text{e}})$ is maintained, independent of the particular choice of $h_{\alpha}$ based on either Ref. \cite{Franta2019} or Ref. \cite{Aspnes1983}.
Consequently, the choice of $h_{\alpha}$ does not impact our data interpretation given in the main text.

%
%
\section{\label{S-sec:COMSOL-membrane}COMSOL Multiphysics\textsuperscript{\circledR} simulation of the experimental situation for a one-laser Raman thermometry measurement on a silicon membrane}
%
%

Depending on the laser wavelength $\lambda$ used during our one-laser Raman thermometry (1LRT), we either encounter surfacic or more volumetric heating conditions, which must be modeled adequately.\\
For $\lambda = \SI{325}{\nano\meter}$, where $h_{\alpha}$ amounts to approximately $\SI{10}{\nano\meter}$ \cite{Franta2019}, the heat source density $Q^{\text{2D}}(r)$ is assumed to be two-dimensional in the COMSOL Multiphysics\textsuperscript{\circledR} model, describing a surfacic heating situation: 
\begin{equation}\label{eq:hsd_membrane_325nm}
    Q^{\text{2D}}(r) = \frac{\displaystyle P_{\text{abs}}}{\displaystyle \pi\,w_{\text{e}}^2}e^{-\frac{\displaystyle (r - r_0)^2}{\displaystyle w_{\text{e}}^2}}
\end{equation}
Here, $w_{\text{e}}$ is the focal laser spot size, $P_{\text{abs}}$ the absorbed power, $r_0$ the position of the heat spot, and $r = \sqrt(x^2 + y^2)$ the radial coordinate in the plane of the membrane.\\
For $\lambda =\,$ 532, 561, and \SI{660}{\nano\meter}, we use a three-dimensional heat source density $Q^{\text{3D}}(r)$ in the model, in which the heated volume depth equals the membrane thickness $d$, because $h_{\alpha}$ always exceeds $d$.
As a result, we obtain an additional factor $1/d$ in the formula for the heat source density in this case:
\begin{equation}\label{eq:hsd_membrane_660nm}
    Q^{\text{3D}}(r) = \frac{\displaystyle P_{\text{abs}}}{\displaystyle \pi\,w_{\text{e}}^2\, d}e^{-\frac{\displaystyle (r - r_0)^2}{\displaystyle w_{\text{e}}^2}}
\end{equation}
To extract the temperature $T_{\text{m}}$ from the simulation, that corresponds to the experimentally probed temperature, we use the following formula:
\begin{equation}\label{eq:Tm_membrane}
    T_m = \frac{\displaystyle \int\int\int_{V} T(r,z) e^{-\frac{\displaystyle (r - r_0)^2}{\displaystyle w_{\text{e}}^2}}\,dV}{\displaystyle \int\int\int_{V}e^{-\frac{\displaystyle (r - r_0)^2}{\displaystyle w_{\text{e}}^2}}\,dV}
\end{equation}
Here, $T(r,z)$ are the simulated local temperatures that depend on the radial coordinate $r$ and the coordinate $z$ that is perpendicular to the membrane surface.
These local temperatures are weighted by the radial intensity distribution of our temperature probe laser.
The temperature probe volume $V$ is considered to have a depth of either $h_{\alpha}$ (for $\lambda = \SI{325}{\nano\meter}$) or $d$ (for all other wavelength) for our membrane samples.
No dependence of $h_{\alpha}$ is implemented in Eqs. \ref{eq:hsd_membrane_325nm}, \ref{eq:hsd_membrane_660nm}, \ref{eq:Tm_membrane}, because multiple internal reflections and the role of the interface roughness render this implementation challenging.
Instead the amount of reflected and absorbed laser power were directly measured for each membrane and each $\lambda$ value to determine the $P_{\text{abs}}$ values.
This approach can also be found in literature \cite{ChavezAngel2014}.

%
%
\section{\label{S-sec:2LRT}Extracting the effective thermal conductivity from a two-laser Raman thermometry measurement}
%
%

\begin{figure*}
    \includegraphics[width=\linewidth]{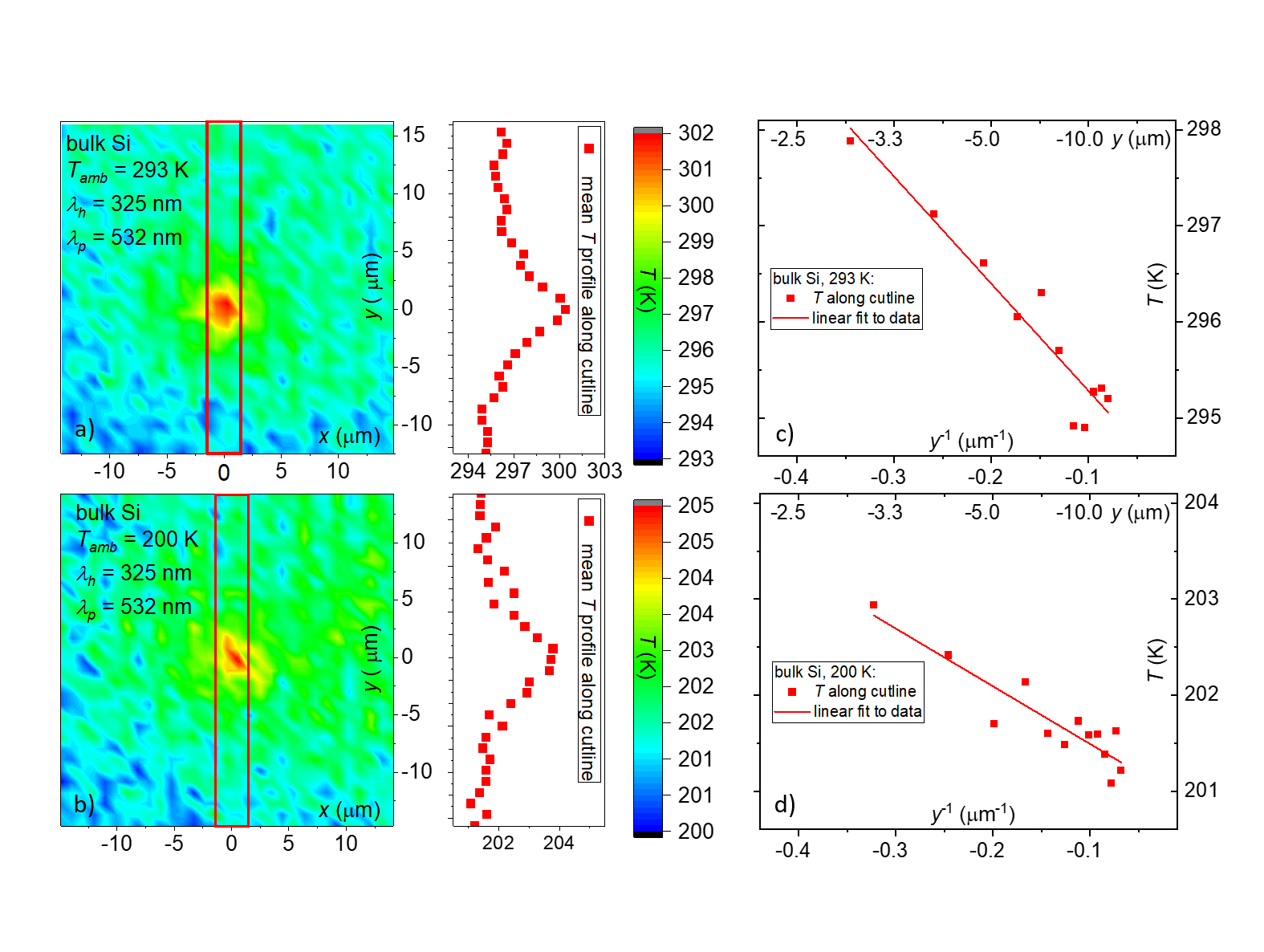}
    
    \caption{\textit{Extraction of \ke values from 2LRT measurements on bulk silicon with $\lambda_{\text{h}} = \SI{325}{\nano\meter}$ and $\lambda_{\text{p}} = \SI{532}{\nano\meter}$.
    a) and b): 2LRT temperature maps $T(x,y)$ recorded with a step size of $\approx \SI{1}{\micro\meter}$ at $T_{\text{amb}} = \SI{293}{\kelvin}$ (a) and $T_{\text{amb}} = \SI{200}{\kelvin}$ (b).
    The red rectangles indicate the cutline along which the mean temperature profile is evaluated. c) and d): Mean temperature profile along the cutlines plotted over the reciprocal distance $y^{-1}$ for $T_{\text{amb}} = \SI{293}{\kelvin}$ (c) and $T_{\text{amb}} = \SI{200}{\kelvin}$ (d) with a linear fit applied to the data to extract the \ke value according to S-Eq. \ref{S-eq:k-2LRT}. } }
    \label{S-fig:2LRT}
\end{figure*}

During two-laser Raman thermometry (2LRT), one laser with wavelength $\lambda_{\text{h}}$, constant power ($P_{\text{abs}}$), and constant position on the sample serves as the heat source, while a second laser with wavelength $\lambda_{\text{p}}$ is used to probe local temperatures.
This, so-called, probe laser is scanned over the sample and to record a spatially resolved temperature map $T(x,y)$.
$T(x,y)$ is obtained from the Raman mode shift $\nu$ with the temperature calibration factor $\frac{\displaystyle \Delta T}{\displaystyle \Delta \nu}$ (compare S-Sec. I). 
Two exemplary $T(x,y)$ maps are shown in S-Fig. \ref{S-fig:2LRT}a and b for bulk silicon at \SI{293}{\kelvin} and at \SI{200}{\kelvin}, respectively.
From such $T(x,y)$ maps the mean temperature profile along a cutline (marked in red in S-Fig. \ref{S-fig:2LRT}a and b) through the heat spot can be extracted to determine the effective thermal conductivity $\kappa_{\text{eff}}$.
We use a mean temperature profile where the temperatures from three individual scan lines contribute. 
A formula for $\kappa_{\text{eff}}$ can be derived from Fourier's law for a bulk material:
\begin{align}
    \vec{J} = -\kappa \vec{\nabla}T \quad &\Leftrightarrow \quad  \frac{P_{abs}}{2\pi r^2} = -\kappa \frac{\partial T}{\partial r}\notag\\
    &\Rightarrow\quad   \frac{P_{abs}}{2\pi} \int_{r_0}^{r} \frac{1}{\tilde{r}^2} \, d\tilde{r} = -\int_{T_0}^{T} \kappa\, d\tilde{T}\notag\\
    & \Rightarrow \quad T(r) = \frac{P_{abs}}{2\pi\kappa_0}\left(\frac{1}{r} - \frac{1}{r_0}\right) + T_0\notag\\
    \frac{\partial T(r)}{\partial r^{-1}} = \frac{P_{abs}}{2\pi\kappa_0} \quad & \Rightarrow \quad \kappa_0 = \frac{P_{abs}}{2\pi} \left(   \frac{\Delta T(r)}{\Delta r^{-1}}   \right)^{-1} \approx \kappa_{\text{eff}}
    \label{S-eq:k-2LRT}
\end{align}
With the heat flux density $\vec{J}$, the distance from the heat spot to another point in the bulk sample $r$, the thermal conductivity $\kappa$, the temperature independent thermal conductivity $\kappa_0$, the temperature $T$, the temperature $T_0 = T(r_0)$ at some fixed distance from the heat spot $r_0$, and the absorbed power $P_{\text{abs}}$. 
In the first line of Eq. \ref{S-eq:k-2LRT}, spherical coordinates and $\vec{J}$ being parallel to $\vec{r}$ were used.
Consequently $\vec{J}$ was replaced by $P_{\text{abs}}$ divided by the surface area of a half-sphere ($2\pi r^2$).
In the second line the variables $r$ and $T$ were separated for subsequent integration.
In the third line it was assumed that $\kappa$ is independent of $T$, which is true only for sufficiently small temperature changes.
As a result $\kappa_0$ replaces $\kappa$.
Since $T(r)$ depends linear on $r^{-1}$, $\kappa_0$ can be expressed as a function of the slope $(\Delta T(r)/\Delta r)^{-1}$ (compare S-Figs. \ref{S-fig:2LRT}c and d).
Finally, we use S-Eq. \ref{S-eq:k-2LRT}, to calculate $\kappa_{\text{eff}}$.
Therefore, only temperatures at a distance $r > \SI{3}{\micro\meter}$ are considered, where $T(r)$ depends linear of $r^{-1}$ and $\kappa_0 \approx \kappa_{\text{eff}}$ is a good approximation. 
In S-Figs. \ref{S-fig:2LRT}c and d, the average temperature is plotted over $-y^{-1}$ on a linear scale for $T_{\text{amb}} = \SI{293}{\kelvin}$ (c) and $T_{\text{amb}} = \SI{200}{\kelvin}$ (d).
With a linear fit to the data and with S-Eq. \ref{S-eq:k-2LRT}, one obtains $\kappa_{\text{eff}}(T = \SI{293}{\kelvin}) = 117_{-11}^{+13}\,\SI{}{\watt\per\meter\per\kelvin}$ and $\kappa_{\text{eff}}(T = \SI{200}{\kelvin}) = 213_{-26}^{+33}\,\SI{}{\watt\per\meter\per\kelvin}$.
Furthermore the effect of heat-induced stress at a distance of $r \geq \SI{3}{\micro\meter}$ is assumed to be negligible and no further correction of the temperatures extracted from the Raman shift $\nu$ is needed.

%
%
\section{\label{S-sec:stress}Considering the impact of stress on the Raman shift}
%
%

To correct the calculated \ke values for any impact of heat-induced stress, we first simulate the heating situation with the COMSOL Multiphysics\textsuperscript{\circledR} software for our experimental $P_{\text{abs}}$ values for both $T_{\text{amb}}$ values (\SI{200}{\kelvin} and \SI{200}{\kelvin}).
For $T_{\text{amb}} = \SI{293}{\kelvin}$ we assume $\kappa_{\text{sim}} = \SI{156}{\watt\per\meter\per\kelvin}$ according to Ref. \cite{Glassbrenner1964} and $h_{\alpha} = \SI{1.1}{\micro\meter}$ according to Ref. \cite{Franta2019}.
For $T_{\text{amb}} = \SI{200}{\kelvin}$ we use $\kappa_{sim} = \SI{200}{\watt\per\meter\per\kelvin}$ and $h_{\alpha} = \SI{1.5}{\micro\meter}$ \cite{Franta2019}.
With the \textit{Solid Mechanics} module in COMSOL Multiphysics\textsuperscript{\circledR} the local heat-induced stress distribution $\eta(r,z)$ is calculated for each specific heating situation, depending on the laser spot size $w_e$.
Next, a weighted mean $\eta(P_{\text{abs}})$ is calculated.
Therefore the local $\eta(r,z)$ values are weighted by the laser intensity distribution (Gaussian distribution in radial direction and decay with the Beer-Lambert law inside the bulk material).
This weighting is the same as we apply it to calculate the simulated measured temperatures $T_{\text{m}}$ for a bulk material from Eq. 3 from the main text.
Our calculations show that for any $w_e$ value $\eta \propto P_{\text{abs}}$.
Correspondingly, from the (linear) slope $\frac{\displaystyle \Delta \eta}{\displaystyle \Delta P_{\text{abs}}}$ that we obtain from the simulation, we can calculate an apparent temperature rise over the absorbed power $\frac{\displaystyle \Delta \Tilde{T}}{\displaystyle \Delta P_{\text{abs}}}$ with the following formula:

\begin{equation}
    \frac{\displaystyle \Delta \Tilde{T}}{\displaystyle \Delta P_{\text{abs}}} = \frac{\displaystyle \Delta \eta}{\displaystyle \Delta P_{\text{abs}}} \cdot \frac{\displaystyle \Delta T}{\displaystyle \Delta \nu} \cdot \frac{\displaystyle \Delta \nu}{\displaystyle \Delta \eta}
\end{equation}

Herein, the factor $\frac{\displaystyle \Delta T}{\displaystyle \Delta \nu}$ arises from the temperature calibration (see S-Sec. I) and the hydrostatic pressure coefficient $\frac{\displaystyle \Delta \nu}{\displaystyle \Delta \eta}$ yields $\SI{0.0045}{\per\centi\meter\per\mega\pascal}$ for bulk silicon and $\SI{0.0032}{\per\centi\meter\per\mega\pascal}$ for bulk germanium \cite{Mernagh1991}.
Based on the proportionality $\kappa \propto \left(\frac{\displaystyle \Delta T_{\text{rise}}}{\displaystyle \Delta P_{\text{abs}}} \right)^{-1}$ (compare, e.g., Ref. \cite{Lax1977}) that holds for a bulk material, the corrected $\kappa_{\text{eff}}$ values can be obtained from the non-corrected values $\kappa_{\text{nc}}$ with the following formula:
\begin{equation}
    \kappa_{\text{eff}} = \kappa_{\text{nc}} \cdot \left( 1 - \frac{\displaystyle \Delta \Tilde{T}/\Delta P_{\text{abs}}}{\displaystyle \Delta T_{\text{rise}} / \Delta P_{\text{abs}}} \right)^{-1}
\end{equation}
By this procedure we obtain an upper boundary approximation for the heat-induced stress.
Because the material can expand freely towards the vacuum at the sample surface but is compressed by the surrounding colder material in the other directions, the pressure situation is indeed not isotropic but rather triaxial with a bisotropic in-plane component.
Still, we assume the pressure to be hydrostatic and use the corresponding pressure coefficient.
As a result the extracted $\kappa_{\text{eff}}$ is a lower boundary approximation of the effective thermal conductivity.

%
%
\section{\label{S-sec:Ge} The laser spot size dependence of the effective thermal conductivity in bulk germanium}
%
%

\begin{figure}
    \includegraphics[width=12cm]{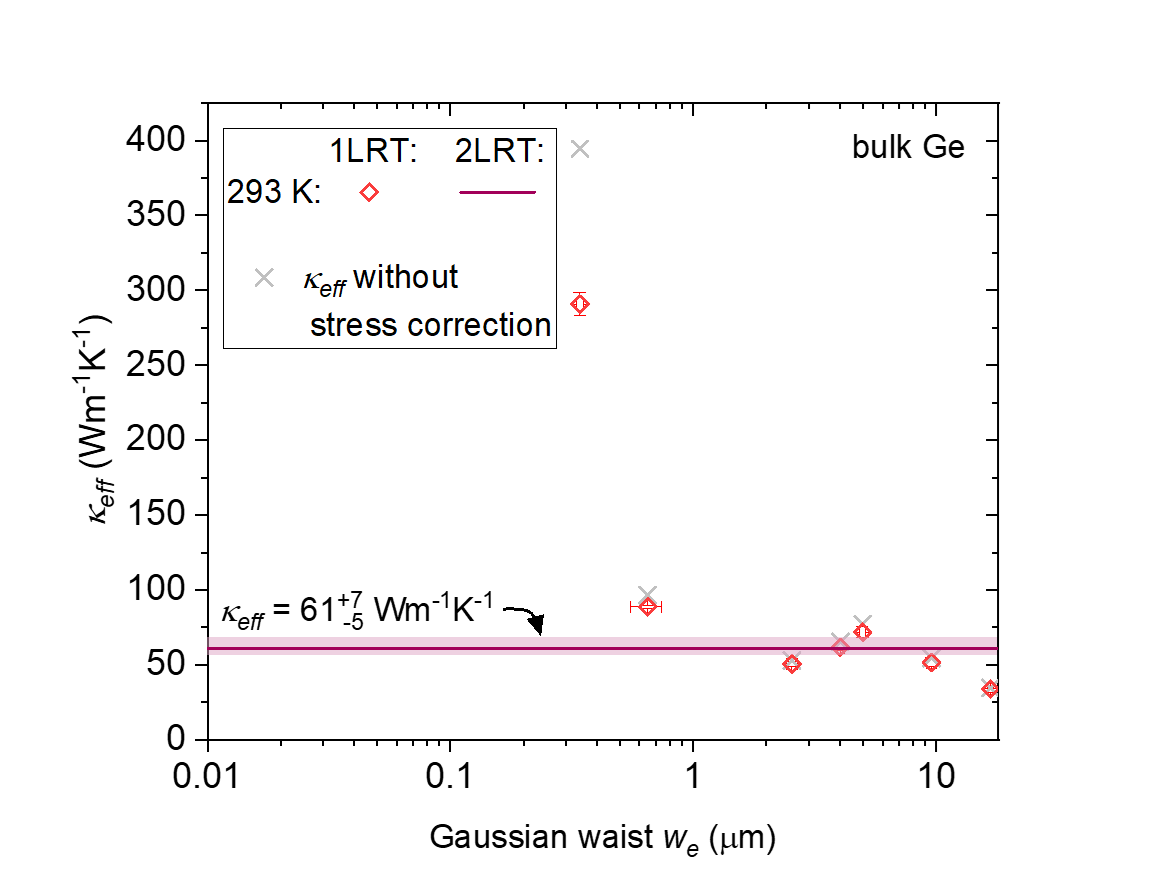}
    
    \caption{\textit{Effective thermal conductivity trend $\kappa_{\text{eff}}(w_{\text{e}})$ extracted from 1LRT measurements ($\lambda_{\text{h}} = \lambda_{\text{p}} = \SI{532}{\nano\meter}$) and corresponding \ke value from a 2LRT measurement ($\lambda_{\text{h}} = \SI{325}{\nano\meter}$ and $\lambda_{\text{p}} = \SI{532}{\nano\meter}$) at $T_{\text{amb}} = \SI{293}{\kelvin}$ for our bulk germanium sample. The symbols show the results from 1LRT measurements, once without any additional stress correction (grey crosses) and once with the effect of heat-induced stress taken into account (red diamonds). The \ke value from a 2LRT measurement is shown with the solid line where the shaded region indicates the uncertainty.}}
    \label{S-fig:Ge}
\end{figure}

The 1LRT measurements under the variation of $w_{\text{e}}$ that we performed for our bulk silicon sample ($T_{\text{amb}} = \SI{293}{\kelvin}$ and $\lambda_{\text{h}} = \lambda_{\text{p}} = \SI{532}{\nano\meter}$) as described in Sec. IIB1 of the main text were repeated under the same experimental conditions for a bulk germanium sample (sample details given in Sec. IIA of the main text).
At \SI{532}{\nano\meter} we assume $h_{\alpha} = \SI{18}{\nano\meter}$ for bulk germanium at room temperature as often found in literature \cite{Jellison1992, Aspnes1983, Nunley2016}.
The resulting \ke values obtained from the data analysis described in Sec. IIIA are shown in S-Fig. \ref{S-fig:Ge} with and without a stress correction (see S-Sec. V) applied to the \ke values.
Similar to the $\kappa_{\text{eff}}(w_{\text{e}})$ trend for bulk silicon, we observe a significant increase of the stress corrected \ke for the smallest $w_{\text{e}}$ value exceeding the \ke value from the 2LRT measurement by a factor of $\approx 4.8$. 
However, compared to our results for bulk silicon, the rise of $\kappa_{\text{eff}}(w_{\text{e}})$ towards smaller $w_{\text{e}}$ values seems more abrupt in bulk germanium.
This deviation is caused by differences in the accumulated thermal conductivity \ka of the two materials.
One can expect that the $l_{\text{ph}}$ ranges of the phonons that contribute to the thermal in bulk germanium are different compared to bulk silicon.
Therefore, these findings in bulk germanium substantiate our observation that 1LRT can serve as a novel tool to gather information about the $l_{\text{ph}}$ values of thermal phonons. 

Furthermore, due to the low $h_{\alpha}$ value at $\lambda_{\text{p}} = \SI{532}{\nano\meter}$ in bulk germanium, we potentially have to consider the formation of native oxide on the germanium surface.
This native oxide can have a thickness of a few nanometer ($\approx 2.3$ \SI{}{\nano\meter}) \cite{Nunley2016}, which is not negligible compared to $h_{\alpha}$.
The native oxide can  be expected to be amorphous and therefore exhibit a low $\kappa$ value, which could lead to an overall reduction of the measured \ke values.

%
%
\section{\label{S-sec:k-cum-membranes}Accumulated thermal conductivity trends for a 2000-nm-thick and a 200-nm-thick silicon membrane from ab initio calculations}
%
%

\begin{figure}[h]
    \includegraphics[width=\linewidth]{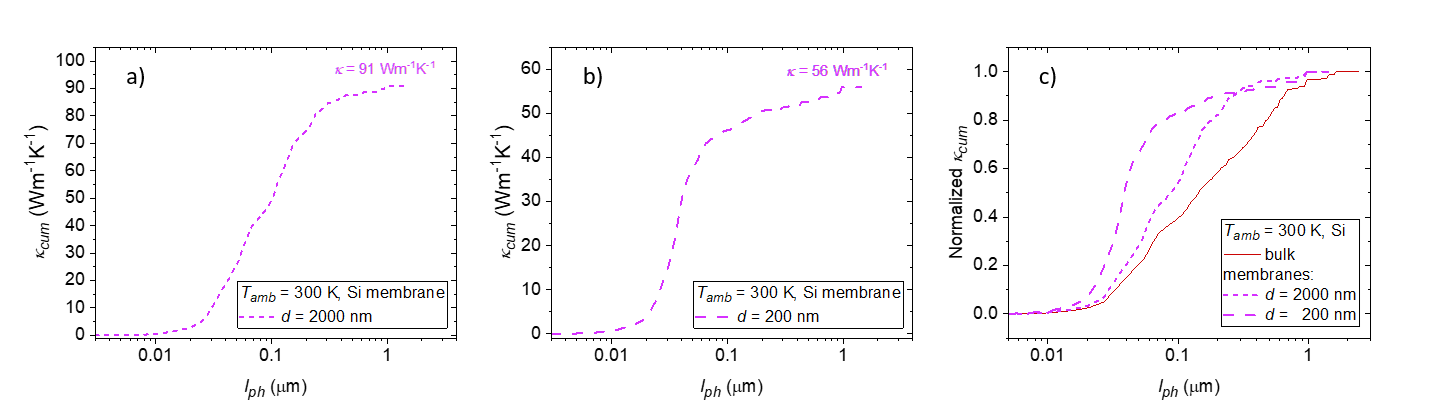}
    
    \caption{\textit{$\kappa_{\text{cum}}(l_{\text{ph}})$ trends from ab initio calculations for different silicon samples at $T_{\text{amb}} = \SI{300}{\kelvin}$.
    a) and b): Results for silicon membranes with a thickness of $d = \SI{200}{\nano\meter}$ (a) and $d = \SI{2000}{\nano\meter}$ (b).
    c): \ka trends from a) and b) compared to the normalized \ka trend of bulk silicon.
    All \ka values were normalized to the corresponding total accumulated $\kappa$ values.}}
    \label{S-fig:BTE-membrane}
\end{figure}

From our \textit{ab initio} calculations we can obtain the accumulated thermal conductivity \ka as a function of the phonon mean free path $l_{\text{ph}}$.
In S-Figs. \ref{S-fig:BTE-membrane}a and b we show these $\kappa_{\text{cum}}(l_{\text{ph}})$ trends for a 2000-nm-thick (a) and a 200-nm-thick (b) silicon membrane at $T_{\text{amb}} = \SI{300}{\kelvin}$.
The total value $\kappa$ that is reached for the convergence of \ka to a constant value, yields \SI{91}{\watt\per\meter\per\kelvin} (S-Fig. \ref{S-fig:BTE-membrane}a) and \SI{56}{\watt\per\meter\per\kelvin} (S-Fig. \ref{S-fig:BTE-membrane}b) for $d = \SI{2000}{\nano\meter}$ and $d = \SI{200}{\nano\meter}$, respectively.
S-Fig. \ref{S-fig:BTE-membrane}c compares the $\kappa_{\text{cum}}(l_{\text{ph}})$ trends of the two silicon membranes with the $\kappa_{\text{cum}}(l_{\text{ph}})$ trend of bulk silicon.
All $\kappa_{\text{cum}}(l_{\text{ph}})$ trends were normalized to their corresponding total $\kappa$ values.

%
%
\section{\label{S-sec:spot-sizes}Methods to determine the laser spot size}
%
%

All laser spot sizes $w_e$ used to measure the $\kappa_{\text{eff}}(w_{\text{e}})$ trends for bulk silicon and germanium were measured with the knife-edge method.
The resulting $w_{\text{e}}$ values are summarized in S-Tab. \ref{S-tab:ke-summary-bulk}.
In our case the free-standing edge of a photonic membrane grown on a silicon substrate is used as the knife-edge due to its outstanding quality (roughness in the sub-5-nm-regime).
In Ref. \cite{Elhajhasan2023} details about this particular sample and the scanning process across the edge can be found.
In this work we use the intensity profile of the silicon Raman mode across the sample edge rather than the photoluminescence signal of a photonic membrane \cite{Elhajhasan2023}.
From the spatial intensity profile, the laser spot size can be extracted by fitting a gaussian-broadened step function to the data.

The $w_{\text{e}}$ values used for the COMSOL Multiphysics\textsuperscript{\circledR} simulation described in S-Sec. \ref{S-sec:COMSOL-membrane} and Sec. III of the main text are listed in Tab. I of the main text.
Here, the $w_{\text{e}}$ value for $\lambda = \SI{325}{\nano\meter}$ was again measured with the knife-edge method.
For $\lambda = $532, 561 and \SI{660}{\nano\meter}, we measured the laser beam radius $w_{\text{e,in}}$ in front of the microscope objective and calculated the focal $w_{\text{e}}$ values with the following formula:
\begin{equation}
    w_{\text{e}} = \frac{ \lambda \cdot f}{\pi \cdot w_{\text{e,in}}}
    \label{s-eq:w_e-from-w_in}
\end{equation}
Here, $f$ is the effective focal length of the microscope objective in use at a given $\lambda$ value, which is known from the supplier.
We found that our calculated values of $w_{\text{e}}$ are in good agreement with previous spot size measurements performed in our labs.
Nevertheless, for our HeCd laser we measured $w_{\text{e}}$, because this laser is a multi-mode laser, resulting in a non-ideal laser beam profile in contrast to our single-mode lasers ($\lambda =$ 532, 561, and \SI{660}{\nano\meter}).

%
%
\section{\label{S-sec:k-values}Overview on effective thermal conductivity values obtained from one-laser Raman thermometry measurements}
%
%
In Tab. \ref{S-tab:ke-summary-bulk} we show a summary of the \ke values (with and without considering heat-induced stress) that were obtained from 1LRT measurements on bulk silicon at $T_{\text{amb}} = \SI{293}{\kelvin}$ and $\SI{200}{\kelvin}$ under the variation of the laser spot size $w_{\text{e}}$.
The extraction of the \ke and the procedure to correct theses values for the effect of heat-induced stress are described in Sec. III of the main work and S-Sec. V of the present Supplemental Material, respectively.

Tab. \ref{s-tab:ke-summary-membranes} summarizes the \ke values obtained from 1LRT measurements at $T_{\text{amb}} = \SI{293}{\kelvin}$ on silicon membranes when $h_{\alpha}$ is varied by tuning $\lambda$.
For any fixed $\lambda$ value, the corresponding $h_{\alpha}$ value is calculated based on Ref. \cite{Franta2019}. We wish to note that when the nominal $h_{\alpha}$ value exceeds $d$, the light penetration depth is limited by the membrane thickness.

    \begin{table}[h]
    \caption{\label{S-tab:ke-summary-bulk} \textit{Summary of $\kappa_{\text{eff}}$ values obtained from 1LRT measurements on bulk silicon with $\lambda_{\text{h}} = \lambda_{\text{p}} = \SI{532}{\nano\meter}$, $h_{\alpha} = \SI{1.1}{\micro\meter}$, and varying $w_{\text{e}}$ values.
    The sub- and superscripts of the \ke values indicate the error.}}
    \begin{ruledtabular}
     \begin{tabular}{l l l l l l l l} 
        $w_{\text{e}}$ [\SI{}{\micro\meter}] & 0.34(1) & 0.65(9) & 2.5(1) & 4.03(15) & 5.0(1) & 9.6(5) & 16.7(4)\\
        \hline
        $T_{\text{amb}}$ [K] & \multicolumn{7}{c}{$\kappa_{\text{eff}}(w_{\text{e}})$ [\SI{}{\watt\per\meter\per\kelvin}] considering the effect of heat-induced stress}\\
        \hline
        293 &   728$_{-26}^{+28}$ & 435$_{-15}^{+16}$ & 255$_{-16}^{+18}$ & 197$_{-11}^{+12}$ & 168$_{-2}^{+2}$ & 177$_{-8}^{+9}$ & 138$_{-7}^{+8}$   \\
        200 &   1385$_{-49}^{+53}$ & 1261$_{-220}^{+338}$ & 553$_{-49}^{+60}$ & 424$_{-8}^{+8}$ & 375$_{-10}^{+11}$ & 214$_{-8}^{+8}$ & 166$_{-9}^{+10}$ \\
        \hline
         $T_{\text{amb}}$ [K] & \multicolumn{7}{c}{$\kappa_{\text{eff}}(w_{\text{e}})$ [\SI{}{\watt\per\meter\per\kelvin}] without any additional correction}\\
        \hline
        293 &   826$_{-30}^{+32}$ & 472$_{-16}^{+17}$ & 272$_{-17}^{+19}$ & 208$_{-12}^{+13}$ & 175$_{-2}^{+2}$ & 185$_{-8}^{+9}$ & 143$_{-7}^{+8}$   \\
        200 &   1544$_{-55}^{+59}$ & 1422$_{-248}^{+381}$ & 591$_{-52}^{+64}$ & 447$_{-8}^{+8}$ & 394$_{-10}^{+11}$ & 220$_{-8}^{+9}$ & 170$_{-9}^{+10}$ \\

    \end{tabular}
    \end{ruledtabular}
    \end{table}

    \begin{table}[h]
    \caption{\label{s-tab:ke-summary-membranes}\textit{Summary of $\kappa_{\text{eff}}$ values obtained from 1LRT measurements on silicon membranes of different thickness $d$ for $T_{\text{amb}} = \SI{293}{\kelvin}$ and varying $h_{\alpha}$ values.
    The sub- and superscripts of the \ke values indicate the error.}}
    \begin{ruledtabular}
     \begin{tabular}{l l l l l} 
        $\lambda$ [\SI{}{\nano\meter}] & 325 & 532 & 561 & 660\\
        $h_{\alpha}$ [\SI{}{\nano\meter}] & 10 & 1100 & 1500 & 4200 \\
        \hline
        $d$ [nm] & \multicolumn{4}{c}{$\kappa_{\text{eff}}(h_{\alpha})$ [\SI{}{\watt\per\meter\per\kelvin}]} \\
        \hline
        2000 &   283$_{-21}^{+24}$ & 271$_{-4}^{+4}$ & 150$_{-9}^{+10}$ & 126$_{-4}^{+4}$ \\
        200 &   26$_{-1}^{+2}$ & 10.0$_{-0.1}^{+0.1}$ & 8$_{-1}^{+1}$ & 27$_{-1}^{+1}$  \\

    \end{tabular}
    \end{ruledtabular}
    \end{table}

\newpage
\bibliography{Bibliography}